\documentclass[authoryear,3p]{elsarticle}

\usepackage{latexsym,euscript,exscale,epsfig,graphicx}
\usepackage{amsmath,amssymb}
\usepackage{subfigure}
\usepackage{natbib}

\def\bfn{{\bf n}}

\def\bfu{{\bf u}}

\def\bfx{{\bf x}}

\def\bfC{{\mathbb{C}}}

\def\bfE{{\mathbb{E}}}

\def\bfJ{{\mathbb{J}}}

\def\bfK{{\mathbb{K}}}

\def\bfL{{\mathbb{L}}}
\def\bfM{{\mathbb{M}}}

\def\bfQ{{\bf Q}}
\def\bft{{\bf t}}

\def\bfe{\mbox{\boldmath $\varepsilon$}}
\def\bfg{\mbox{\boldmath $\gamma$}}

\def\bfrho{\mbox{\boldmath $\rho$}}
\def\bfeta{\mbox{\boldmath $\eta$}}
\def\bfs{\mbox{\boldmath $\sigma$}}
\def\bftau{\mbox{\boldmath $\tau$}}

\def\cbfs{\mbox{\boldmath $\check{\sigma}$}}
\def\cbfe{\mbox{\boldmath $\check{\varepsilon}$}}
\def\hbfs{\mbox{\boldmath $\hat{\sigma}$}}
\def\hbfe{\mbox{\boldmath $\hat{\varepsilon}$}}

\def\eps{\varepsilon}

\def\pr{\parallel}
\def\pp{\perp}

\def\eeq{\varepsilon_{e}}
\def\oe{\overline{\varepsilon}}
\def\oeeq{\overline{\varepsilon}_{e}}

\def\e0{\varepsilon_0}
\def\obfe{\overline{\bfe}}

\def\ceeq{\check{\varepsilon}_{e}}

\def\obfs{\overline{\bfs}}

\def\os{\overline{\sigma}}
\def\seq{\sigma_{e}}
\def\oseq{\overline{\sigma}_{e}}

\def\s0{\sigma_0}
\def\sef{\widetilde{\sigma}_0}
\def\hspa{\hat{\sigma}_{\parallel}}
\def\hspe{\hat{\sigma}_{\perp}}
\def\hseq{\hat{\sigma}_{e}}
\def\cseq{\check{\sigma}_{e}}
\def\r{^{(r)}}
\def\l0{\lambda}
\def\m0{\mu}
\def\bfLo{\bfL}
\def\bfMo{\bfM}

\def\Weff{\widetilde{w}}
\def\Ueff{\widetilde{u}}
\def\Meff{\widetilde{\bfM}}
\def\Leff{\widetilde{\bfL}}
\def\etaeff{\widetilde{\bfeta}}
\def\taueff{\widetilde{\bftau}}
\def\leff{\widetilde{\lambda}}
\def\meff{\widetilde{\mu}}
\def\keff{\widetilde{\kappa}}
\def\feff{\widetilde{g}}
\def\geff{\widetilde{h}}

\def\p{\partial}

\def\hepa{\hat{\varepsilon}_{\parallel}}
\def\hepe{\hat{\varepsilon}_{\perp}}
\def\heeq{\hat{\varepsilon}_e}

\newlength{\StressFieldWidth}
\newlength{\plotWidth}
\newlength{\ScaleWidth}
\newlength{\FieldBoxWidth}
\journal{International Journal of Solids and Structures}

\begin{document}

\begin{frontmatter}

\title{Infinite-contrast periodic composites with strongly nonlinear behavior: effective-medium theory versus full-field simulations}

\author[meam,ep]{M. I. Idiart\corref{cor}} \ead{mii23@cam.ac.uk}
\author[cea,ep]{F. Willot\fnref{willot}}   \ead{francois.willot@ensmp.fr}
\author[cea]{Y.-P. Pellegrini}             \ead{yves-patrick.pellegrini@cea.fr}
\author[meam,ep]{P. Ponte Casta\~neda}     \ead{ponte@seas.upenn.edu}

\address[meam]{Department of Mechanical Engineering and Applied Mechanics, University of Pennsylvania, Philadelphia, PA 19104-6315, U.S.A.}
\address[ep]{Laboratoire de M\'{e}canique des Solides, C.N.R.S. UMR7649, D\'{e}partement de M\'{e}canique, \'{E}cole Polytechnique, 91128 Palaiseau Cedex, France.}
\address[cea]{CEA, DAM, DIF, F-91297 Arpajon, France.}

\cortext[cor]{Corresponding author. Now at: \'{A}rea Departamental Aeron\'{a}utica, Facultad de Ingenier\'{i}a - Universidad Nacional de La Plata, Calles 1 y 47, La Plata B1900TAG, Argentina. Tel.: +54 221 423 6679; fax: +54 221 423 6678.}
\fntext[willot]{Now at: MINES Paristech, CMM- Centre de Morphologie Math\'ematique, Math\'ematiques et Syst\`emes, 35 rue Saint Honor\'e, F-77300 Fontainebleau, France.}

\begin{abstract}
This paper presents a combined numerical-theoretical study of the macroscopic behavior and local field distributions in a special class of two-dimensional periodic composites with viscoplastic phases. The emphasis is on strongly nonlinear materials containing pores or rigid inclusions. Full-field numerical simulations are carried out using a Fast-Fourier Transform algorithm [H. Moulinec, P. Suquet, C. R. Acad. Sci. Paris II 318, 1417 (1994)], while the theoretical results are obtained by means of the `second-order' nonlinear homogenization method [P. Ponte Casta\~neda, J. Mech. Phys. Solids 50, 737 (2002)]. The effect of nonlinearity and inclusion concentration is investigated in the context of power-law (with strain-rate sensitivity $m$) behavior for the matrix phase under in-plane shear loadings. Overall, the `second-order' estimates are found to be in good agreement with the numerical simulations, with the best agreement for the rigidly reinforced materials. For the porous systems, as the nonlinearity increases ($m$ decreases), the strain field is found to localize along shear bands passing through the voids (the strain fluctuations becoming unbounded) and the effective stress exhibits a singular behavior in the dilute limit. More specifically, for small porosities and fixed nonlinearity $m>0$, the effective stress decreases linearly with increasing porosity. However, for ideally plastic behavior ($m = 0$), the dependence on porosity becomes non-analytic. On the other hand, for rigidly-reinforced composites, the strain field adopts a tile pattern with bounded strain fluctuations, and no singular behavior is observed (to leading order) in the dilute limit.
\end{abstract}

\begin{keyword}
Porous media \sep Reinforced composites \sep Viscoplasticity \sep Homogenization
\end{keyword}

\end{frontmatter}

\section{Introduction}
\label{sec:INTRO}

The overall (visco-) plastic flow of composites exhibiting strong constitutive nonlinearities and large heterogeneity contrasts is particularly difficult to model due to the tendency of the strain (-rate) field to localize in shear bands running through the composite along `minimal paths' \citep{Drucker66, Duxbury06}. As full-field numerical simulations of such material systems demand significant computational power, especially when the microstructure is random, nonlinear effective-medium theories, also known as homogenization methods, are particularly attractive modeling tools. Most of the available theories rely on the use of a `linear comparison medium' of one sort or another (see, for instance, reviews by \citet{PCS98,Willis00}). Two approximations are typically involved in this class of theories. The first one consists in linearizing the nonlinear behavior of the individual constituents following some appropriate linearization procedure, and the second one consists in estimating the overall response of the resulting linear comparison medium. For example, \cite{Hill65} proposed an incremental procedure making use of the tangent moduli of the phases evaluated at the average field in the phases and of the self-consistent linear homogenization method to estimate the macroscopic response of elastoplastic polycrystals, whereas \cite{Hutch76} proposed the use of the secant modulus, also evaluated at the phase averages, together with the self-consistent method to estimate the response of viscoplastic polycrystals. The linearization schemes in both of these approaches were physically plausible, but {\it ad hoc}. \cite{PC92} introduced new variational principles where the trial field was the moduli of a suitably chosen `linear comparison composite.' By choosing constant moduli in the phases, this procedure leads to improved analytical estimates \citep{PC91} for nonlinear composites, and as shown by \cite{Suquet95}, the resulting linearization can be interpreted in terms of the secant moduli of the phases evaluated at the second moments of the fields in the phases. A more general procedure, making use of a `linear thermoelastic composite' and having the property of delivering estimates that are exact to second order in the contrast, was developed by \cite{PC02a}. In addition, a related procedure, making use of a similar linearization, but exploiting instead the properties of an assumed Gaussian distributions for the fields, was advanced by \cite{Pell01}.

Recent comparisons with numerical simulations for two-phase materials \citep{Idiart06b,Rekik07, Idiart07b} and polycrystals \citep{Lebensohn04a,Lebensohn04b,Lebensohn07} indicate that, within that class of `linear comparison' theories, the so-called `second-order' theory of \citet{PC02a} delivers the most accurate predictions for the macroscopic response as well as for low-order statistics of the local fields. This theory has been found to capture some of the essential features associated with strain localization. For instance, it correctly predicts the anisotropic and unbounded character of the strain fluctuations in ideally plastic, particle-reinforced {\it random} composites \citep{Idiart06b,Idiart07b}, and the non-analytic dependence on porosity of the yield stress of {\it random} porous media with vanishingly small porosities \citep{PC02a,Idiart07b}.

The `second-order' theory makes use of a linear comparison composite (LCC) with anisotropic phases, where the direction and strength of anisotropy is determined by the means and variances of the mechanical fields in each phase, and the constitutive nonlinearity. This work is aimed at understanding how the `second-order' theory --- and `linear comparison' theories more generally --- copes with strain localization in nonlinear composites by exploiting the constitutive anisotropy in the underlying LCC. To that end, this work focuses on a special class of \textit{two}-dimensional \textit{periodic} composites for which numerical simulations are simple enough so that a thorough and systematic comparison is feasible. The paper builds on the recent works of \cite{Willot08a,Willot08b}, which provided similar analyses and comparisons for strongly anisotropic linear elastic composites. As will be seen below, the anisotropy of the LCC in the context of the second-order estimates for the nonlinear composites in some sense plays the role of the nonlinearity, with strong nonlinearity leading to strong anisotropy in the LCC. The full-field numerical simulations are carried out using the Fast-Fourier Transform (FFT) algorithm of \citet{Moulinec94} in the form recently proposed by \citet{Willot07}. We focus on porous materials and rigidly-reinforced composites, which are the cases of extreme heterogeneity contrast, and therefore good case studies. In addition, particular attention will be given to the dilute limits, which serve to illustrate the significant differences in behavior between these two extreme cases.

\section{Preliminaries on periodic composites}
\label{sec:EFF}

\subsection{Local and effective behavior}
\label{sec:local}

\begin{figure}[t]
    \begin{center}
    \includegraphics*[width=2.5in]{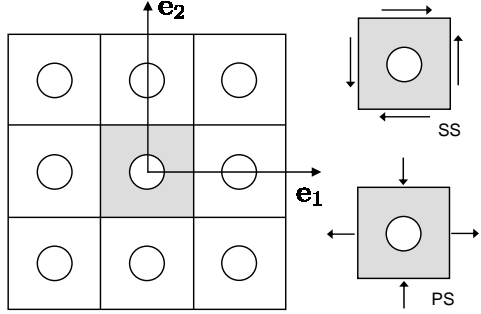}
  \end{center}
  \caption{Microstructure of the periodic material with reference axes and unit cell, together with a schematic representation of `simple shear' (SS) and `pure shear' (PS) macroscopic loadings.}
    \label{fig:micro}
\end{figure}

The materials considered are made up of a continuous matrix phase ($r=1$) containing cylindrical inclusions ($r=2$) with circular cross section, aligned with the $x_3$-axis and periodically distributed in the transverse plane $x_1$-$x_2$, as shown in Fig. \ref{fig:micro}. The inclusion phase will be taken as either vacuous or rigid. The constitutive behavior of the matrix phase is taken to be characterized by an \textit{isotropic}, \textit{incompressible} strain potential $w^{(1)}$, such that the stress and strain tensors are related by
\begin{equation}
    \bfs = \frac{\p w^{(1)}}{\p \bfe}(\bfe), \quad w^{(1)}(\bfe) = \phi(\eeq),
    \label{mat_const}
\end{equation}
where the von Mises equivalent strain is defined in terms of the deviatoric part of the strain tensor by $\eeq=\sqrt{(2/3) \bfe_d \cdot \bfe_d}$, and tr$(\bfe)=0$. This constitutive relation can be used within the context of the deformation theory of plasticity, where $\bfe$ and $\bfs$ represent the infinitesimal strain and stress, respectively. Relation (\ref{mat_const}) applies equally well to viscoplastic materials, in which case $\bfe$ and $\bfs$ represent the Eulerian strain rate and Cauchy stress, respectively. A relation completely equivalent to (\ref{EffConst}) results from a dual formulation, which makes use of a stress potential $u^{(1)}$, such that
\begin{equation}
 \bfe = \frac{\p u^{(1)}}{\p \bfs}(\bfs), \quad u^{(1)}(\bfs) = \psi(\seq),
 \label{mat_const_u}
\end{equation}
where the von Mises equivalent stress is given in terms of the deviatoric stress tensor by $\seq = \sqrt{(3/2) \bfs_d\!:\!\bfs_d}$.

Let $\langle\cdot\rangle$ denote the volume average over the composite. The \textit{effective behavior} of the composite is defined as the relation between the average stress $\obfs=\left\langle \bfs \right\rangle$ and the average strain $\obfe=\left\langle \bfe \right\rangle$. When the ratio between the characteristic length scale of the unit cell to that of the specimen tends to zero, there exists (under certain technical assumptions) a \textit{homogenized} behavior, which can be characterized by
\begin{equation}
    \obfs = \frac{\p \Weff}{\p \obfe}(\obfe), \quad \obfe = \frac{\p \Ueff}{\p \obfs}(\obfs),
    \label{EffConst}
\end{equation}
where  $\Weff$ and $\Ueff$ are the \textit{effective strain and stress potentials}, defined by
\begin{equation}
    \Weff (\obfe) = \mathop {\min }\limits_{{\bfe} \in {\cal K}({\obfe})}
                   \langle w(\bfx,\bfe) \rangle, \quad
                    \Ueff (\obfs) = \mathop {\min}\limits_{\scriptstyle {\bfs \in {\mathcal S}(\obfs)}\hfill}\
                   \langle u(\bfx,\bfs) \rangle,
    \label{w-eff}
\end{equation}
with $\mathcal K(\obfe)$ and $\mathcal S(\obfs)$ denoting the set of `kinematically admissible' strain and `statically admissible' fields (see \citet{PCS98} for details). Thus, the problem of finding the effective behavior of the composite reduces to that of computing the effective potential $\Weff$ or $\Ueff$.

For definiteness, it is assumed that the behavior of the matrix follows a power law of the form
\begin{equation}
    \phi(\eeq) = \frac{\s0 \e0}{1+m} \left( \frac{\eeq}{\e0} \right)^{1+m}, \quad \psi(\seq) = \frac{\s0 \e0}{1+n} \left( \frac{\seq}{\s0} \right)^{1+n},
    \label{power_law}
\end{equation}
where $\s0$ and $\e0$ are the flow stress and the reference strain, respectively, $n=1/m$ denotes the nonlinearity, and $m$ denotes the so-called
strain-rate sensitivity, and is such that $0\leq m \leq 1$. This model is particularly appropriate to study a wide range of material behaviors, such as the time-independent plastic deformation of metals as well as their time-dependent viscous deformation (i.e., high temperature creep). The limiting values $m=1$ and $m=0$ correspond to linear and rigid-ideally plastic behaviors, respectively.

We restrict the analysis to \textit{isochoric}, \textit{plane-strain} macroscopic deformations (or purely deviatoric stress). Because of the intrinsic anisotropy of its microstructure, the effective behavior of the composite material will depend on the loading angles $\bar \theta$ and $\bar \psi$ that are defined by the orientation of the principal axes of the deformation and loading, respectively, with the symmetry axes of the microstructure. We will focus on the two ``extreme'' cases of loading orientation: ($i$) `pure shear' (PS) corresponding to the case where the principal axes of $\obfe$ and $\obfs$ coincide with the Cartesian axes introduced in Fig. \ref{fig:micro}, and ($ii$)  `simple shear' (SS) corresponding  to the case where the principal axes of these tensors form $45^\circ$ with respect to the Cartesian axes, as depicted in Fig. \ref{fig:micro}. It then follows from the homogeneity of the potentials (\ref{power_law}) that, for these specific macroscopic loadings, the effective potentials (\ref{EffConst}) can be written as
\begin{equation}
    \Weff(\obfe) = \frac{\sef \e0}{1+m} \left( \frac{\oeeq}{\e0} \right)^{1+m}, \quad
    \Ueff(\obfs) = \frac{\sef \e0}{1+n} \left( \frac{\oseq}{\sef} \right)^{1+n},
    \label{power_law_eff}
\end{equation}
where $\oeeq=(2/\sqrt{3})\sqrt{\oe_{12}^2+(1/4)(\oe_{11}-\oe_{22})^2}$ and $\oseq=\sqrt{3} \sqrt{\os_{12}^2+(1/4)(\os_{11}-\os_{22})^2}$ are the
macroscopic equivalent strain and stress, respectively, and $\sef$ denotes the \textit{effective flow stress} of the material, which depends on the specific macroscopic loading (SS or PS), the strain-rate sensitivity $m$ and the inclusion volume fraction $f$.

\subsection{Statistics of the local fields}
\label{stats}

In addition to the effective behavior, homogenization methods can deliver useful information about the field distributions in the form of low-order statistics such as the first and second moments of the fields in each constituent phase \citep{Idiart07}. For the macroscopic loadings considered here, it is natural to express the deviatoric parts of the local fields as
\begin{equation}
    \bfe_d = \frac{\sqrt{3}}{2} \left( \eps_{\text{PS}}\, \mathsf{e}_{\text{PS}} + \eps_{\text{SS}}\, \mathsf{e}_{\text{SS}} \right), \quad
    \bfs_d = \frac{1}{\sqrt{3}} \left( \sigma_{\text{PS}}\, \mathsf{e}_{\text{PS}} + \sigma_{\text{SS}}\, \mathsf{e}_{\text{SS}} \right),
    \label{components}
\end{equation}
where the unit second-order tensors $\mathsf{e}_{\text{PS}}$ and $\mathsf{e}_{\text{SS}}$ are given in terms of the Cartesian axes of Fig. \ref{fig:micro} by
\begin{equation}
    \mathsf{e}_{\text{SS}} = {\bf e}_1 \otimes {\bf e}_2 + {\bf e}_2 \otimes {\bf e}_1, \quad
    \mathsf{e}_{\text{PS}} = {\bf e}_1 \otimes {\bf e}_1 - {\bf e}_2 \otimes {\bf e}_2.
    \label{modes}
\end{equation}
The prefactors in (\ref{components}) are such that $\eeq^2 = \eps_{\text{SS}}^2 + \eps_{\text{PS}}^2$ and similarly for the stress.
We also introduce the fourth-order projection tensors (see \cite{Willot08a}):
\begin{equation}
    \bfE^{\text{SS}} = \frac{1}{2} \mathsf{e}_{\text{SS}} \otimes \mathsf{e}_{\text{SS}}, \quad  \bfE^{\text{PS}} = \frac{1}{2} \mathsf{e}_{\text{PS}} \otimes \mathsf{e}_{\text{PS}}.
    \label{E-tensors}
\end{equation}
Let $\langle\cdot\rangle\r$ denote the volume average over phase $r$. We are interested in the first moments (phase averages), denoted by $\obfe\r = \langle \bfe \rangle\r$ and $\obfs\r = \langle \bfs \rangle\r$, and the second moments $\langle \bfe \otimes \bfe \rangle\r$ and $\langle \bfs \otimes \bfs \rangle\r$ of the field distributions. A measure of the field fluctuations within each phase is given by the phase covariance tensors
\begin{equation}
    \bfC_{\bfe}\r = \langle \bfe \otimes \bfe \rangle\r - \obfe\r \otimes \obfe\r,
    \label{C_eps}
\end{equation}
and similarly for the stress. For power-law composites like the ones considered here, it can be shown that the local strain and stress fields are homogeneous functions of degree 1 in $\oeeq$ and $\oseq$, respectively. In addition, it follows from symmetry considerations that, for `aligned' loadings in the sense described above, the phase averages of the local fields should be co-axial with the macroscopic averages. Then, the deviatoric part of the phase averages can be written as
\begin{equation}
        \obfe_d\r = \frac{\oeeq\r}{\oeeq} \obfe_d, \quad \obfs_d\r = \frac{\oseq\r}{\oseq} \obfs_d,
        \label{averages}
\end{equation}
where the ratios $\oeeq\r/\oeeq$ and $\oseq\r/\oseq$ depend on the strain-rate sensitivity, the porosity, and the loading direction (PS or SS). In turn, it also follows from symmetry considerations that the corresponding second moments should be `aligned' with the macroscopic averages, in the sense that one of their eigentensors is co-axial with $\obfe$ and $\obfs$. Then, the standard deviations ($SD^{(r)}(\cdot) = \sqrt{\langle (\cdot)^2 \rangle\r - (\langle \cdot \rangle\r)^2}$) of the field components defined in (\ref{components}) are obtained by projecting the covariance tensors (\ref{C_eps}) onto the tensors (\ref{E-tensors}):
\begin{equation}
    SD\r(\eps_{\text{SS}}) = \sqrt{\frac{2}{3} \bfE^{\text{SS}}\!::\!\bfC\r_{\bfe} }, \quad
    SD\r(\sigma_{\text{SS}}) = \sqrt{\frac{2}{3} \bfE^{\text{SS}}\!::\!\bfC\r_{\bfs} },
    \label{Traces}
\end{equation}
and similarly for the PS components. For convenience, when discussing the results, we will refer to the SS and PS components of the strain/stress tensors
as `parallel' and `perpendicular' to the macroscopic strain/stress tensors in the case of SS loadings, and conversely in the case of PS loadings.

\subsection{Duality between porous materials and rigidly-reinforced composites}
\label{sec:dual}

\begin{table}[ht]
\caption{Duality relations between power-law, two-dimensional materials with pores and rigid inclusions. The tensor $\bfQ$ represents a $\pi/4$ counterclockwise rotation in the plane.}
\begin{center}
        \begin{tabular}{|c|cccc|}
        \hline
        \parbox{2cm}{\vspace{3mm} Porous \vspace{3mm}}           & $1/m$ & $\sef^{-1/m}$ & $\bfQ^T\cdot \bfe\cdot \bfQ$ & $\bfQ^T\cdot\bfs\cdot \bfQ$ \\
        \hline
        \parbox{2cm}{\vspace{1mm} Rigidly- \\ reinforced \vspace{1mm}} & $m$   & $\sef$      & $\bfs$                       & $\bfe$ \\
        \hline
        \end{tabular}
\end{center}
\label{table:duality}
\end{table}

The methods to be described in Sections \ref{sec:HOM} and \ref{sec:FFT} are general and can be directly applied to porous materials and rigidly-reinforced composites. In this work, however, we exploit the duality relations of \citet{Francfort01} for incompressible, nonlinear, two-dimensional elasticity (see Proposition 5.3 in that Ref.), which allow us to obtain results for rigidly-reinforced materials directly from corresponding results for porous materials. In the case of a power-law matrix, results for a rigidly-reinforced composite with a strain-rate sensitivity $m$ can be obtained from corresponding results for a porous material with strain-rate sensitivity $1/m$ using the relations given in Table \ref{table:duality}. Note that we are therefore required to solve for porous materials with strain-rate sensitivities larger than 1. Thus, the methods need to be implemented for porous materials only.

\subsection{Second-order variational method}
\label{sec:HOM}

The so-called `second-order' method is a fairly general method for estimating the effective potentials of nonlinear composites, introduced by \citet{PC02a}, which delivers estimates that are exact to second order in the heterogeneity contrast. The central idea of this method is to introduce a \textit{linear comparison composite} (LCC), with the same microstructure as the nonlinear composite, whose constituent phases are identified with appropriate linearizations of the given nonlinear phases, determined through a variational procedure. This allows the use of the many different methods already available to estimate the effective potentials of \textit{linear} composites to generate corresponding estimates for the effective potentials of \textit{nonlinear} composites. Estimates for the effective behavior then follow from relations (\ref{EffConst}), and estimates for the field statistics follow from similar identities \citep{Pellegrini01,Idiart07}.

The `second-order' method makes use of a LCC with local elasticity tensors given by generalized secant moduli that are intermediate between the standard secant and tangent moduli of the nonlinear phases, employed in earlier methods. In general, these elasticity tensors are \textit{anisotropic}, even if the nonlinear phases are isotropic. The direction and strength of anisotropy are determined by the means and variances of the mechanical fields in the LCC, and the constitutive nonlinearity. For the particular class of material systems and loading conditions considered in this work, the matrix phase in the LCC is incompressible and characterized by two different shear moduli $\lambda$ and $\mu$, one for each of the shear deformation modes defined in (\ref{modes}). The degree of anisotropy is thus given by the ratio $k=\lambda/\mu$ between the two moduli, with the extreme values $k=1$ and $k=0$ corresponding to isotropic and strongly anisotropic phases, respectively. Linear materials of this type have been studied in detail in \cite{Willot08a}. The linear estimates of the Hashin-Shtrikman type derived in that paper were found to be very accurate, relative to FFT simulations,  for all values of $k$ and inclusion concentrations in the range $0\leq f \leq 0.3$. Those estimates are used here to determine the effective behavior of the LCC.

The emphasis in this paper, however, is on the results rather than the methodology. For this reason, the details of the  `second-order' method, which has already been discussed in some detail in the context of composites with random microstructures \citep{PC02a,Idiart07}, are summarized together with the resulting estimates in Appendix \ref{app:expr}. In view of the high accuracy of the linear estimates utilized, any discrepancies between the `second-order' estimates and the numerical simulations observed in Sections \ref{sec:porous} and \ref{sec:rigid} below, in the range $0\leq f \leq 0.3$, can be mostly attributed to the error introduced in the linearization.

Finally, it is noted that the strain and stress formulations of the `second-order' method --based on linearizations of (\ref{mat_const}) and (\ref{mat_const_u}), respectively -- deliver different predictions: the estimates exhibit a \textit{duality gap} (see \citet{PC02a,Idiart06a}). Thus, two different sets of `second-order' estimates will be reported, based on the strain (W) and stress (U) formulations.

\subsection{Numerical Fast Fourier Transform method}
\label{sec:FFT}

The FFT approach used in this work is well-known \citep{Michel01,Moulinec03,Moulinec04}. For this reason, we shall limit ourselves here to a few remarks. The need to deal with cases of infinite contrast and strongly nonlinear behavior (small or zero nonlinearity exponent, $m\ll 1$) prompted us, mainly for convenience, to consider the modified version of the augmented Lagrangian method \citep{Michel01} proposed by \cite{Willot07}. 

The $L\times L$ pixel medium is discretized with resolutions $L=256$, $512$ and $1024$. The incompressibility of the material is approached by a bulk compressibility modulus $\kappa=1000$, with elastic shear modulus $\mu = 1$. The plastic yield stress was $\sigma_0=1$. Loading is carried out by increasing the driving overall deformation in steps of approximately $\Delta\langle\bfe_\parallel\rangle = 0.1$, up to $\langle\bfe_\parallel\rangle=2$ to $4$. Such high values were found necessary to obtain reliable averages and second moments of the strain. The quantity $\eta^2 \equiv \langle||\nabla \cdot \bfs||^2\rangle/\obfs\!:\!\obfs \to 0$ is used to monitor convergence at each loading step, where iterations are stopped for $\eta$ of order $10^{-4}$ (or even $10^{-5}$ for the smallest inclusion volume fractions in the stiffest cases) and/or when no notable progress occurs in the decrease of $\eta$. In the process, records are made of the iteration number $n$, and of the fields of interest, $x^{(n)}$. From three such records (including the most converged one), theoretical converged limits $x^{(\infty)}$ are estimated by a power-law extrapolation of the type $x^{(n)}(L) = x^{(\infty)}(L) + \alpha(L) n^{-\beta(L)}$ from which $\alpha$, $\beta$, and $x^{(\infty)}$ are determined. A similar extrapolation procedure on $L$ is then used to deduce the infinite-resolution limit $x^{(\infty)}(L\to\infty)$ from the converged results at the three resolutions.
      
The modified version of the method alluded to above, detailed in \cite{Willot07}, consists in replacing the continuum Green function by a discrete counterpart. The discretization scheme enforces compatibility and equilibrium at each pixel, consistently with the system discreteness. Probably for this reason, the modified algorithm was observed to enjoy faster convergence rates, especially for the stiffest cases\footnote{For instance, with the above convergence criterion, and in the more demanding situation of a pixelwise disordered medium, as was considered in \cite{Willot07}, the number of iterations to convergence is roughly proportional to $\eta^{-0.7}$ with the discretized Green function, and to $\eta^{-1}$ with the continuous one.}. No specific investigation of convergence properties was carried out in the present work, but the typical number of iterations needed here per loading step is of order $10^3$, except when $m\ll 1$ where it can climb up to $10^6$. It is noted however that in terms of CPU, solving problems with $m>0$ is much more time-consuming than with $m=0$. We refer the reader to Eq. (42) in \cite{Michel01}, which has no analytical solution for $m>0$. As a consequence, it has to be solved numerically at each material point and each iteration, whereas an analytical solution is directly implemented for perfectly
plastic materials.

A few cases were checked by comparing to calculations using the continuous Green function. Quite generally, excellent agreement was found between both methods \textit{save in one remarkable case}: in the perfectly-plastic limit with rigid inclusions, markedly different solutions are obtained. This difference in behavior presumably originates from the short-range spatially-dispersive (``smoothing") features of the discretized Green function \citep{Willot07}, though the precise mechanism at play remains to been investigated. More importantly, the solutions from the modified approach turn out to be the most relevant ones to the present theoretical study. This point is further discussed in Sec.\ \ref{sec:rigidfieldmaps}.

\section{Results for porous materials}
\label{sec:porous}

\subsection{Field maps}

Full-field distributions in porous materials under pure shear (PS) and simple shear (SS) loadings have been obtained by the FFT method described in Sec. \ref{sec:FFT}. For porosities $f \lesssim 0.3$, the simulations show that the reduced displacement $\bfu^{*} = \bfu - \obfe\cdot\bfx$ within the unit cell organizes in a periodic array of counter-rotating vortices or convection cells. For the case of a linear matrix ($m=1$), the patterns can be found in Table I of \cite{Willot08a} (maps B and E). As the nonlinearity increases, the patterns remain similar in character but exhibit stronger displacement gradients, and therefore more localized strains.

\begin{table*}[!ht]
	\includegraphics*[width=16.4cm]{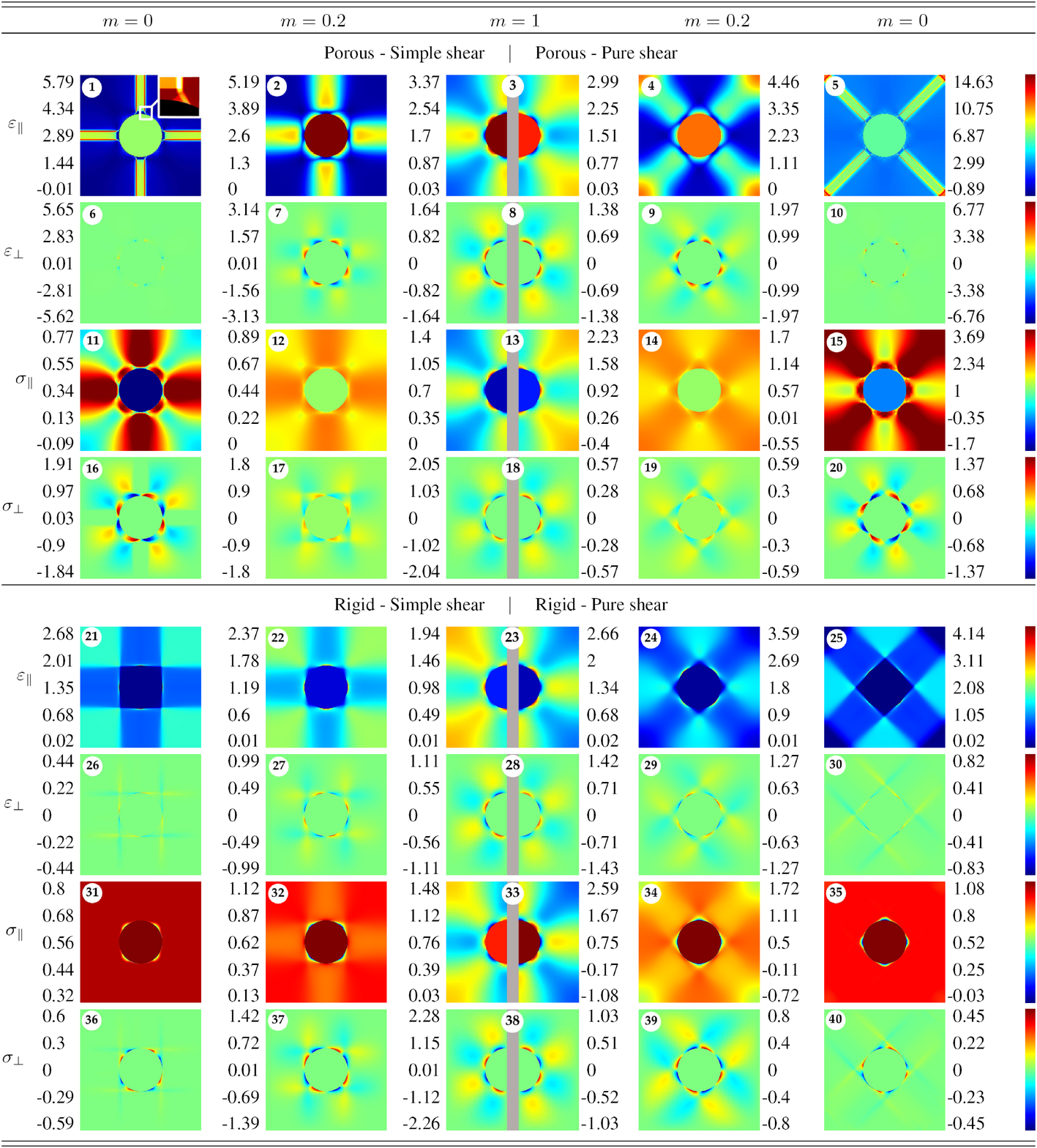}
	\caption{Maps of the `parallel' ($\pr$) and `perpendicular' ($\perp$) components of the shear strain and stress fields, in porous and rigidly-reinforced power-law materials under simple shear and pure shear loadings, for $f=0.1$ and several values of the strain-rate sensitivity ($m=0$, $0.2$, $1$). The strain and stress fields are normalized by the equivalent macroscopic strain $\oeeq$ and stress $\oseq$, respectively.}
	\label{table:maps}
\end{table*}

The corresponding strain and stress distributions are shown in Table \ref{table:maps}, respectively normalized by the equivalent macroscopic strain $\oeeq$ and stress $\oseq$. Maps of the `parallel' ($\pr$) and `perpendicular' ($\perp$) components (see end of Section \ref{stats} for definition) are displayed for $f=0.1$ and several values of the strain-rate sensitivity. Each map is labelled by a number (1-20), and goes along its own field scale on its right, in correspondance with the color scale at the extreme right of the rows. As anticipated above, the distributions of both components of the strain (maps 1-10) exhibit smooth variations in the linear case but become progressively more localized in bands as the nonlinearity increases. Across such bands, the tangential component of the displacement field varies significantly. More precisely, if the vectors $\bft$ and $\bfn$ denote, respectively, the directions tangential and normal to the band, then $\eps_{tn}$ increases with decreasing band width, but not $(\eps_{nn} - \eps_{tt})/2$. Indeed, the maps show that the deformation tends to localize in bands running across the specimen along directions that are parallel to those of maximum macroscopic shear: $0^\circ - 90^\circ$ and $\pm 45^\circ$ with respect to the reference axes of Fig. \ref{fig:micro}, under SS (maps 2 and 7) and PS (maps 4 and 9) loading, respectively. As will be seen in the next subsection, the fact that these bands have some preferential orientation dictated by the macroscopic loading leads to a strong anisotropy of the strain fluctuations with increasing nonlinearity.

In the strongly nonlinear ideally plastic limit ($m \rightarrow 0$), the variation of the tangential displacement across the localization bands becomes discontinuous, and consequently the `parallel' strain becomes unbounded along well-defined straight shear bands (maps 1 and 5). The paths followed by these shear bands are such that the macroscopic (dissipation) strain energy is minimal \citep{Duxbury06}.
However, equilibrium in ideally plastic materials requires that shear bands meet free boundaries at $\pm 45^\circ$ with respect to the normal to the boundary \citep{Kachanov04}. Consequently, the bands have to bend in regions sufficiently close to the boundary of the pores (see insert in map 1), which causes the `perpendicular' component of the strain (maps 6 and 10) to become very large in those regions\footnote{In the simulations, the strain values within the shear bands are large but finite; however, these values are found to increase with increasing mesh resolution, which indicates that the exact solution involves unbounded strains.}. A closer look at the shear bands reveals a certain structure of finite width set by the diameter of the pores, but further work is required to estimate this width. Interestingly, while in the SS case the two orthogonal families of bands intersect each other `inside' the pores (map 1), in the PS case they do so at the corners of the unit cell (map 5). These are points where the `parallel' strain takes the highest values. In fact, these are points in the matrix phase where four displacement vortices meet, leading to large and highly localized displacement gradients, which are strongly enhanced by the constitutive nonlinearity.

On the other hand, the stress field (maps 11-20) exhibits smooth variations for all $m$. In the ideally plastic limit, the `parallel' component of the stress (maps 11, 15) takes the highest values in the regions where the above-mentioned shear bands develop, as expected.

\begin{table}[!ht]
	\includegraphics*[width=16.4cm]{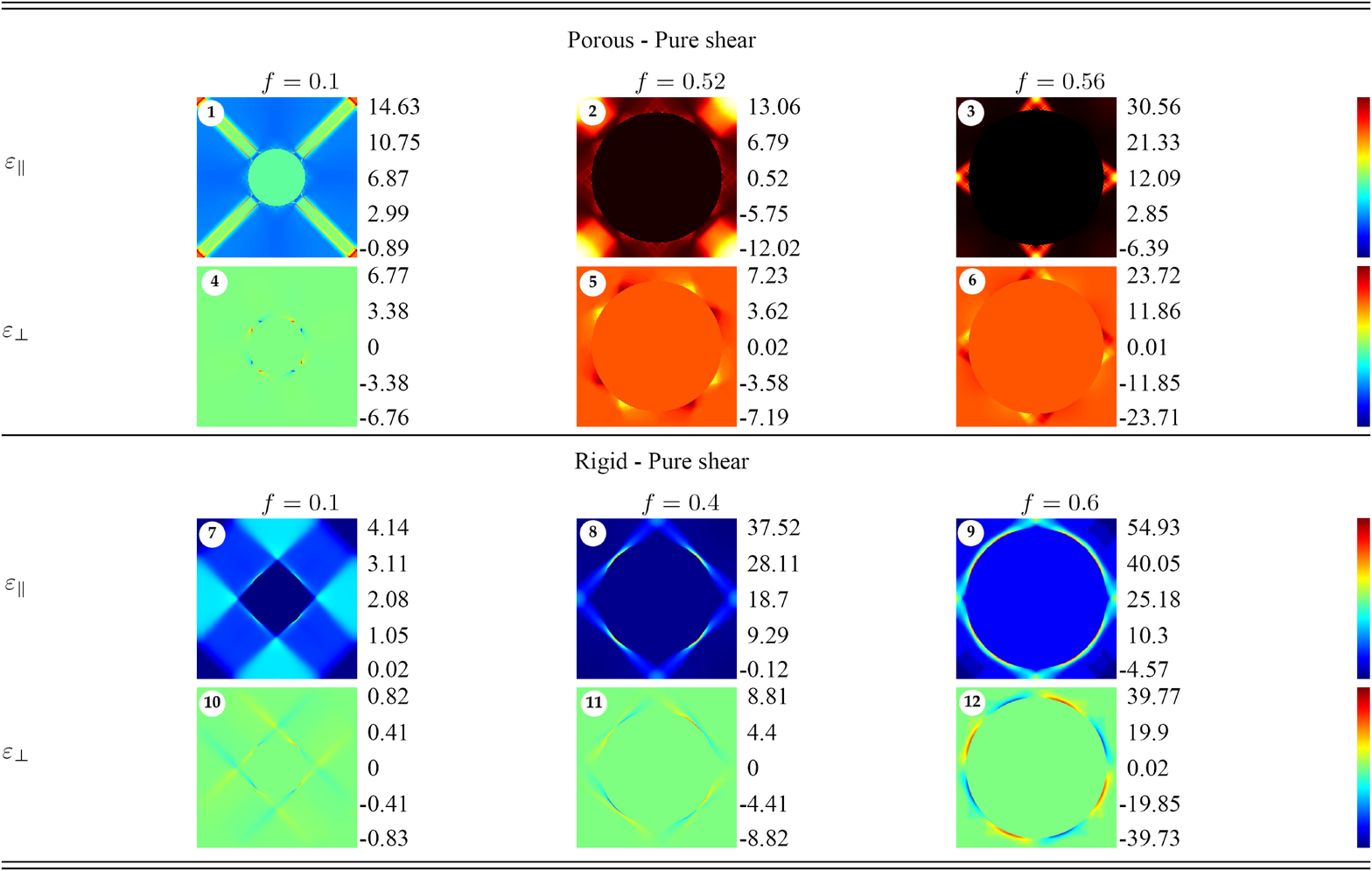}
	\caption{Change of shear band patterns with concentration $f$ in porous and rigidly-reinforced ideally plastic materials under pure shear (PS) loading. The strain field is normalized by the equivalent macroscopic strain $\oeeq$.}
	\label{table:maps-f}
\end{table}

Finally, it is noted that the `minimal paths' followed by shear bands in PS change their configuration for porosities $f > f_c \approx 0.52$, see maps 1-6 in Table \ref{table:maps-f}. While for $f<f_c$ the two families of shear bands intersect at the corners of the unit cell, for $f>f_c$ they intersect at the mid-edge points of the unit cell. This is because the (dissipation) strain energy, and consequently the total length of the shear bands, must go to zero as $f$ approaches the close-packing threshold $f_{cp}$. Only the latter configuration satisfies this requirement. As will be seen in the next subsection, this change in shear band pattern is manifested at the macroscopic level as an inflexion point (or kink) in the effective flow stress for PS as a function of $f$ (see Fig. \ref{fig:ideal-porous}a).

\subsection{Effective behavior and field statistics}

\begin{figure*}[t]
\centerline{
\begin{tabular}{c@{\hspace{2mm}}c@{\hspace{2mm}}c}
  \vspace{-2mm}
  \includegraphics*[width=\plotWidth]{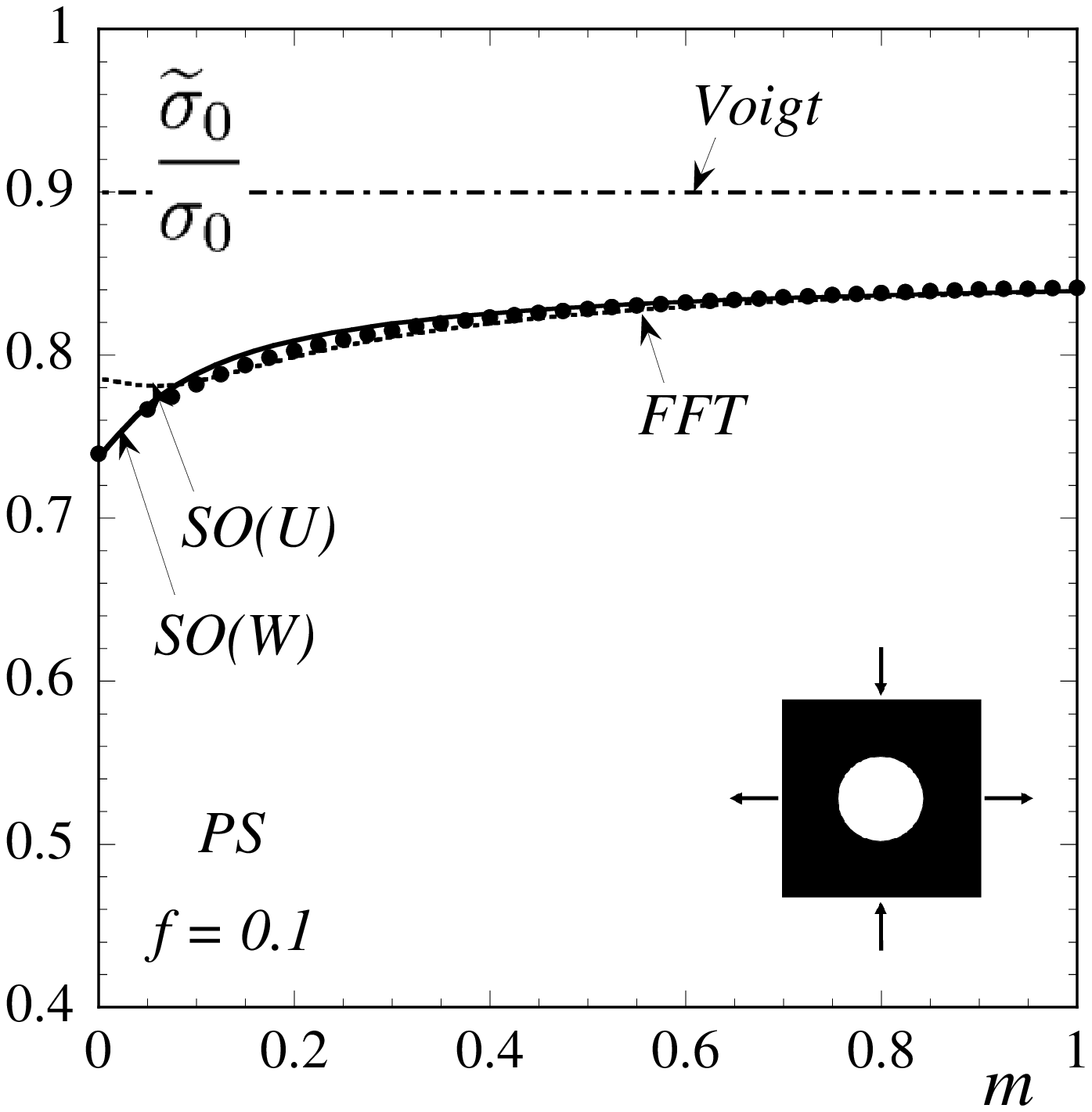} & \includegraphics*[width=\plotWidth]{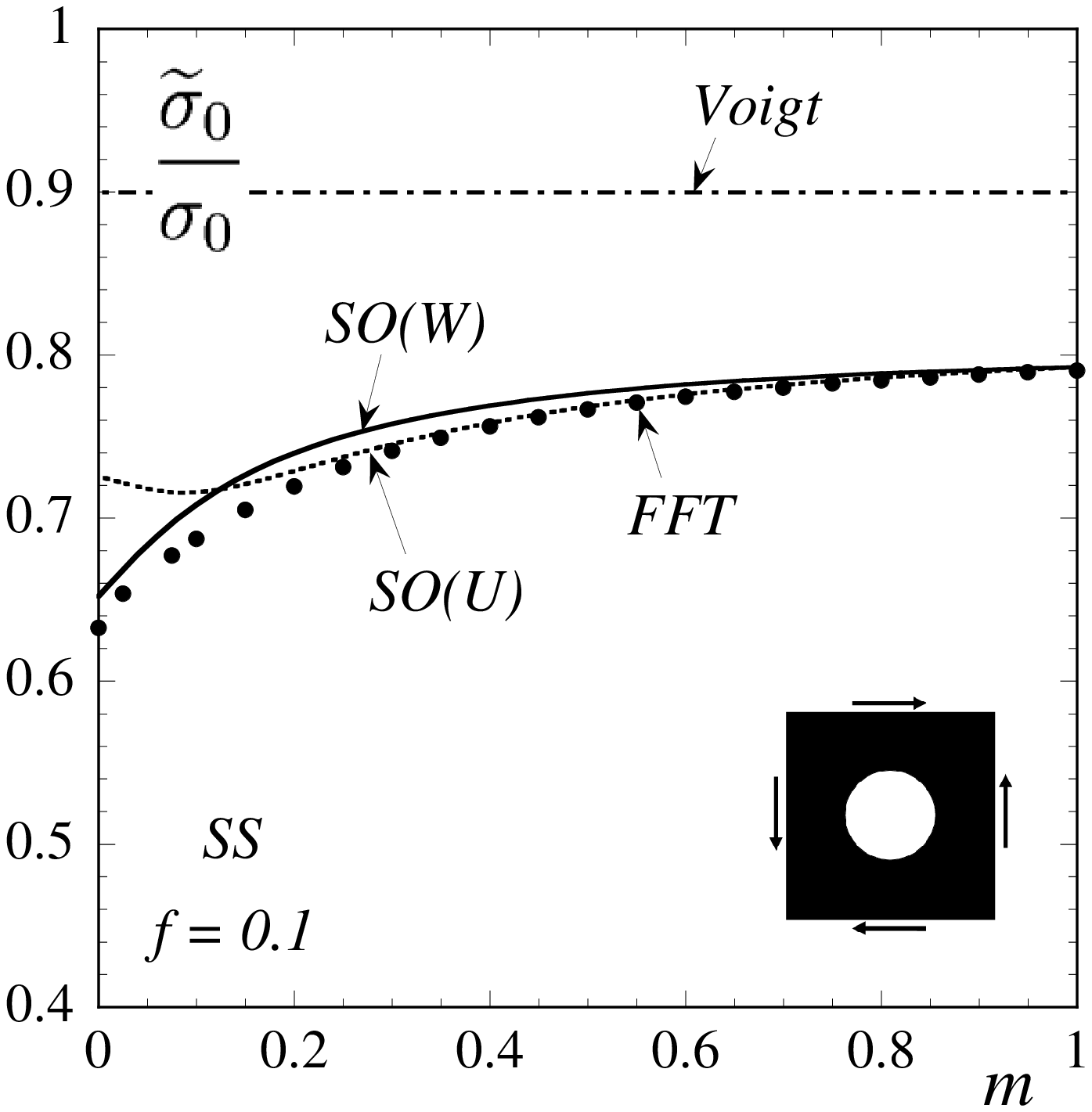} & \includegraphics*[width=\plotWidth]{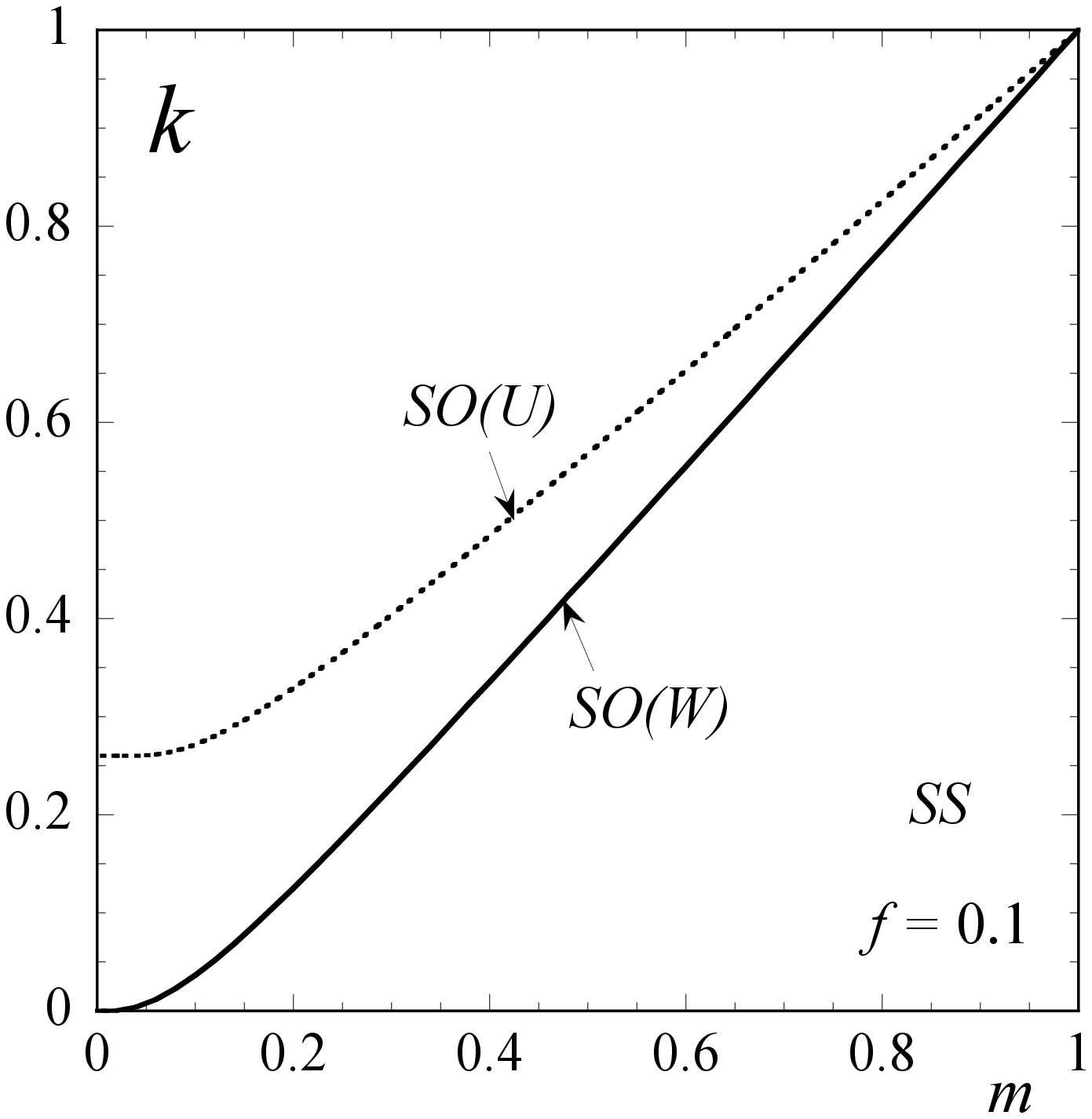} \\
  (a) & (b) & (c) \\
  \vspace{-2mm}
    \includegraphics*[width=\plotWidth]{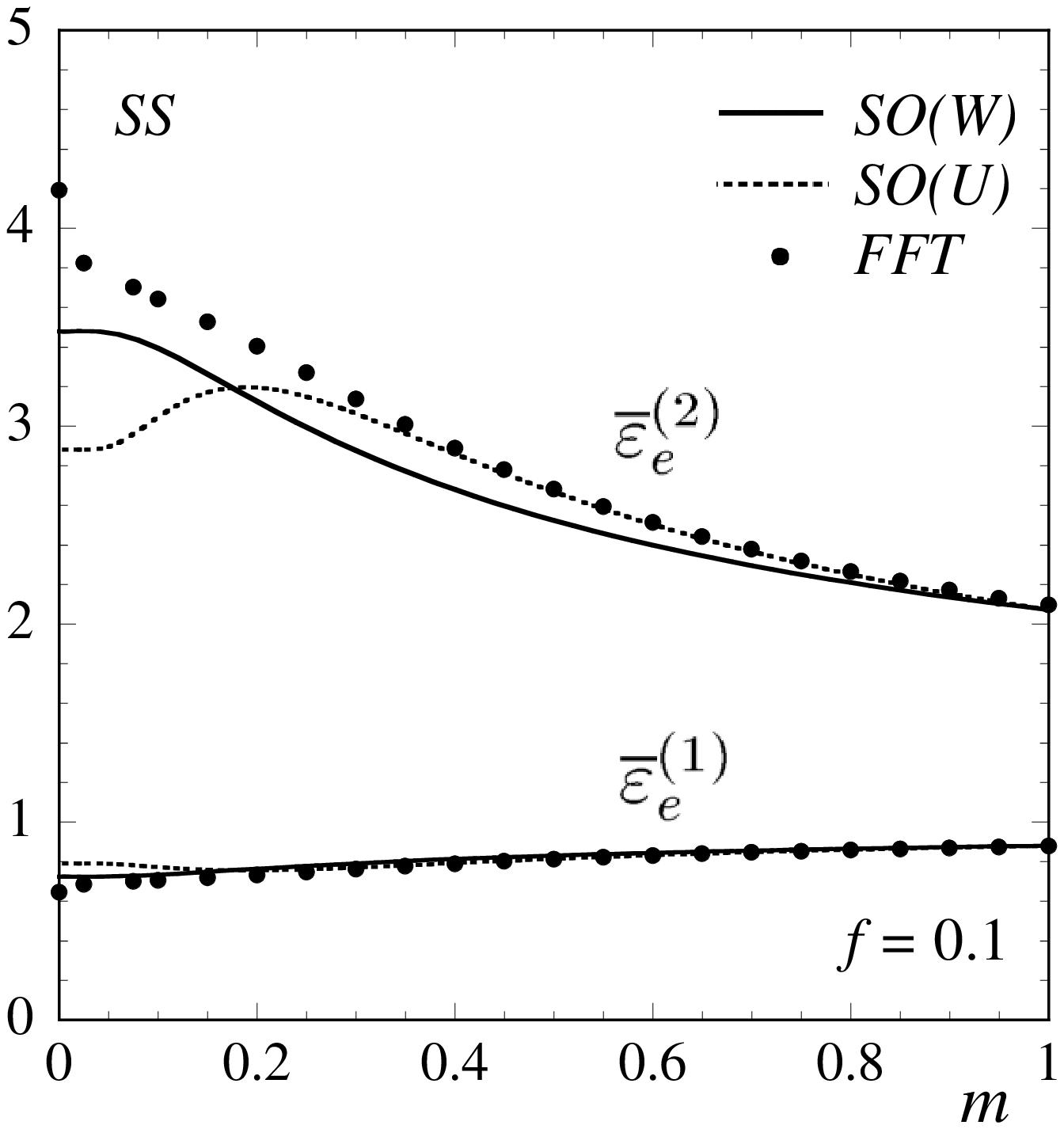}   & \includegraphics*[width=\plotWidth]{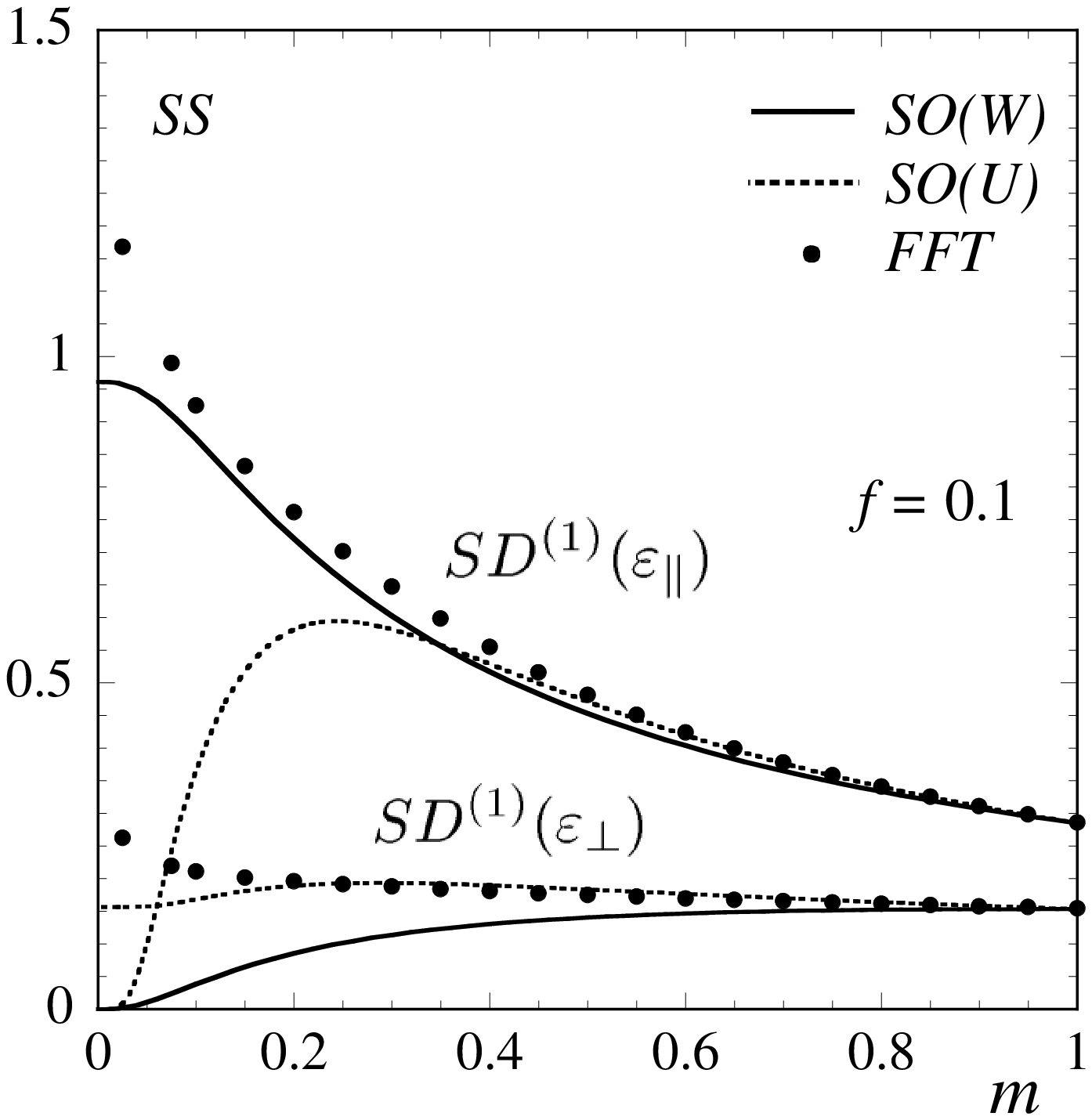} & \includegraphics*[width=\plotWidth]{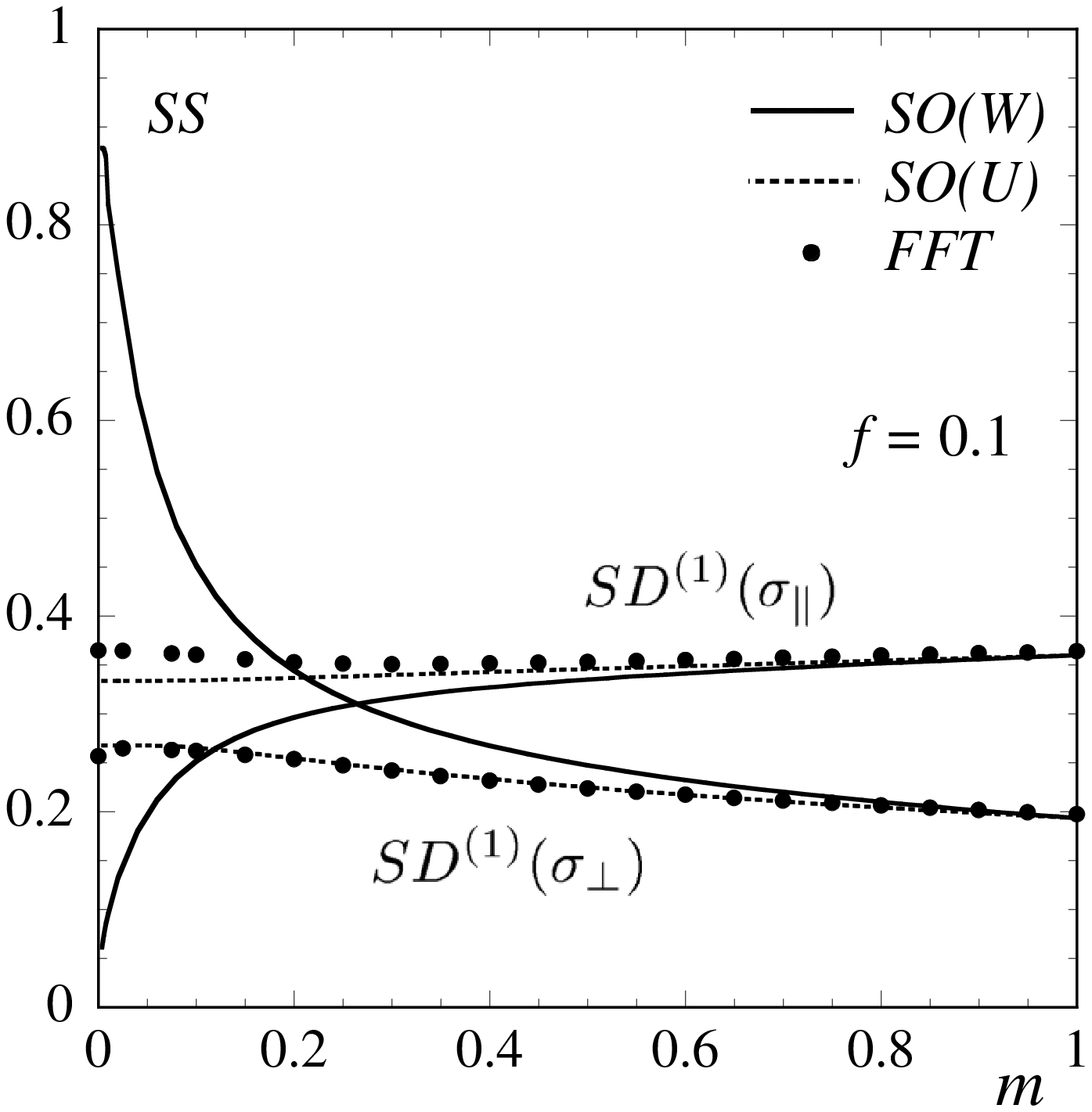} \\
  (d) & (e) & (f)
\end{tabular}
}
\caption{Comparisons between the `second-order' (SO) estimates and full-field FFT numerical calculations for porous, power-law materials vs. strain-rate sensitivity $m$: effective flow stress $\sef$, normalized by the flow stress of the matrix phase $\s0$, for (a) `pure shear' (PS) and (b) `simple shear' (SS) loadings; (c) anisotropy ratio $k$ of linear comparison matrix associated with SO estimates for SS; (d)-(f) field statistics for SS, normalized by the macroscopic equivalent strain $\oeeq$ and stress $\oseq$. The porosity is $f=0.1$.}
\label{fig:eff-porous}
\end{figure*}

The effect of strain-rate sensitivity $m$ on the `second-order' (SO) and FFT predictions is explored in Fig. \ref{fig:eff-porous}, for a moderate porosity ($f=0.1$). Parts (a) and (b) show plots for the effective flow stress $\sef$ under  `pure shear' (PS) and `simple shear' (SS) loadings, respectively, normalized by the flow stress of the matrix $\s0$. As anticipated in Section \ref{sec:HOM}, the strain (W) and stress (U) versions of the SO estimates, as given by expressions (\ref{sef}) and (\ref{sef_u}), are not identical; however, this duality gap is very small for most values of $m$. Good agreement is observed between the SO estimates and the numerical simulations for all values of $m$. The predictions are softer for SS than under PS loading. This anisotropy increases with increasing nonlinearity and is entirely due to the anisotropic arrangement of the phases: it reflects the fact that most of the energy is stored (dissipated) along the localization bands described above, which are shorter in SS (see maps 1 and 5 in Table \ref{table:maps}). The SO estimates are able to capture this interplay between geometrical arrangement and constitutive nonlinearity of the phases in the effective behavior, in contrast to the classical Voigt bound, which depends only on the volume fraction of the phases.

A key improvement in the `second-order' method over earlier `linear comparison' methods is the use in the linearization scheme of generalized secant moduli whose anisotropy can depend on microstructural features and loading conditions, in addition to constitutive nonlinearity. Fig. \ref{fig:eff-porous}c shows the anisotropy ratios $k$ associated with the SO estimates for SS, which solve equations (\ref{eqk}) and (\ref{eq_u}). (The corresponding plots for PS are similar and therefore omitted.) Indeed, a non-trivial dependence of $k$ on $m$ is observed, such that the anisotropy strength is increasingly stronger with increasing nonlinearity, the compliant mode being set by the direction of macroscopic load ($k < 1$). In contrast, linearization schemes based on the secant and tangent moduli restrict the anisotropy to be $k=1$ and $k=m$, respectively, and as a result lead to predictions that are less accurate in general, especially for field statistics (see comparisons in \cite{Idiart06b} for random composites).

Predictions for the field statistics under SS loading are shown in Figs. \ref{fig:eff-porous}d-f. (The corresponding results for PS are similar and therefore omitted.) In these plots, the strain and stress quantities are normalized by the macroscopic equivalent strain $\oeeq$ and stress $\oseq$, respectively. The FFT results show that as the nonlinearity increases, the pores undergo larger deformations and the strain fluctuations in the matrix increase, while the stress fluctuations remain fairly constant. The strong anisotropic growth of strain fluctuations is a direct consequence of the strain localizing along bands with a preferred orientation determined by the loading, as observed in Table \ref{table:maps} (maps 1-10). In the ideally plastic limit, the strain fluctuations become \textit{unbounded}\footnote{The FFT results for these quantities are finite, but increase with increasing mesh resolution.}. The SO estimates, given by expressions (\ref{sav})-(\ref{ssec}) and (\ref{eav_u})-(\ref{ssec_u}), are in good agreement with the numerical results in the range $0.3 \lesssim m \leq 1$. Within that range, the anisotropy ratios of the LCC matrix lie in the range $0.3 \lesssim k \leq 1$, see Fig. \ref{fig:eff-porous}c, and the local fields in the LCCs, reported in \cite{Willot08a}\footnote{Table II in \cite{Willot08a} shows stress maps for various values of $k$; due to linearity, strain maps are the same up to a change of scale given by the matrix moduli.}, mimic fairly well the nonlinear fields. For smaller values of $m$, however, the linear fields cannot mimic as accurately the localized character of the nonlinear strain field, and the SO predictions for the field fluctuations deteriorate significantly. In particular, the SO predictions for the strain fluctuations remain bounded as $m \rightarrow 0$. We note in passing that strain statistics are more accurately predicted by the strain-based SO(W) estimates, while stress statistics are more accurately predicted by the stress-based SO(U) estimates.

\begin{figure}[t]
\centerline{
\begin{tabular}{c@{\hspace{2mm}}c@{\hspace{2mm}}c}
    \vspace{-2mm}
  \includegraphics*[width=\plotWidth]{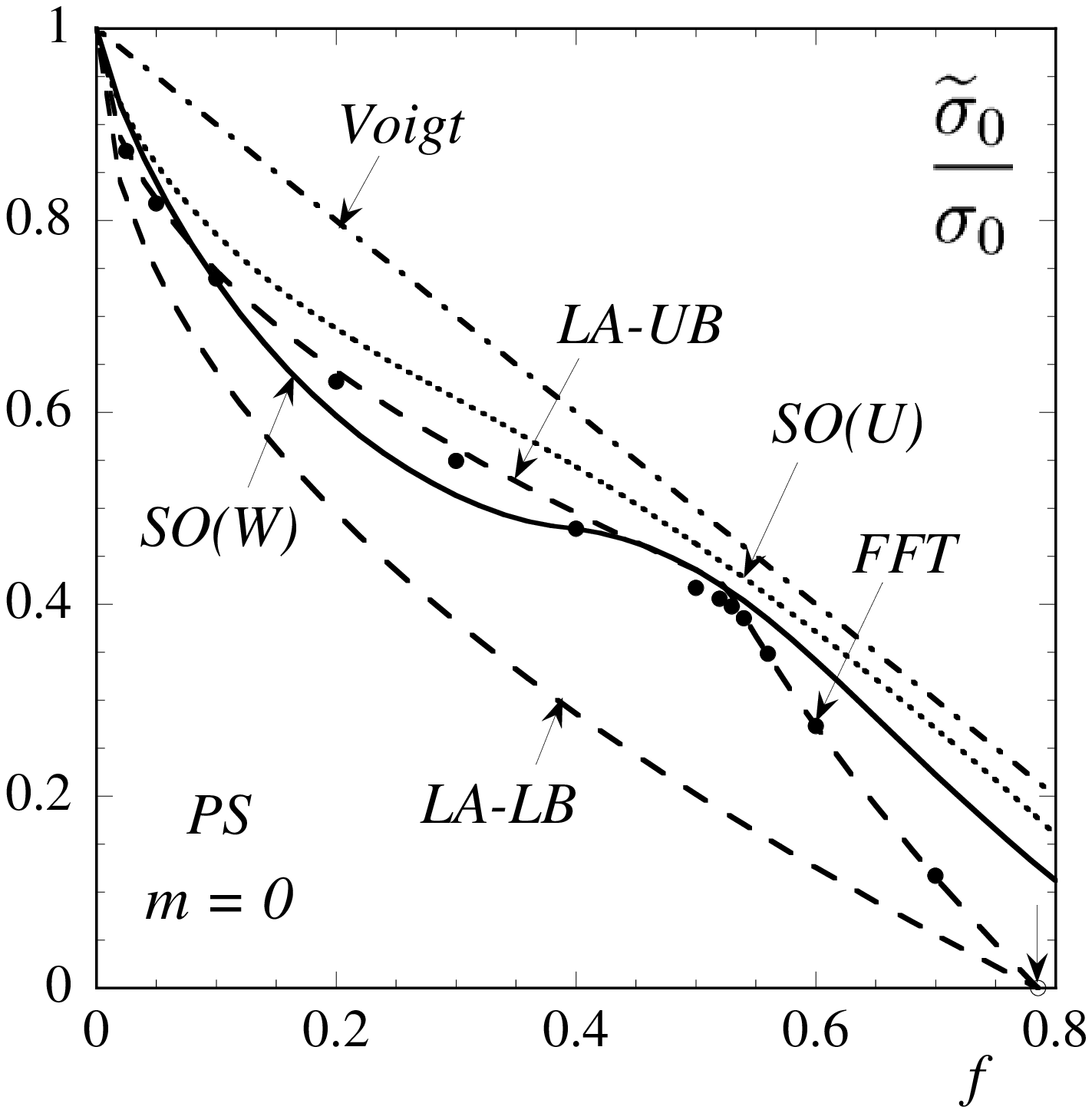} & \includegraphics*[width=\plotWidth]{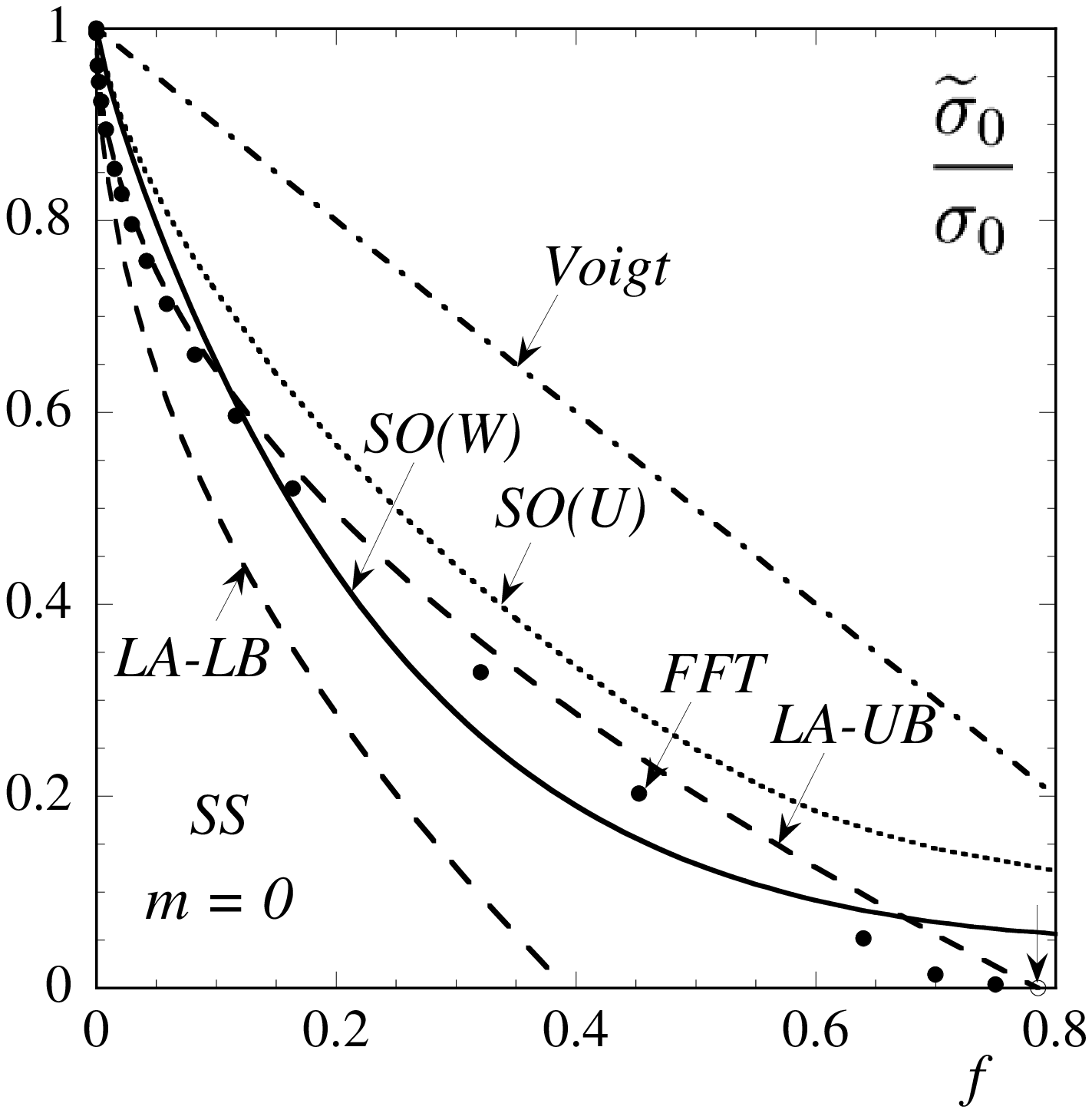} & \includegraphics*[width=\plotWidth]{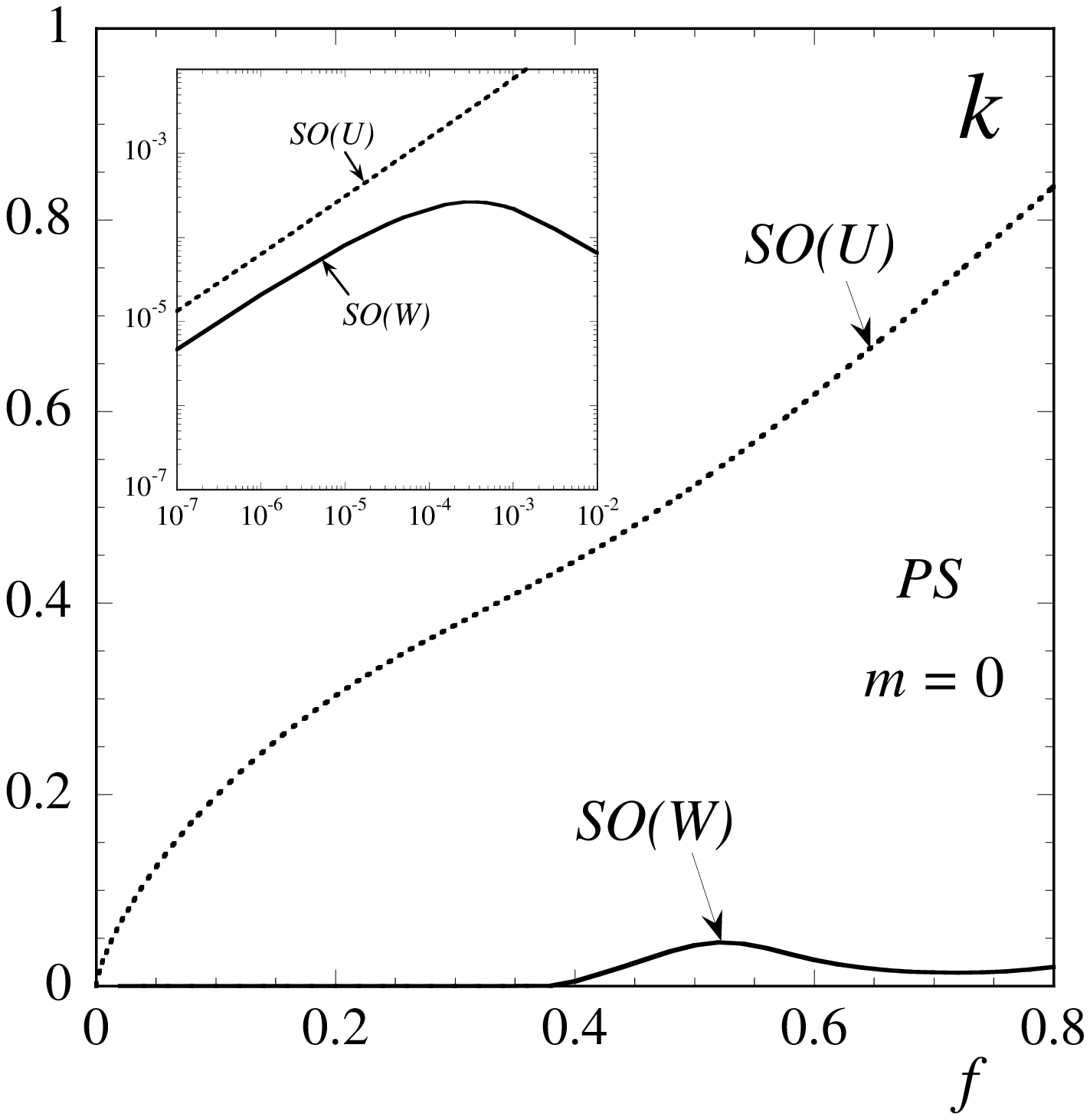}\\
    (a) & (b) & (c)
\end{tabular}
}
\caption{Comparisons between the `second-order' (SO) estimates and full-field FFT numerical calculations for porous, ideally plastic materials vs. porosity $f$: effective flow stress $\sef$ for (a) `pure shear' (PS) and (b) `simple shear' (SS) loadings; (c) anisotropy ratio $k$ of linear comparison matrix associated with SO estimates for PS.}
\label{fig:ideal-porous}
\end{figure}

The effect of porosity in ideally plastic solids ($m=0$) is explored in Fig. \ref{fig:ideal-porous}. Parts (a) and (b) show plots for $\sef$, and include the rigorous upper (UB) and lower (LB) bounds derived in Appendix \ref{app:la} by limit analysis (LA). The FFT results lie close to the LA upper bound, and are consistent with a close-packing threshold given by $f_{cp}=\pi/4 \approx 0.785$, the porosity at which neighboring pores come into contact with each other. As anticipated in the previous subsection, the FFT results for PS exhibit an inflexion point (or a kink) at $f_c \approx 0.52$, as a consequence of a change in the shear band pattern, see Table \ref{table:maps-f}. The LA upper bound given by (\ref{bounds_ps}) also exhibits this feature. It is quite remarkable that both versions of the SO estimates are able to capture, at least qualitatively, the correct anisotropic dependence of $\sef$ on porosity, including the inflexion point in the PS case. At large porosities, however, both SO estimates become rather stiff and remain finite at the close-packing threshold. This inaccuracy is inherited from the linear estimates used for the LCC, which become inadequate at porosities close to $f_{cp}$ (see \cite{Willot08a}). Improved SO estimates for large porosities could thus be obtained by using more appropriate/accurate linear estimates.

\subsection{Dilute limit}

\begin{figure}[t]
\centerline{
\begin{tabular}{c@{\hspace{1cm}}c}
    \vspace{-2mm}
  \includegraphics*[width=\plotWidth]{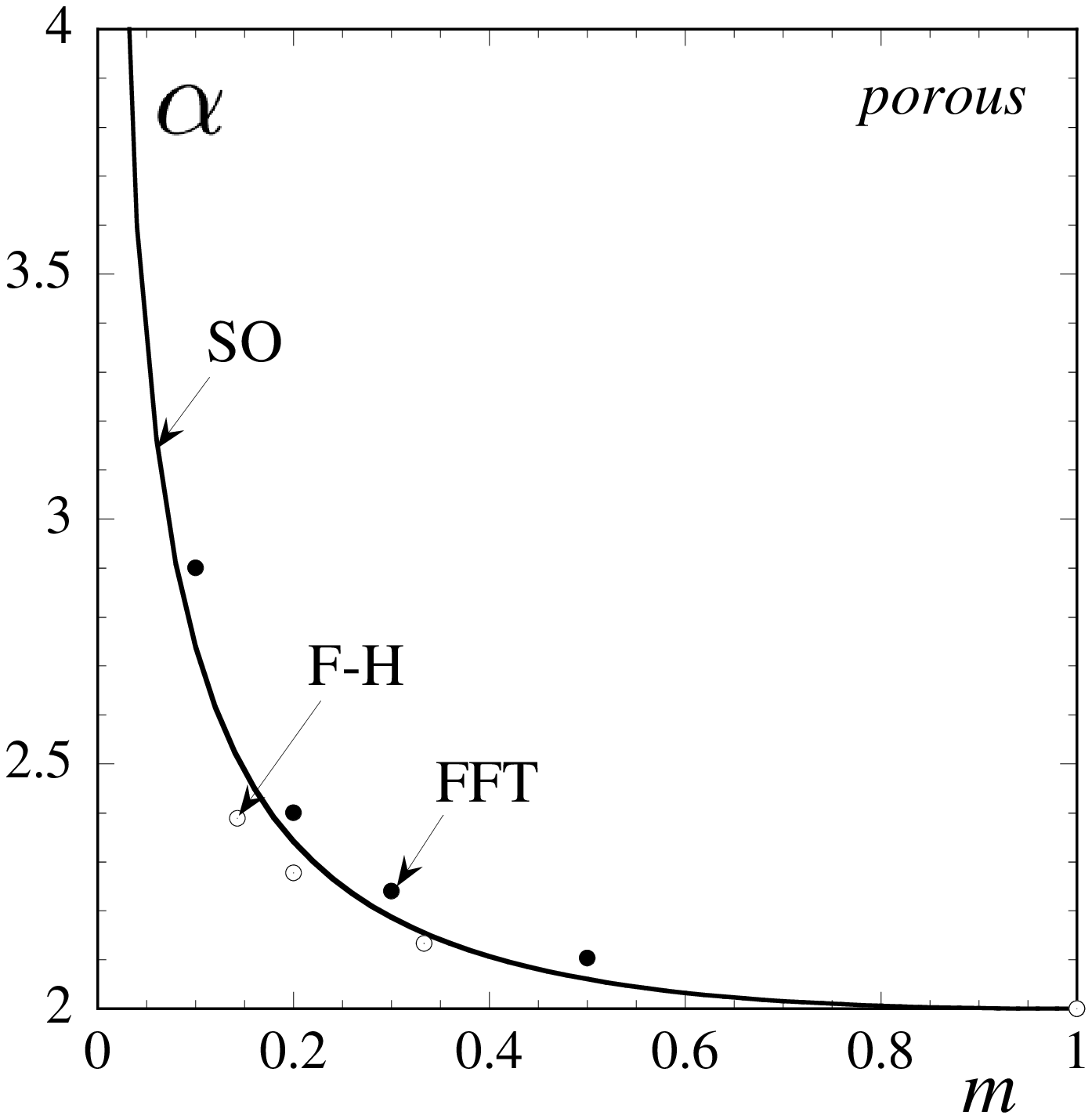} & \includegraphics*[width=\plotWidth]{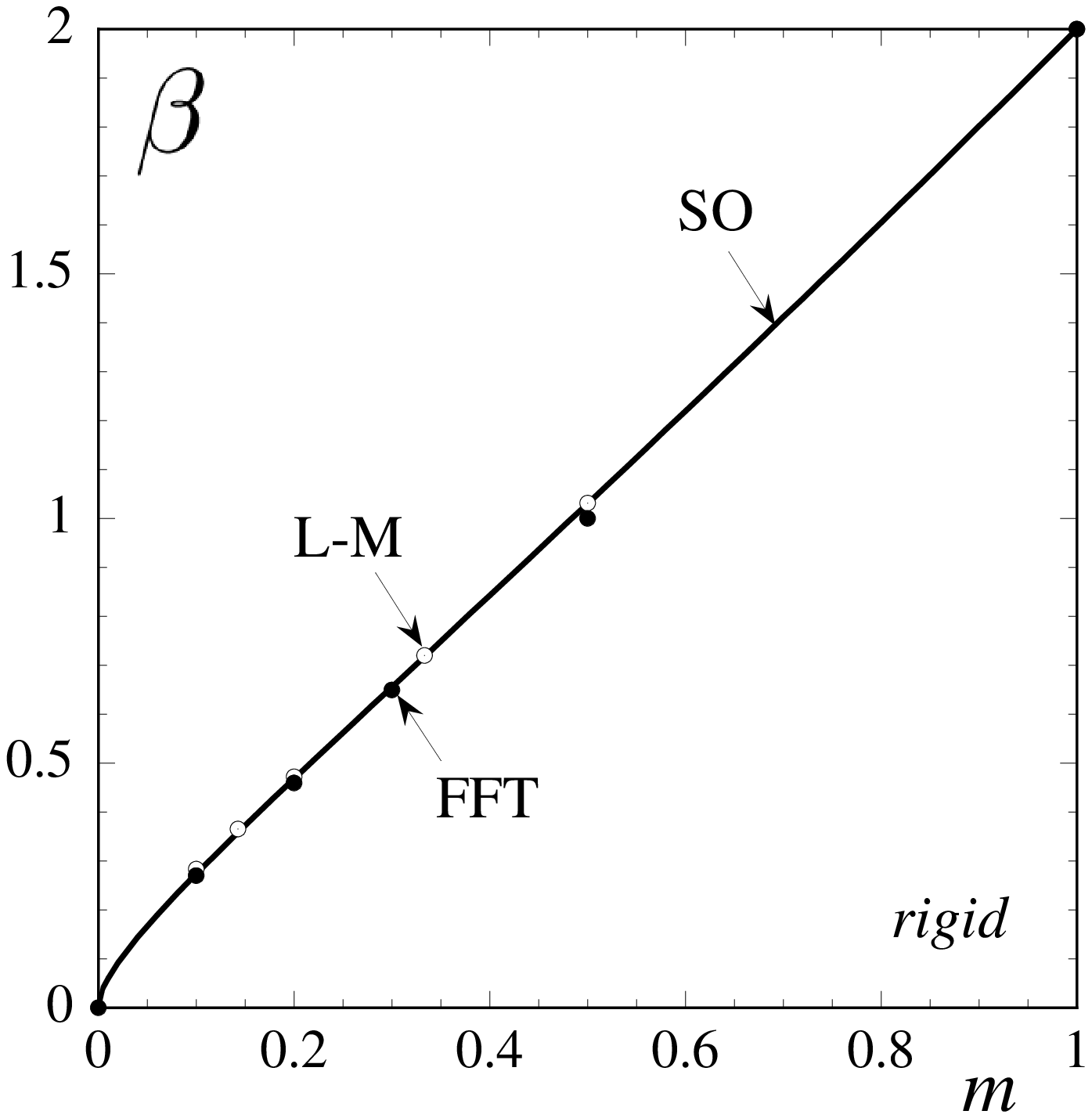} \\
    (a) & (b)
\end{tabular}
}
\caption{Comparisons between the `second-order' (SO) estimates and full-field FFT numerical simulations for power-law materials with dilute inclusion concentration: (a) porous materials and (b) rigidly-reinforced composites. Also included for comparison are the results of \cite{Fleck86} and \cite{Lee92}.}
\label{fig:dilute}
\end{figure}

Consider power-law materials with vanishingly small porosity levels ($f \rightarrow 0$) and fixed rate-sensitivity $m > 0$. For these materials, it is widely accepted that the effective behavior is insensitive to the actual distribution of porosity. Following \cite{Duva84}, \cite{Fleck86} proposed the ansatz
\begin{equation}
 \frac{\sef}{\s0} \approx 1 - \alpha(m) f,
 \label{so_dilute_m}
\end{equation}
and determined the coefficient $\alpha(m)$ numerically by considering an isolated circular pore in an infinite matrix. The present study confirms the ansatz (\ref{so_dilute_m}): both the FFT results and the SO estimates exhibit a linear correction at vanishingly small porosities. The predicted behavior is thus isotropic. In particular, the SO estimate for $\alpha(m)$ can be obtained by expanding (\ref{sef}) and (\ref{sef_u}), see Appendix \ref{app:expr} for details; the result is
\begin{equation}
    \alpha(m) = \frac{(1 + \sqrt{m})^2 } { 2\sqrt{m}}
    \label{alpha}
\end{equation}
for both versions. A comparison with the numerical results for $\alpha$ is shown in Fig. \ref{fig:dilute}a (Fig. \ref{fig:dilute}b will be discussed in Section \ref{sec:rigid}). As can be seen in the figure, the numerical and theoretical results coincide for $m=1$, agreeing with the classical result of \cite{Eshelby57}, and are in good agreement for all other values of $m$. Both the FFT simulations and the SO estimates predict vanishingly small strain fluctuations in the matrix phase as $f \rightarrow 0$, which is consistent with non-interacting pores in the dilute limit. As $m \rightarrow 0$, however, $\alpha(m)$ becomes unbounded, and the range of validity of (\ref{so_dilute_m}) vanishes.

Indeed, for ideally plastic solids ($m \rightarrow 0$) with small but fixed porosity $f >0$, the limit analysis bounds derived in Appendix \ref{app:la} demand that
\begin{equation}
 \frac{\sef}{\s0} = 1 - \alpha_0 f^{1/2},
 \label{bounds_dilute}
\end{equation}
where the coefficient $\alpha_0$ depends on the specific loading condition (PS or SS), see expressions (\ref{bounds_ss}) and (\ref{bounds_ps}). Thus, the dependence of $\sef$ on $f$ in the dilute limit is now \textit{non-analytic}, and the behavior is anisotropic. This is a consequence of the developing strain localization, see maps 1-10 in Table \ref{table:maps}, which introduces long-range interactions between pores. As a result, the effect of the actual distribution of porosity on the effective behavior persists in the dilute limit, and a crossover between the regular regime (\ref{so_dilute_m}) and a singular regime (\ref{bounds_dilute}) ensues. One can attempt to estimate the crossover threshold $f_c(m)$ between these regimes by equating expressions (\ref{so_dilute_m}) and (\ref{bounds_dilute}), to find
\begin{equation}
 f_c(m)=\alpha_0^2/\alpha(m)^2\sim 4\alpha_0^2 m\qquad\text{for}\qquad m\ll 1.
 \label{porous_crossover}
\end{equation}
However, preliminary results for sufficiently small $f$ and $m$ --not shown here-- indicate that the crossover threshold may be much lower. A detailed study of the crossover will be reported elsewhere.

The FFT results for dilutely voided ideally plastic materials are consistent with (\ref{bounds_dilute}), with the coefficient $\alpha_0$ being very close to that of the upper bound -- a numerical fit to the results yields $\alpha_0 = 0.826$ and $\alpha_0 = 1.185$ in PS and SS, respectively. The upper bound coefficient is given in terms of the ratio between the length of the bands $l_b$ and the side length of the unit cell $l$ by $\alpha_0 = (2/\sqrt{\pi}) (l_b/l)^{-1}$. The ratio is $l_b/l=1$ for SS, and $l_b/l=\sqrt{2}$ for PS, see Figs. \ref{fig:la_displ}a,b. The porous material is thus weaker under SS than under PS because the underlying shear bands are shorter in the first case. In turn, the exponent $1/2$ in (\ref{bounds_dilute}) arises because the energy is dissipated at \textit{one}-dimensional straight shear bands running through \textit{two}-dimensional pores \citep{Drucker66}. As a result, the effective behavior remains anisotropic in the dilute limit. In addition, the strain fluctuations in the matrix phase remain unbounded as $f \rightarrow 0$, which is consistent with a collective behavior of the pores in the dilute limit.

`Second-order' estimates for dilutely voided ideally plastic materials are obtained by expanding (\ref{sef}) and (\ref{sef_u_0}), resulting in the expression
\begin{equation}
    \frac{\sef}{\s0} \approx 1 - \alpha_{so} f^{2/3},
 \label{so_dilute}
\end{equation}
where the coefficient is independent of the loading, and is given by $\alpha_{so} = 1/2$ in the strain version and $\alpha_{so} = 3/2^{5/3}$ in the stress version. In addition, the SO estimates predict non-vanishing strain fluctuations in the matrix phase as $f \rightarrow 0$. Thus, remarkably, the SO estimates also predict a crossover between the regular regime (\ref{so_dilute_m}) and the singular regime (\ref{so_dilute}), although the exponent in the singular regime is incorrect. In this connection, it is recalled that SO estimates for 2D \textit{random} porous materials exhibit exactly the same dilute limit \citep{PC02b,Pastor02}. The $2/3$ exponent thus appears to be independent of the distribution of porosity.

It is noted, however, that the linearization in both versions of the SO estimates is such that $k \rightarrow 0$ as $f \rightarrow 0$, see insert in Fig. \ref{fig:ideal-porous}c. In this limit, the LCC effective moduli (\ref{Leff}) exhibit a crossover behavior (see \cite{Willot08a}): if $k \lesssim \pi f$, the medium is strongly anisotropic, the strain in this linear medium localizes in bands connecting the pores, and the dilute limit depends on the distribution of porosity; if on the other hand $k \gtrsim \pi f$, the anisotropy is weak and the pores behave as isolated inclusions. Making use of the asymptotic expansions of \cite{Willot08a}, it is shown in Appendix \ref{app:expr} that $k \sim f^{2/3}$ for both versions. Thus, the dilute SO estimates (\ref{so_dilute}) do not depend on the actual distribution of porosity because they make use of the `uncorrelated' regime of the linear estimates. However, the fact that $k \rightarrow 0$ suggests that the `correlated' regime could be exploited by suitably modifying the linearization scheme. In this connection, it should be noted that the estimates (\ref{so_dilute}) make use of an ad-hoc prescription for certain reference tensors entering in the linearization, see expressions (\ref{eref}) and (\ref{sref}). The above findings hint that alternative prescriptions may be developed resulting in dilute SO estimates that depend on the distribution of porosity (in the singular regime), and that would yield improved estimates more generally.

\section{Results for rigidly-reinforced composites}
\label{sec:rigid}

\subsection{Field maps}
\label{sec:rigidfieldmaps}

Full-field distributions in rigidly-reinforced materials are shown in Table \ref{table:maps} (maps 21-40), for three values of the strain-rate sensitivity ($m=1, 0.2, 0$) and a moderate reinforcement concentration ($f=0.1$). Unlike in the porous case, the strain does \textit{not} localize in shear bands as the nonlinearity increases. Instead, an inclusion screening effect develops with increasing nonlinearity: the strain becomes \textit{small} (compared to $\oeeq$) along bands of width equal to one inclusion diameter, running across the specimen along the directions of maximum macroscopic shear ($0^\circ$ and $90^\circ$ with respect to the Cartesian axes) and seeking out the rigid inclusions, see maps 21-25 in Table \ref{table:maps}. Thus, the macroscopic deformation is mainly accommodated by a fairly uniform strain present in the remaining regions outside these bands. The corresponding stress distributions, on the other hand, are seen to become progressively more uniform as the nonlinearity increases, see maps 31-40. In the ideally plastic limit, both components of the stress are uniform almost everywhere, exhibiting variations only in small regions around the inclusions (maps 31, 35, 36, 40). In fact, the matrix phase is at yield almost everywhere, and this translates into no reinforcement effect in the effective behavior due to the rigid inclusions, as will be seen below.

Under PS loading, the two families of bands intersect each other not only at inclusion sites but also at regions in the matrix phase (corners of the unit cell), giving rise to the formation of square regions where the strain is almost null (see map 25). No such regions
develop under SS shear (see map 21). This difference is responsible for the strong anisotropy in the effective behavior observed in Fig. \ref{fig:eff-rigid} below for high reinforcement concentrations. Indeed, while in the SS case the strain pattern remains essentially unchanged for any reinforcement concentration below the close-packing threshold $f_{cp} = \pi/4 \approx 0.78$, in the PS case the strain pattern has to change abruptly for reinforcement concentrations larger than $f=\pi/8 \approx 0.39$, the value of $f$ at which the
square regions come into contact with each other. This transition is shown in Table \ref{table:maps-f}, maps 7-12. For $f > 0.4$, the strain tends to concentrate in a rim around the inclusions, being very small in most of the matrix phase. As will be seen below, this results in a significant increase in the effective flow stress for that range of reinforcement concentrations. See also \cite{Willot08b} for analogous effects in the linear anisotropic porous case.

It should be emphasized at this point that, due to the lack of \textit{strict} convexity of the potentials (\ref{mat_const}) and (\ref{mat_const_u}) when $m=0$, the strain field in an ideally plastic composite may not be unique \citep{Suquet81} (the effective stress, however, is uniquely defined, \cite{Bouchitte91}). In fact, it is easy to verify that, since the matrix of the reinforced materials considered here is at yield almost everywhere, alternative strain fields to the `diffuse' fields shown in Table \ref{table:maps} are possible and would consist of infinitely thin straight shear bands running through the matrix at $0^{\circ}$-$90^{\circ}$ for SS, and $\pm 45^{\circ}$ for PS (provided $f<\pi/8$). In this connection, it was found that the same augmented Lagrangian method, separately applied with the discretized Green function (see Section \ref{sec:FFT}), and with the `continuous' Green function (as used in the original FFT algorithm of \cite{Michel01}) converge to completely different `diffuse' and localized solutions for the strain field, respectively (whereas the stress field, is independent of the choice of the Green function at high resolution). With the `continuous' green function, thin straight shear bands tangent to the inclusion-matrix interface are observed (see also \cite{Moulinec98}). In any event, as will be seen below, the ideally plastic `second-order' estimates derived in this work, which follow from a power-law regularization, predict strain statistics entirely consistent with the `diffuse' strain fields of Table \ref{table:maps}, and \textit{not} with the localized solution.

\subsection{Effective behavior and field statistics}
\label{sec:rig-eff}

\begin{figure*}[t]
\centerline{
\begin{tabular}{c@{\hspace{2mm}}c@{\hspace{2mm}}c}
  \vspace{-2mm}
    \includegraphics*[width=\plotWidth]{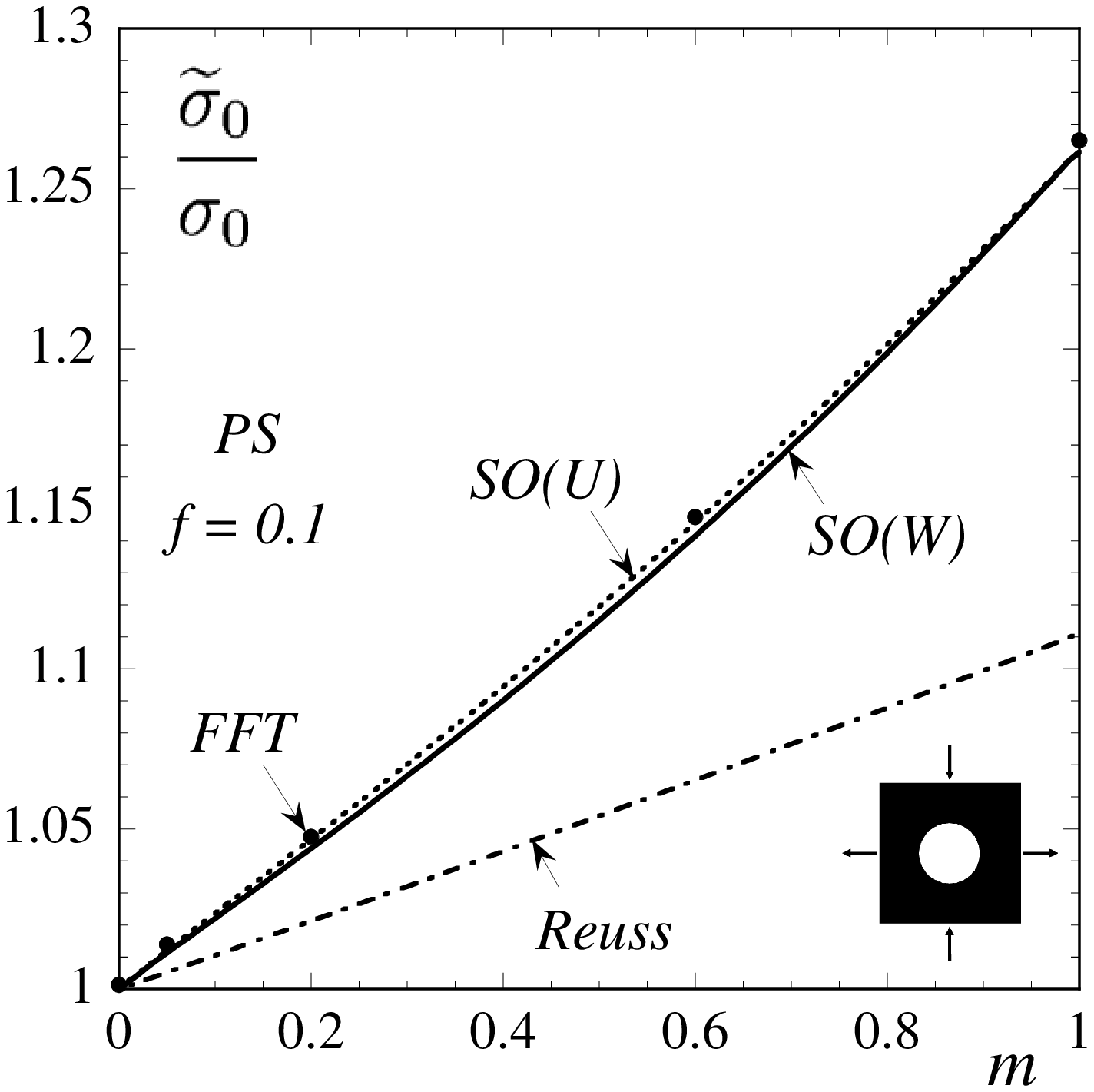} & \includegraphics*[width=\plotWidth]{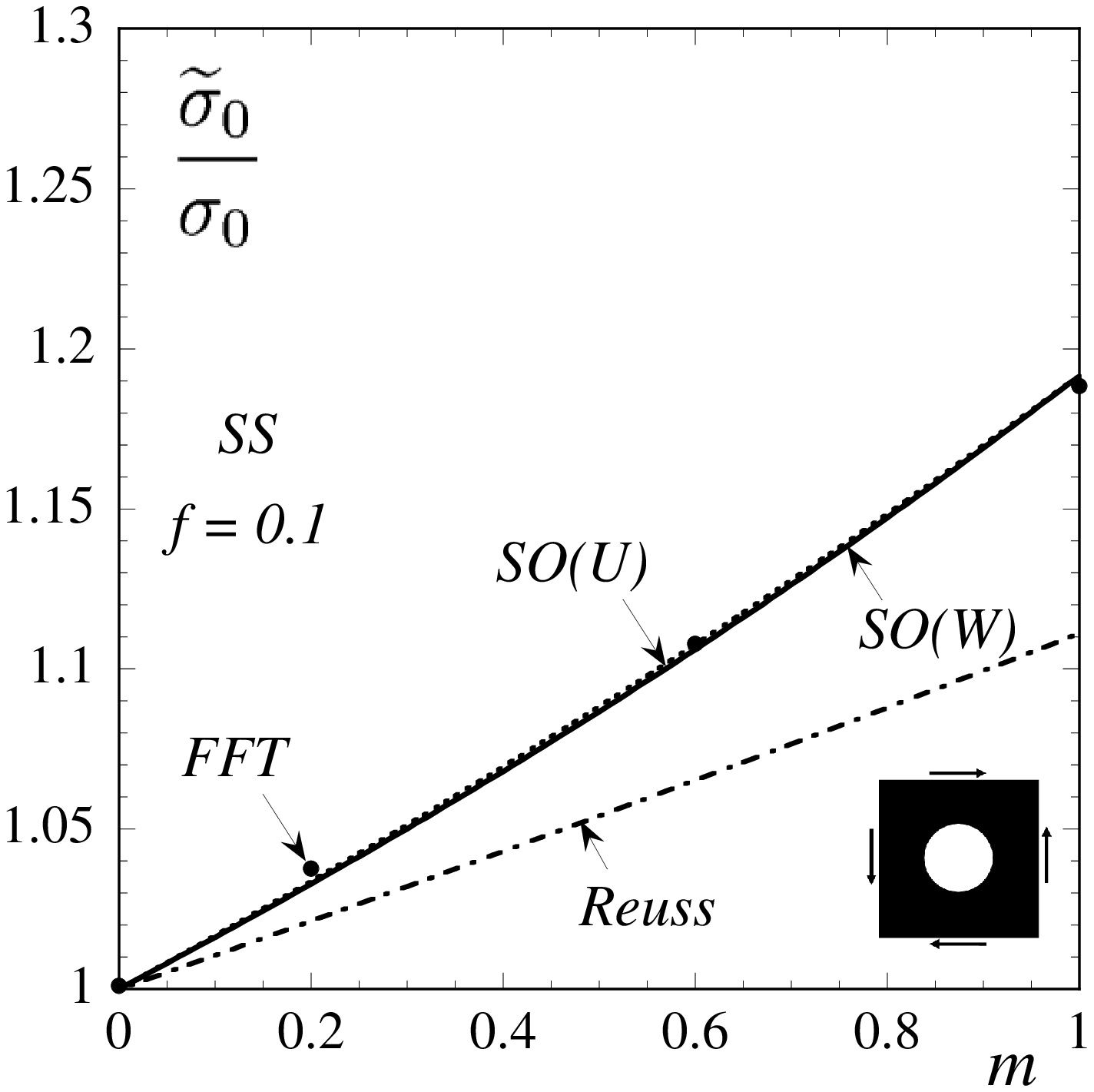} & \includegraphics*[width=\plotWidth]{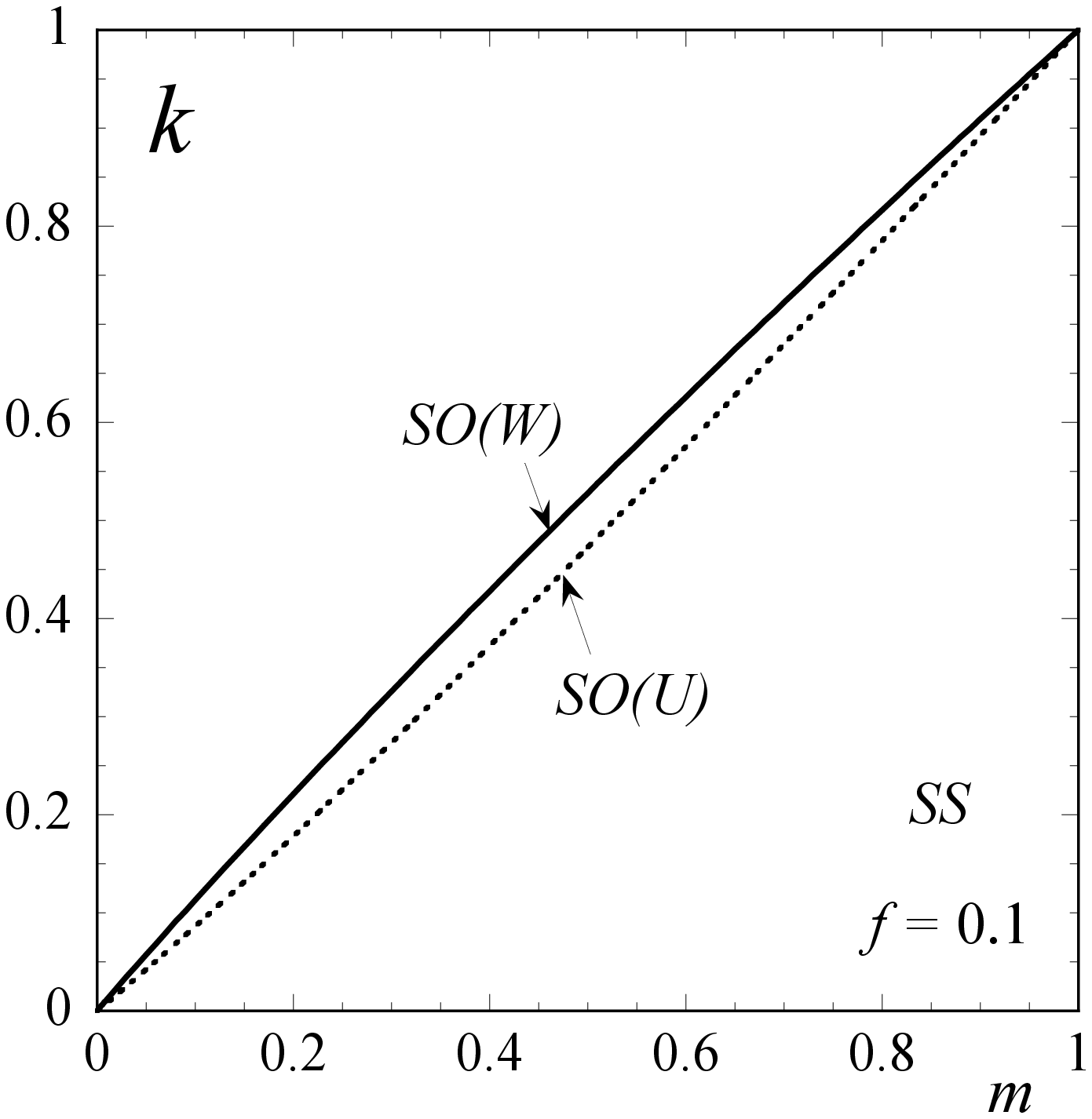} \\
    (a) & (b) & (c) \\
    \includegraphics*[width=\plotWidth]{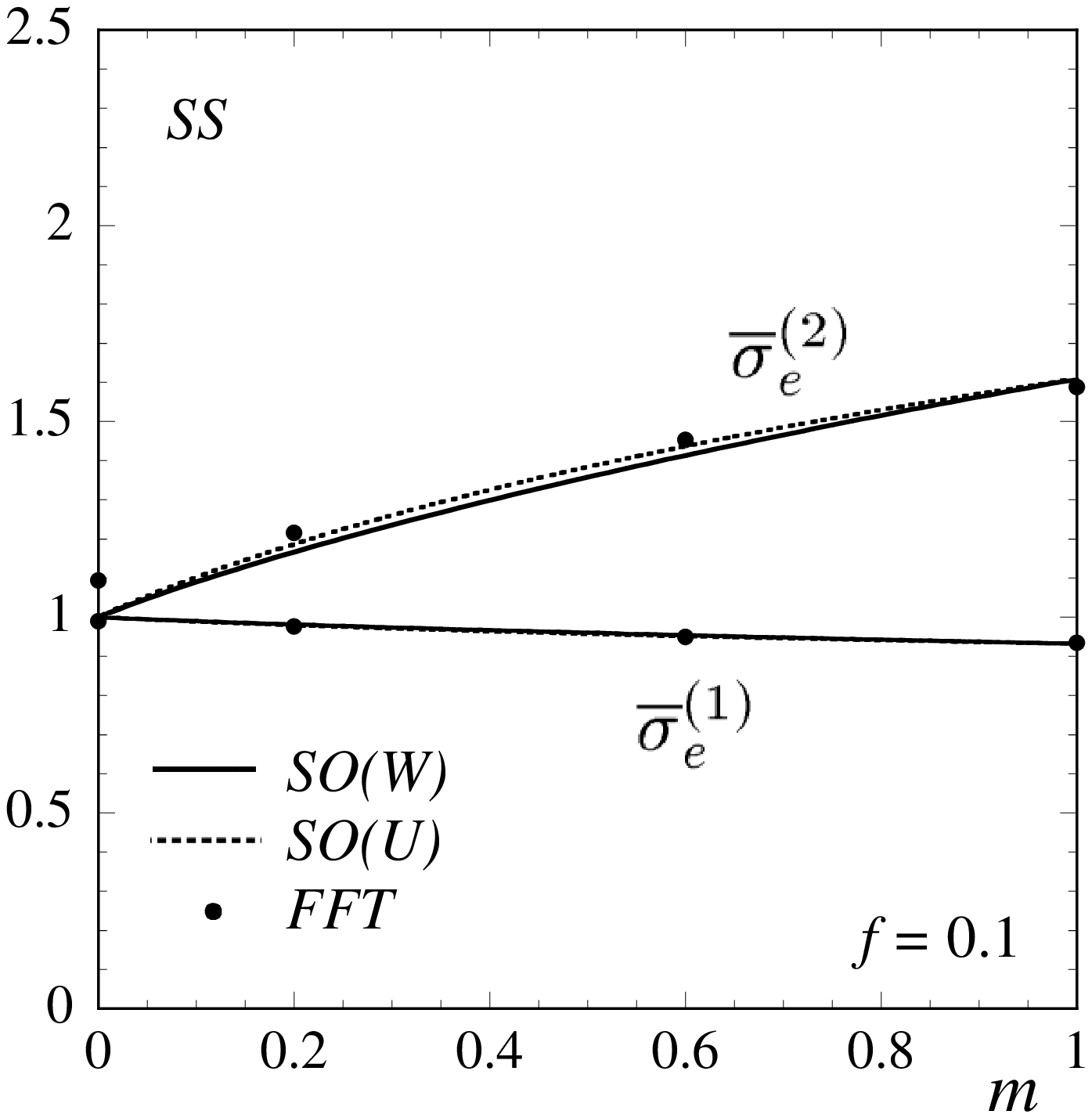} & \includegraphics*[width=\plotWidth]{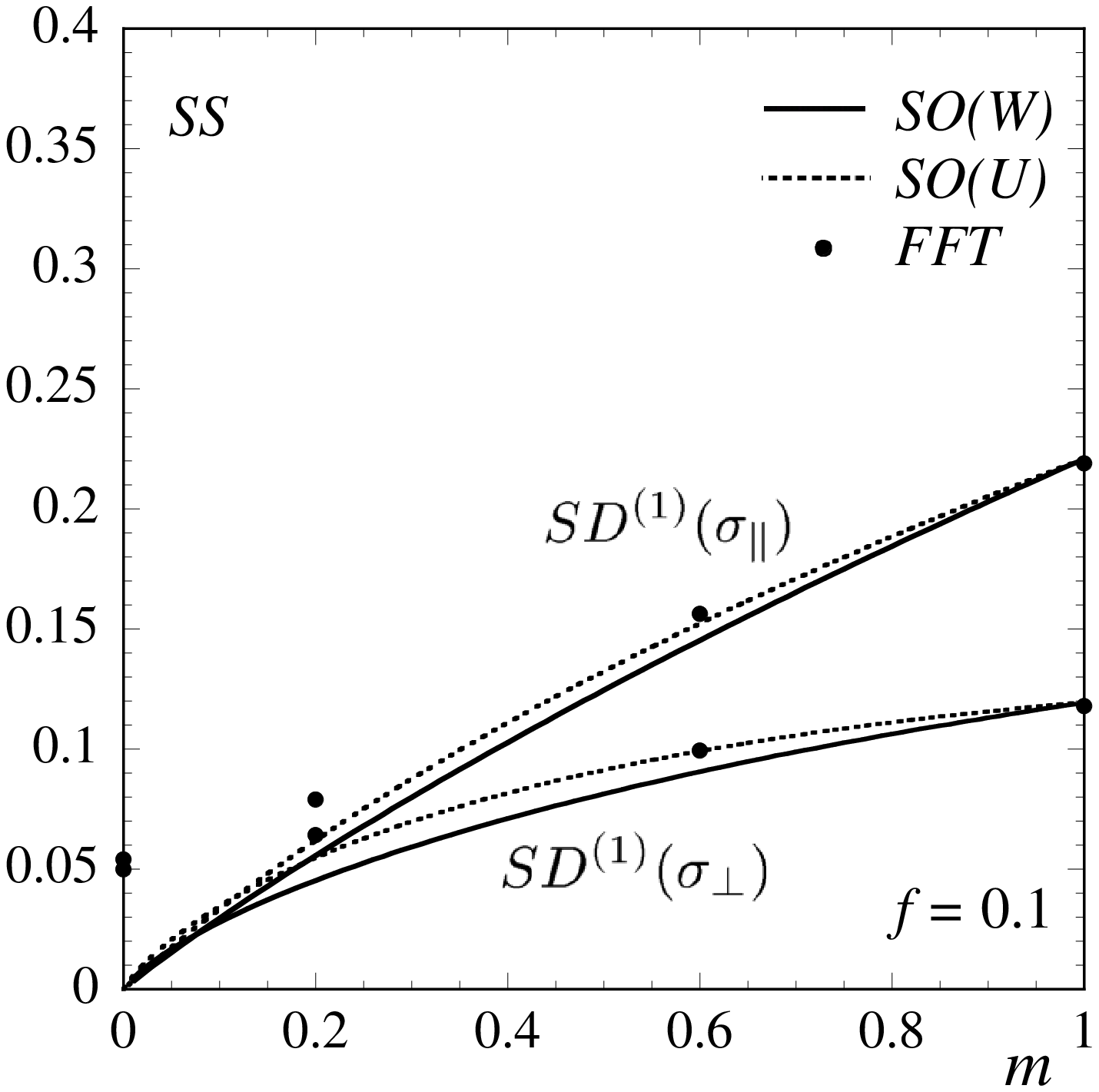} & \includegraphics*[width=\plotWidth]{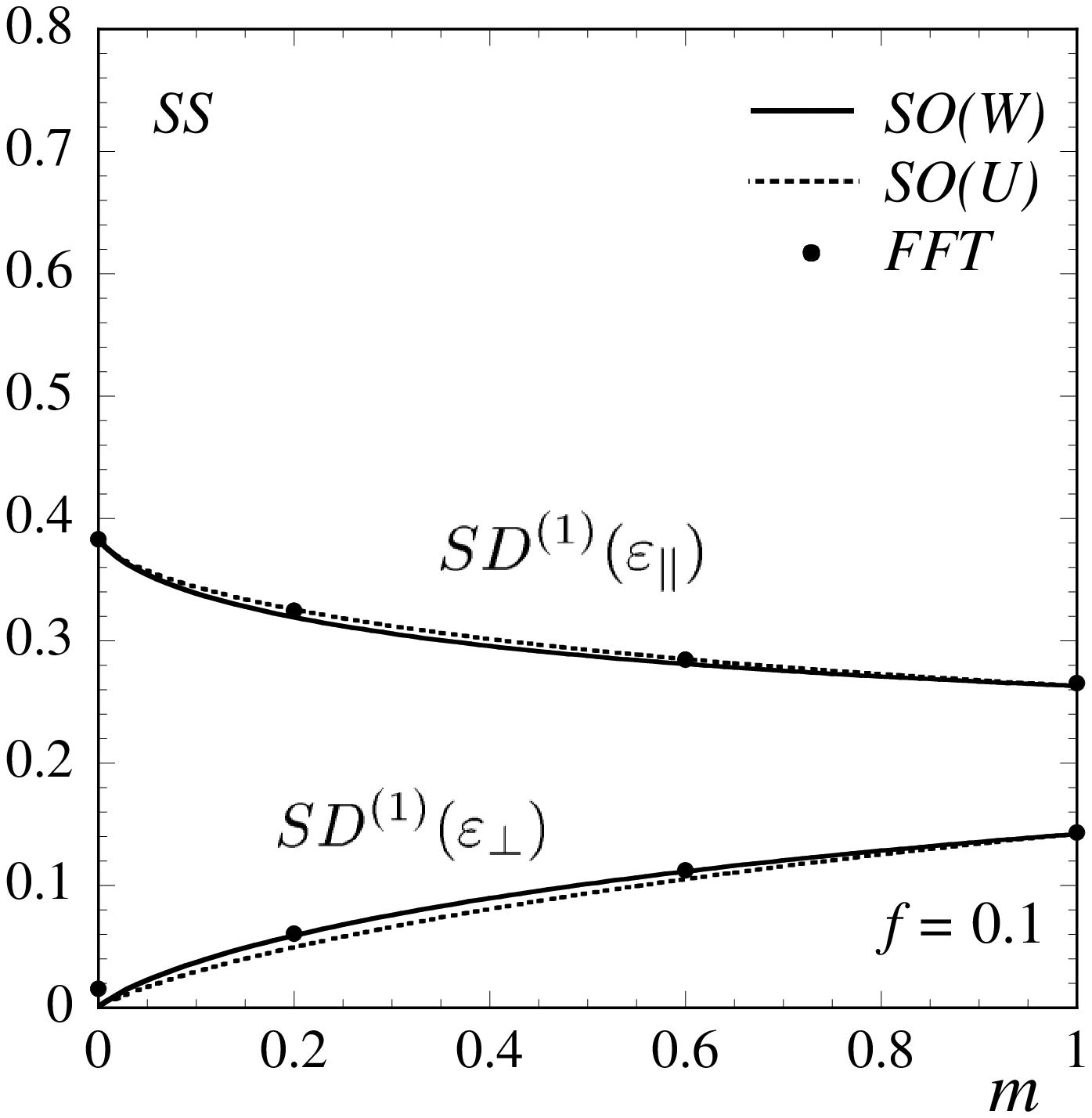} \\
    (d) & (e) & (f)
\end{tabular}
}
\caption{Comparisons between the `second-order' (SO) estimates and full-field FFT numerical calculations for rigidly-reinforced, power-law materials vs. strain-rate sensitivity $m$: effective flow stress $\sef$ for (a) `pure shear' (PS) and (b) `simple shear' (SS) loadings; (c) anisotropy ratio $k$ of linear comparison matrix associated with SO estimates for SS; (d)-(f) field statistics for SS. The reinforcement concentration is $f=0.1$.}
    \label{fig:eff-rigid}
\end{figure*}

The effect of matrix nonlinearity $m$ on the `second-order' (SO) and FFT predictions is explored in Fig. \ref{fig:eff-rigid} for a moderate reinforcement concentration ($f=0.1$). Parts (a) and (b) show plots for the effective flow stress $\sef$ under  `pure shear' (PS) and `simple shear' (SS) loadings, respectively, normalized by the flow stress of the matrix $\s0$. Both versions of the SO estimates are found to be in good agreement with the FFT results, for all values of the strain-rate sensitivity $m$ and both loading conditions. The estimates agree with the FFT results in that $\sef$ decreases monotonically with increasing nonlinearity (i.e., decreasing $m$), and in that the composite material is `stronger' under PS than under SS loadings, for all values of $m$ different than zero.

In the ideally plastic limit ($m \rightarrow$ 0), the limit analysis upper bound derived in Appendix \ref{app:la} coincides with the classical Reuss lower bound, and therefore give the exact result which is $\sef/\s0=1$ for both loading conditions. Indeed, this is the result obtained with the FFT calculations. Both versions of the SO estimates are found to coincide with the exact result in this strongly nonlinear limit, thus predicting no reinforcement effect due the presence of rigid inclusions. In this connection, it is recalled that this is the exact result whenever the macroscopic deformation can be accommodated by straight shear bands passing through the matrix phase and avoiding the inclusions (see \citet{Drucker66} and \citet{Taliercio92}). In the SS case, this is possible for any reinforcement concentration smaller than the close-packing threshold $f_{cp}$, while in the PS case this is not possible for $f > \pi/8$ and therefore a finite reinforcement effect is expected in that case.

Predictions for the field statistics under SS loading are shown in Figs. \ref{fig:eff-rigid}e-f, normalized by $\oeeq$ and $\oseq$. Unlike in the porous case, the SO estimates are in good agreement with the FFT results for all values of $m$. This is mainly because the strain field in the LCCs can now mimic fairly well the nonlinear strain field, even in the ideally plastic limit. Indeed, as $m\rightarrow 0$ the anisotropy ratio $k \rightarrow 0$, see Fig. \ref{fig:eff-rigid}c, and the strain distributions in such LCCs, reported in \cite{Willot08a}\footnote{Table II of \cite{Willot08a} shows fields for porous materials with a different definition for $k$, and require suitable reinterpretation. Thus, maps 3-5 and 8-10 in that table can be interpreted as strain fields in rigidly-reinforced composites under PS, while maps 16-18 and 21-23 can be interpreted as corresponding strain fields under SS.}, are similar to those of maps 21-30 in Table \ref{table:maps}. The nonlinear predictions show decreasing inclusion stress concentration and matrix stress fluctuations with increasing nonlinearity. This is consistent with the stress field becoming more homogeneous throughout the composite as $m \rightarrow 0$, see maps 31-40 in Table \ref{table:maps}. The small discrepancies between the SO estimates and the FFT results in the ideally plastic limit are due to the small local variations of stress around the inclusions observed in the numerical simulations. The predictions also show that the strain fluctuations remain bounded in the ideally plastic limit, unlike in porous materials. This is because the strain distribution is essentially piecewise uniform rather than localized in shear bands (cf. maps 21-30 and maps 1-10 in Table \ref{table:maps}). In this connection, it is interesting to note that recent FFT simulations of rigidly-reinforced composites with \textit{random} microstructures \citep{Moulinec03,Idiart06b} have shown shear band strain localization and unbounded `parallel' strain fluctuations as $m \rightarrow 0$, in agreement with the `second-order' predictions for that class of materials \citep{Idiart06b,Idiart07b}. It is thus remarkable that the `second-order' method, being only a homogenization theory, is able to correctly predict the completely different character of strain distribution arising in periodic versus random reinforced composites, as reflected not in the macroscopic behavior but rather in higher-order statistical information.

\begin{figure}[t]
\centerline{
\begin{tabular}{c@{\hspace{2mm}}c@{\hspace{2mm}}c}
    \vspace{-2mm}
    \includegraphics*[width=\plotWidth]{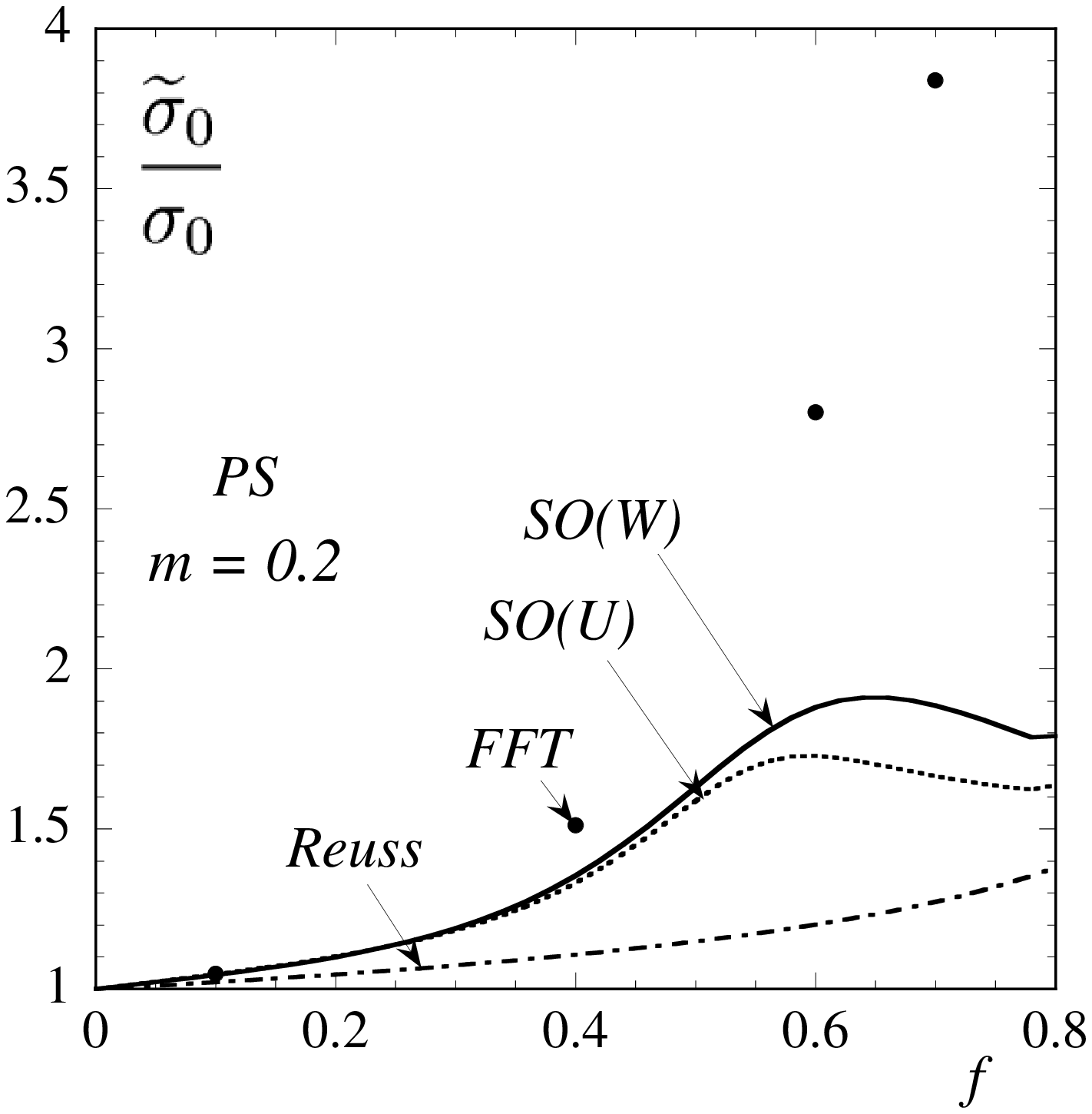} & \includegraphics*[width=\plotWidth]{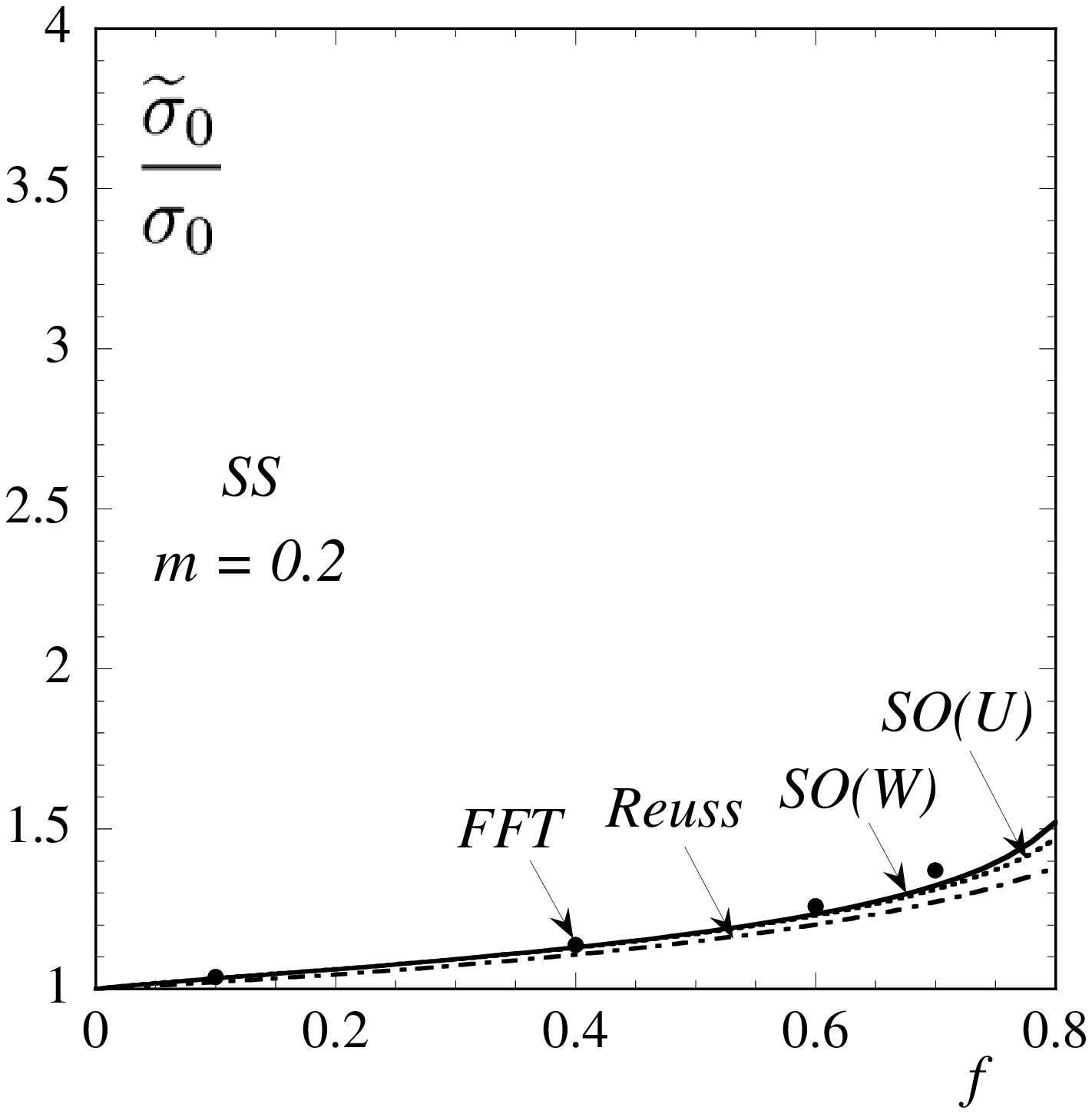} & \includegraphics*[width=\plotWidth]{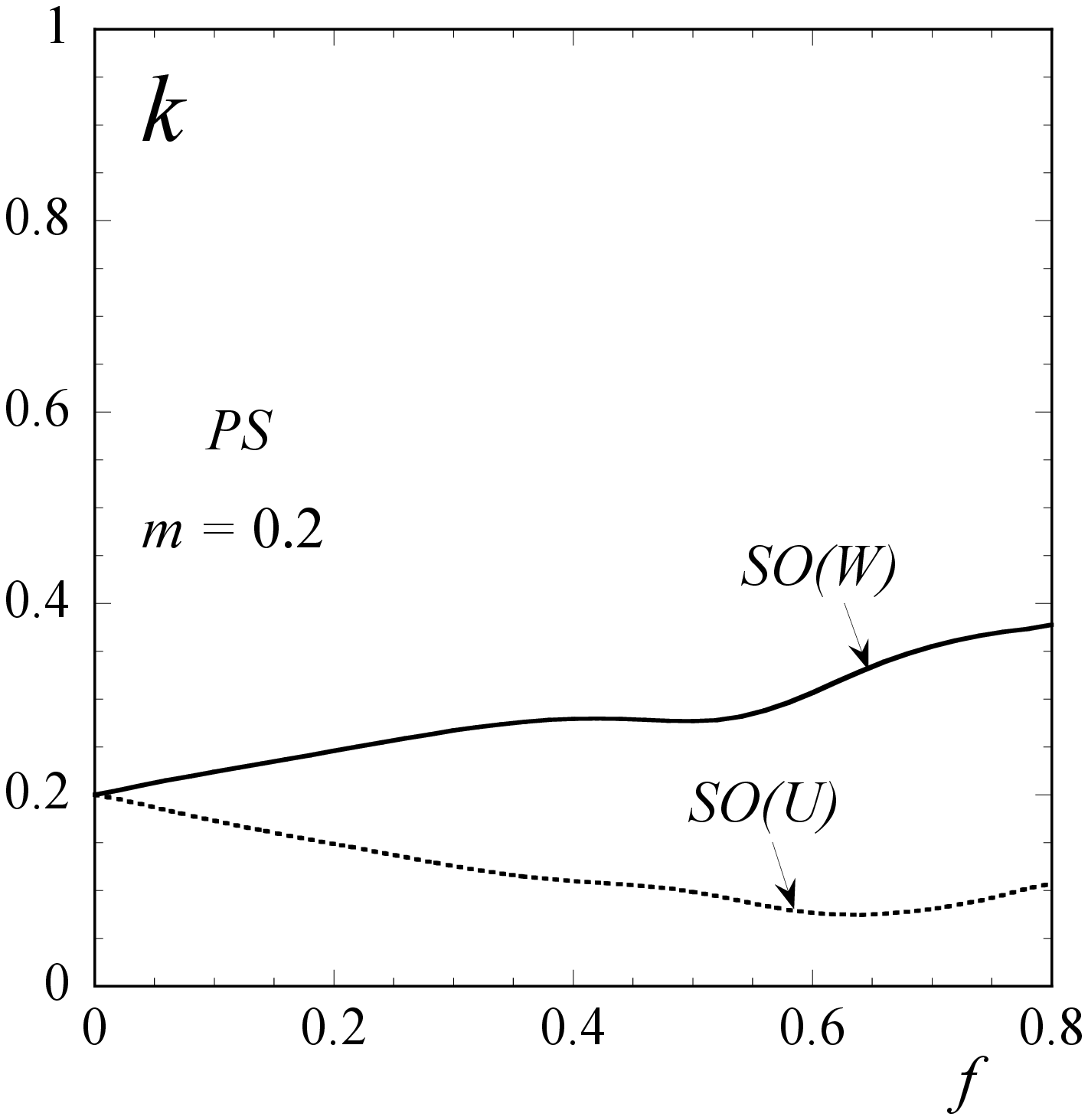} \\
    (a) & (b) & (c)
\end{tabular}
}
\caption{Comparisons between the `second-order' (SO) estimates and full-field FFT numerical calculations for rigidly-reinforced, power-law materials vs. reinforcement concentration $f$: effective flow stress $\sef$ for (a) `pure shear' (PS) and (b) `simple shear' (SS) loadings; (c) anisotropy ratio $k$ of linear comparison matrix associated with SO estimates for PS. The strain-rate sensitivity is $m=0.2$.}
\label{fig:eff-rigid-f}
\end{figure}

The effect of reinforcement concentration is explored in Fig. \ref{fig:eff-rigid-f}, for a moderate nonlinearity ($m=0.2$). Parts (a) and (b) show plots for $\sef$. The FFT results show that the composite becomes `stiffer' as the reinforcement concentration increases, as expected, and that, for moderate and higher concentrations, this effect becomes significantly stronger for PS than for SS loadings. As anticipated in the previous subsection, this nonlinearity-induced anisotropy in the effective behavior is due to the fact that, as the nonlinearity increases, the circular inclusions behave progressively more like square inclusions with an orientation relative to the lattice that depends on the loading orientation (see Table \ref{table:maps}). Due to the orientation of these `square' inclusions, the close-packing effects become important at lower values of $f$ in PS than in SS loadings. This, together with the fact that the linear estimates utilized in the context of the SO estimates become progressively less accurate as the close-packing effects become more important (see \cite{Willot08a}), is the reason why, at large values of $f$, the SO estimates are significantly less accurate for PS than for SS loadings. As already mentioned in the previous section, the SO estimates could be improved for high values of $f$ by using a different linear estimate, appropriate for reinforcement concentrations near the close-packing threshold.

\subsection{Dilute limit}
\label{rigid-dilutelimit}

Consider power-law materials with vanishingly small reinforcement concentrations ($f \rightarrow 0$) and fixed strain-rate sensitivity $m > 0$. As already discussed, for these materials,  the effective behavior is expected to be insensitive to the actual distribution of inclusions, and \cite{Lee92} proposed the ansatz
\begin{equation}
 \frac{\sef}{\s0} \approx 1 + \beta(m) f,
 \label{dilute2}
\end{equation}
and determined the coefficient $\beta(m)$ numerically by considering an isolated circular inclusion in an infinite matrix. The present study confirms the ansatz (\ref{dilute2}): both the FFT results and the SO estimates exhibit a linear correction at vanishingly small $f$. The predicted behavior is thus isotropic. In particular, the SO estimate for $\beta(m)$ can be obtained by expanding (\ref{sef}) and (\ref{sef_u}), noting that $k \rightarrow m$ as $f \rightarrow 0$ (see Fig. \ref{fig:eff-rigid-f}c); the result is
\begin{equation}
    \beta(m) = \frac{1}{2}\sqrt{m} (1 + \sqrt{m})^2
    \label{beta}
\end{equation}
for both versions. This result is in exact agreement with that obtained in the context of \textit{random} rigidly-reinforced composites \citep{PC96,PC02b}. A comparison between the SO estimate and the numerical results for $\beta$ is given in Fig. \ref{fig:dilute}b. The numerical and theoretical results coincide for $m=1$ with the classical result of \cite{Eshelby57}, as they should, and are in excellent agreement for all other values of $m$. Note that, unlike in the porous case, the coefficient (\ref{beta}) vanishes as $m \rightarrow 0$, suggesting an extended range of validity of (\ref{dilute2}) in this limit.

Now consider ideally plastic solids ($m \rightarrow 0$) with fixed reinforcement concentration $0 < f < \pi/8$. As already mentioned in Section \ref{sec:rig-eff}, the limit analysis bounds derived in Appendix \ref{app:la} demand that
\begin{equation}
 \frac{\sef}{\s0} = 1,
 \label{dilute3}
\end{equation}
independent of the value of $f$, in exact agreement with the corresponding FFT results and SO estimates. The dilute expansion of (\ref{dilute3}) is consistent with the ideally plastic limit of (\ref{dilute2}). Thus, in contrast to the porous case, the dilute concentration and small rate-sensitivity regimes in reinforced composites are consistent, at least to first order in $f$. In addition, the FFT simulations and SO estimates predict vanishingly small strain fluctuations in the matrix phase as $f \rightarrow 0$ for both regimes. This suggests that, in dilutely reinforced ideally plastic solids, 
only weak long-range interactions between inclusions develop, and the inclusions effectively behave as if they were isolated at leading order. This in turn may explain why the `second-order' predictions are more accurate for reinforced composites than for porous materials. It is noted, however, that the present analysis does not exclude the existence of a crossover behavior at higher-order corrections. In fact, preliminary results for sufficiently small $f$ and $m$ do suggest that such a crossover exists, and will be reported elsewhere.

\section{Concluding remarks}
\label{sec:CONCLUSIONS}

This paper presented a combined numerical-theoretical study of the macroscopic behavior and local field distributions for a special class of two-dimensional periodic composites with \textit{viscoplastic} phases, with emphasis on the infinite contrast, strongly nonlinear case. The comparisons show good overall agreement between the `second-order' (SO) homogenization estimates and the FFT full-field numerical simulations, not only for the effective behavior but also for the first- and second-order field statistics of the stress and strain fields.

More specifically, the strain field for the rigidly-reinforced materials was found to develop (with decreasing strain-rate sensitivity $m$) a banding pattern consisting of low-strained bands of finite width set by the size of the inclusions. For both pure and simple shear loading, and volume fractions allowing the passage of straight bands through the matrix, the resulting interaction between inclusions is not strong, and the effective flow stress tends to the yield stress of the matrix phase in the ideally plastic limit. In these cases, the SO estimates are extremely accurate. However, for pure shear and volume fractions greater than $\pi/8$, the inclusions block the straight paths and the deformation mechanism changes abruptly leading to a finite strengthening effect for larger volume fractions all the way up to inclusion contact (which the SO estimates only capture approximately). By contrast, the strain field in porous materials tends to localize (again with decreasing strain-rate sensitivity) in bands seeking to pass through the pores in order to minimize the plastic dissipation. In this case, the interaction between the  pores is obviously quite strong and dependent on the geometry of the configuration as well as the relative orientation of the loading (e.g., pure versus simple shear). Again for pure shear, an abrupt change in deformation mechanism is observed for values of the porosity near $50 \%$. In this case, the SO estimates are quite accurate for values of strain-rate sensitivity between 0.2 and 1, and deteriorate for smaller strain-rate sensitivities (although they correctly capture the transition).

One of the main findings of this work is the observation of a crossover-type behavior for the porous case in the dilute concentration ($f$) and small rate-sensitivity ($m$) regimes. Indeed, the FFT simulations show that for finite values of $m>0$, the effective flow stress relative to the flow stress of the matrix is reduced from unity by a factor proportional to the porosity $f$ (cf. (\ref{so_dilute_m})). This is consistent with the classical dilute limit with non-interacting inhomogeneities. However, the coefficient $\alpha$ (see (\ref{alpha})) blows up as $m^{-1/2}$ when $m \rightarrow 0$, suggesting that the range of validity of this expansion tends to zero in the the ideally plastic limit. In fact, first taking the limit as $m \rightarrow 0$, and then considering small porosity $f$ leads to a non-analytic  dependence on the porosity ($1/2$-power dependence on $f$ -- see (\ref{bounds_dilute})), together with non-vanishing strain fluctuations in the matrix phase, which is consistent with a collective-type behavior induced by the strong interactions among the voids as the developing shear bands seek to minimize the overall plastic dissipation by passing through the voids. Between these two asymptotic regimes, there is an `intermediate' asymptotic (i.e., crossover) regime where the type of solution transitions from one regime (i.e., `strongly dilute') to the other (i.e., `strongly nonlinear'). Mathematically, this phenomenon can be linked to the change of character of the governing equations --- from elliptic to hyperbolic (see \cite{CHEN05} and \cite{Willot08b} for analogous results in other systems). In this connection, it should be mentioned that, remarkably, the SO theory is able to predict a crossover behavior for the porous materials, with a very good quantitative prediction in the strongly dilute regime (see Fig. \ref{fig:dilute}$a$ for $\alpha$), but with the wrong exponent ($2/3$  versus $1/2$) in the strongly nonlinear regime. However, the true form of the crossover in this case is still unknown.

For the rigidly-reinforced materials, on the other hand, the two asymptotic regimes were found to be consistent up to leading order. In this case, the strongly dilute regime is linear in $f$ (cf. (\ref{dilute2})) with a coefficient $\beta(m)$ that tends to zero as $m \rightarrow 0$, while the strongly nonlinear regime has a vanishing correction for sufficiently small values of $f$ (cf. (\ref{dilute3})). In addition, the strain fluctuations are vanishingly small in this limit. This suggests that the ideally plastic limit in the case of rigid inclusions is in some sense less singular, which may help explain why the SO theory is able to deliver much more accurate predictions in this case. We note, however, that the analysis provided in this work does not exclude the possibility of a crossover-type behavior in the higher-order corrections, which will be investigated in future work.

It should be emphasized that the combined dilute and ideally plastic limit for porous media is particularly relevant to the ductile fracture initiation problem, where the distribution of porosity is expected to be random. However, for random microstructures, the corresponding numerical simulations are much more challenging than for the periodic microstructures discussed in this work. The only results that the authors are aware of for porous materials with random microstructures are the numerical results (based on limit analysis) of \citet{Pastor02}, which suggest that the dilute exponent in the `strongly nonlinear' regime should be between $1/2$ and $2/3$, at least for a class of 2D microstructures known as composite cylinders, as well as the FFT results of \cite{Willot07a} for `pixel-type' 2D random microstructures that suggest an exponent of $2/3$. On the other hand, the SO estimate of the Hashin-Shtrikman type predicts \citep{PC02b} a crossover behavior with an exponent of $2/3$ in the strongly nonlinear regime for 2D random microstructures. Although the role of disorder on the dilute exponent still remains to be investigated in detail, the relative success --- at least in qualitative terms --- of the SO theory for the periodic case suggests that the corresponding predictions of \cite{PC02b} for the random case should at least be qualitatively correct.

Finally, it should be noted that this study was restricted to isochoric loadings `aligned' with the periodic microstructure. It is emphasized, however, that the homogenization methods utilized can handle completely general loading conditions (see \cite{Bilger05,DIP08b}). In particular, porous materials subjected to non-isochoric loadings, where more complicated localization patterns arise, will be considered elsewhere.

\section*{Acknowledgements}

The work of M.I.I.\ and P.P.C.\ was supported by the National Science Foundation grants CMS-02-01454 and OISE-02-31867. The work of F.W.\ was supported by a CEA Ph.D.\ grant.

\appendix
\section{Second-order estimates for power-law porous materials}
\label{app:expr}

\subsection{Strain formulation}
\label{sec:sow}

When specialized to porous materials with an isotropic matrix phase characterized by (\ref{mat_const}), the `second-order' variational estimate for the effective strain potential is given by \citep{PC02a,Idiart06a}
\begin{equation}
    \Weff(\obfe)\!=\!\mathop{\rm stat}_{\l0,\m0} \left\{ \Weff_L(\obfe; \cbfe^{(1)}\!, \bfL^{(1)}) \!+\!(1-f) v^{(1)}(\cbfe^{(1)}\!, \bfL^{(1)}) \right\},
    \label{SOW}
\end{equation}
where the stat\textit{ionary} operation consists in setting the partial derivative of the argument with respect to the variable equal to zero. In this expression, $\Weff_L$ is the effective potential of a porous LCC with the same microstructure as the nonlinear material and with a matrix characterized by a second-order, Taylor-type expansion of the nonlinear potential (\ref{mat_const})$_2$ about a reference strain tensor $\cbfe^{(1)}$, given by
\begin{equation}
        w^{(1)}_L(\bfe; \cbfe^{(1)}, \bfL^{(1)}) = w^{(1)}(\cbfe^{(1)}) + \frac{\p w^{(1)}}{\p\bfe}(\cbfe^{(1)})\!:\!(\bfe - \cbfe^{(1)}) + \frac{1}{2} (\bfe - \cbfe^{(1)})\!:\!\bfL^{(1)}\!:\!(\bfe - \cbfe^{(1)}),
    \label{wL}
\end{equation}
where $\bfL^{(1)}$ is an incompressible, symmetric, fourth-order tensor (of moduli) of the form
\begin{equation}
        \bfL^{(1)} = 2 \l0\, \bfE_{\pr} + 2 \m0\, \bfE_{\pp}, \quad k = \l0/\m0.
        \label{L0}
\end{equation}
Here, $\l0$ and $\m0$ are two shear moduli, and
\begin{equation}
    \bfE_{\pr} = \frac{2}{3} \frac{\cbfe_d^{(1)}}{\ceeq^{(1)}} \otimes \frac{\cbfe_d^{(1)}}{\ceeq^{(1)}}, \quad \bfE_{\pp} = \bfK - \bfE_{\pr},
    \label{Etensors}
\end{equation}
are two orthogonal, fourth-order, projection tensors with principal axes `aligned' with $\cbfe^{(1)}$, and $\bfK$ denotes the standard, fourth-order, shear projection tensor. It should be noted that, even though the nonlinear potential (\ref{mat_const})$_1$ is isotropic, the tensor of moduli (\ref{L0}) is generally anisotropic. The ratio $k$ constitutes a measure of this anisotropy.

In turn, the `error' function $v^{(1)}$ in (\ref{SOW}), a measure of the degree of nonlinearity of $w^{(1)}$, is defined as
\begin{equation}
    v^{(1)}(\cbfe^{(1)}, \bfL^{(1)})\!=\!\mathop{\rm stat}_{\hbfe^{(1)}}\left\{w^{(1)}(\hbfe^{(1)})\!-\!w_L^{(1)}(\hbfe^{(1)}; \cbfe^{(1)}\!, \bfL^{(1)})
    \right\}.
    \label{VFunct}
\end{equation}

The nonlinear estimate (\ref{SOW}) requires a linear estimate for the effective potential $\Weff_L$ of the porous LCC. Noting that the matrix potential (\ref{wL}) corresponds to a `thermoelastic' material with a `thermal stress' and a `specific heat' given by
\begin{eqnarray}
    \bftau^{(1)} &=& \frac{\p w^{(1)}}{\p\bfe}(\cbfe^{(1)}) - \bfLo^{(1)}\!:\!\cbfe^{(1)},   \label{tau0} \\
    g^{(1)}      &=& w^{(1)}(\cbfe^{(1)}) - \bftau^{(1)}\!:\!\cbfe^{(1)} - \frac{1}{2} \cbfe^{(1)}\!:\!\bfLo^{(1)}\!:\!\cbfe^{(1)}\!, \label{f0}
\end{eqnarray}
the effective potential becomes \citep{Willis81}
\begin{equation}
    \Weff_L(\obfe) = \frac{1}{2}\, \obfe\!:\!\Leff\!:\!\obfe + \taueff\!:\!\obfe + \feff,
    \label{WeffL}
\end{equation}
with $\taueff$ and $\feff$ given in terms of $\Leff$ by
\begin{eqnarray}
    \taueff &=& \Leff\!:\!\bfLo^{(1)\,-1}\!:\!\bftau^{(1)}, \label{teff} \\
    \feff   &=& (1-f)\, g^{(1)} - \bftau^{(1)}\!:\!\bfLo^{(1)\, -1}\!:\!\left[ \Leff - (1-f)\, \bfLo^{(1)} \right]\!:\!\bfLo^{(1)\, -1}\!:\!\bftau^{(1)} \label{feff}.
\end{eqnarray}
In these expressions, $\Leff$ denotes the effective modulus tensor of a porous material with a purely elastic matrix with modulus tensor (\ref{L0}), and completely characterizes (\ref{WeffL}).

For later use, it is recalled that the first and second moments of the strain field $\bfe_L$ in the matrix of the porous LCC, can be extracted from the
potential $\Weff_L$ by means of the identities \citep{Idiart07}
\begin{equation}
        \obfe_L^{(1)} = \frac{1}{1-f} \left. \frac{\p \Weff_L}{ \p \bftau^{(1)} } \right|_{\bfL^{(1)}\!,\, g^{(1)}}, \quad
        \left\langle \bfe_L \otimes \bfe_L \right\rangle^{(1)} = \frac{2}{1-f} \left. \frac{\p \Weff_L }{ \p \bfLo^{(1)} } \right|_{\bftau^{(1)}\!,\,g^{(1)}},
    \label{eeL}
\end{equation}
where the variables that are held fixed have been emphasized. Thus, the variational procedure employed to derive the estimate (\ref{SOW}) dictates an `optimal' choice for the moduli $\l0$ and $\m0$ of the matrix phase in the LCC, which follows from the stationarity condition in (\ref{SOW}). In addition, the reference strain $\cbfe^{(1)}$ must also be specified in order to characterize the potential $w^{(1)}_L$. Unfortunately, it has not yet been possible to find a similar `optimality' condition for this variable, and an ad hoc prescription for $\cbfe^{(1)}$ is therefore required. Hereafter, following \citet{Idiart06a} we identify $\cbfe^{(1)}$ with the macroscopic deviatoric strain, i.e.,
\begin{equation}
        \cbfe^{(1)} = \obfe_d.
        \label{eref}
\end{equation}
In addition to giving sensible results in the case of isotropic composites, this prescription has the advantage of simplicity. It is emphasized, however, that this prescription is probably not `optimal'.

Then, the stationarity conditions in (\ref{SOW}) and (\ref{VFunct}) yield a system of nonlinear algebraic equations for the variables $\hbfe^{(1)}$, $\l0$ and $\m0$. More explicitly, making use of (\ref{mat_const})$_2$, (\ref{L0}) and (\ref{eref}), the conditions resulting from the stationary operation in (\ref{VFunct}) are
\begin{equation}
    3 \l0 \, \left(\hepa^{(1)} - \oeeq \right)\! =\! (\hepa^{(1)}/\heeq^{(1)})\, \phi'(\heeq^{(1)}) - \phi'(\oeeq), \quad
    3 \m0\! =\!(1 /\heeq^{(1)})\,\phi'(\heeq^{(1)}),
    \label{gen_comp}
\end{equation}
where $\hat{\varepsilon}_{\parallel,\perp}^{(1)} = \sqrt{(2/3) \hbfe^{(1)}\!:\!\bfE_{\pr,\pp}\!:\!\hbfe^{(1)}}$ represent the components of $\hbfe^{(1)}$ that are `parallel' and `perpendicular' to the macroscopic strain $\obfe$. The equivalent part of $\hbfe^{(1)}$ is then
$(\heeq^{(1)})^2=(\hepa^{(1)})^2+(\hepe^{(1)})^2$.

In turn, the conditions resulting from the stationarity operation in (\ref{SOW}) are
\begin{equation}
    \label{epe}
    \hepa^{(1)}=\oeeq - \sqrt{ \frac{1}{1-f} \ \frac{2}{3} \frac{\p \Weff_L}{\p \l0}}, \quad
    \hepe^{(1)}=\sqrt{ \frac{1}{1-f} \ \frac{2}{3} \frac{\p \Weff_L}{\p \m0} },
\end{equation}
where a choice of roots has been made \citep{Idiart06a}. The arguments inside the square roots in these relations depend on certain traces of the strain
covariance tensor in the matrix phase of the LCC \citep{PC02a} as identities (\ref{eeL}) show. It then follows from relations (\ref{gen_comp}) that the moduli $\l0$ and $\m0$ depend on the intraphase field fluctuations (in the LCC). This is one of the main reasons for the improvement of the `second-order' method over earlier linear comparison methods which made use of only the first or the second moments of local fields \citep{PCS98}.

Finally, making use of the relations (\ref{gen_comp}), the estimate (\ref{SOW}) reduces to
\begin{equation}
        \Weff(\obfe) = (1-f) \left[ \phi(\heeq^{(1)}) - \phi'(\oeeq) \left( \hepa^{(1)} - \oe_{Le}^{(1)} \right) \right],
        \label{SOW2}
\end{equation}
where $\oe_{Le}^{(1)}$ is the equivalent part of (\ref{eeL})$_1$.

\vspace{2mm}
\noindent\textit{Effective behavior.} An estimate for the effective behavior of the nonlinear porous material is obtained by differentiating
expression (\ref{SOW}) with respect to $\obfe$. For the specific choice of reference tensor (\ref{eref}), this yields \citep{Idiart07}
\begin{equation}
        \obfs = \frac{\p \Weff}{\p \obfe}(\obfe) = \obfs_L + (1-f) \ \bfrho^{(1)},
        \label{somac}
\end{equation}
where $\obfs_L = \Leff\, \obfe + \taueff$ is the macroscopic stress in the LCC, and the (incompressible) tensor $\bfrho^{(1)}$ is given by
\begin{equation}
    \bfrho^{(1)} = \left[ \bfL^{(1)} - \bfL_t^{(1)} \right]\!:\!(\hbfe^{(1)} - \obfe_L^{(1)}) + \frac{4}{3}\frac{\l0 - \m0}{\oeeq^2}\! \left[\left\langle (\bfe_{L_d}\!\!-\obfe_d)\! \otimes\! (\bfe_{L_d}\!-\!\obfe_d) \right\rangle^{(1)}\!\! - \!(\hbfe_d^{(1)}\!\! - \!\obfe_d)\!\otimes\!(\hbfe_d^{(1)}\!\! - \!\obfe_d)\right]\!:\!\obfe_d,
  \label{rhoso}
\end{equation}
where $\bfL_t^{(1)} = \p^2 w^{(1)}(\obfe)/\p \bfe \p \bfe$ is the tangent modulus tensor of the matrix phase evaluated at $\obfe$.
The averaged quantity appearing in the second term is determined from (\ref{eeL}).

\vspace{2mm}
\noindent\textit{Field statistics.} In addition to the effective behavior, estimates for the field statistics can also be extracted from
corresponding estimates for $\Weff$ by means of the procedure proposed by \citet{Idiart07}. Applying this procedure in the context of the `second-order' method, the following estimates for the first and second moments of the fields in each phase are obtained \citep{Idiart07}:
\begin{eqnarray}
    \obfe\r=\obfe_L\r, && \obfs\r = \obfs\r_L + \bfrho\r, \label{sav} \\
    \langle \bfe \otimes \bfe \rangle\r=\langle \bfe_L \otimes \bfe_L \rangle\r,  && \langle \bfs \otimes \bfs \rangle\r=\langle \bfs_L \otimes \bfs_L \rangle\r + 2 \cbfs\r \otimes_s \bfrho\r, \label{ssec}
\end{eqnarray}
where $\cbfs^{(1)} = \p w^{(1)}(\obfe)/\p \bfe$, in the porous phase $\cbfs^{(2)}=\bfrho^{(2)}=\bf0$, and the subscript $s$ in the tensor product
denotes symmetrization. The strain statistics in the LCC are given by expressions (\ref{eeL}), and the corresponding stress statistics
follow from the linear stress-strain relation associated with (\ref{wL}). Thus, while the nonlinear estimates for the strain statistics in the matrix coincide with those quantities in the LCC, the estimates for the stress statistics exhibit certain correction terms which depend on the matrix nonlinearity through the tensor (\ref{rhoso}).

\vspace{2mm}
\noindent\textit{Expressions for power-law materials under `aligned' loadings.} When the macroscopic loading is `aligned' in the sense described in Sec. \ref{sec:local} (see Fig. \ref{fig:micro}), the (compressible) effective tensor $\Leff$ exhibits the same symmetry as the $\bfL^{(1)}$ defined by (\ref{L0}) with (\ref{Etensors}) and (\ref{eref}), i.e.,
\begin{equation}
    \Leff = 2 \keff \bfJ + 2\leff \bfE_{\pr} + 2\meff \bfE_{\pp},
    \label{Leff}
\end{equation}
where $\keff$ is a bulk modulus, $\leff$ and $\meff$ are the `parallel' and `perpendicular' shear moduli, and $\bfJ$ denotes the standard fourth-order hydrostatic identity tensor. In this work, the moduli are determined approximately using the linear estimates of the Hashin-Shtrikman type derived in \cite{Willot08a}. For SS loadings, $\bfE_{\pr,\pp}$ coincide with the tensors $\bfE^\text{SS,PS}$ defined by (\ref{E-tensors}), so that the moduli are given by expressions (22) of \cite{Willot08a}, with $m=\ell=0$; for PS loadings, $\bfE_{\pr,\pp}$ coincide with the tensors $\bfE^\text{PS,SS}$, so that the moduli are given by those same expressions with $\l0$ ($\leff$) and $\m0$ ($\meff$) interchanged.

Evaluating the derivatives of $\Weff_L$ in expressions (\ref{epe}) and (\ref{eeL}), we arrive at the following relations:
\begin{equation}
    \hepa^{(1)}/\oeeq = 1 - \sqrt{ \left( 1+\tau \right)^2\, \hat{r}_{\pr} + \left( 1 - \overline{\eps}_{L\,e}^{(1)}/\oeeq\right)^2}, \quad
    \hepe^{(1)}/\oeeq = |1+\tau|\ \hat{r}_{\pp}^{1/2}, \quad
    \overline{\eps}_{L\,e}^{(1)}/\oeeq = 1 + \left( 1+\tau \right)\, \overline{r},
    \label{eav_e}
\end{equation}

where $\tau = (\bftau^{(1)} \cdot \bftau^{(1)})^{1/2}/(3 \l0 \oeeq)$, and where
\begin{equation}
    \hat{r}_{\pr} = \frac{1}{(1-f)^2} \left\{\!(1-f)\! \left[\frac{\leff}{\l0}\!+\!\left(\frac{\leff}{\l0}\right)'\!\! k \right]\!\! -\!\! \left(\frac{\leff}{\l0}\right)^2\!\right\}, \quad    \hat{r}_{\pp} = -\frac{1}{(1-f)} \left( \frac{\leff}{\l0}\right)'\, k^2, \quad \overline{r}= \frac{1}{(1-f)} \frac{\leff}{\l0} - 1.
\label{hav}
\end{equation}

Here, the prime denotes differentiation with respect to $k$. Making use of (\ref{gen_comp}), one obtains $\tau = (1/k-1) (\hepa^{(1)}/\oeeq)$, which, after some algebraic manipulations, yields
\begin{equation}
    \hepa^{(1)}/\oeeq = 1 - r/(1-k), \quad \hepe^{(1)}/\oeeq = [(1-r)/k]\ \hat{r}_{\pp}^{1/2}, \quad \overline{\eps}_{L\,e}^{(1)}/\oeeq = 1+[(1-r)/k]\ \overline{r},
    \label{eav_e2}
\end{equation}

where the function $r$ has been defined as
\begin{equation}
        r = \left[ 1 + \frac{k}{1-k} \frac{1}{\sqrt{\hat{r}_{\pr}+\overline{r}^2}}\right]^{-1}.
        \label{h}
\end{equation}
Thus, the right hand sides of expressions (\ref{eav_e2}) are explicit in $k$. Then, by taking the ratio of the two conditions
(\ref{gen_comp}) with $\phi$ given by (\ref{power_law})$_1$, we obtain the relation
\begin{equation}
        \frac{ \ln\left[ (1-k) \left(\hepa^{(1)}/\oeeq\right) + k \right] }{ \ln\left(\heeq^{(1)}/\oeeq\right) } = 1-m,
        \label{eqk}
\end{equation}
where $\heeq^{(1)} = \sqrt{(\hepa^{(1)})^2+(\hepe^{(1)})^2}$. This relation, together with (\ref{eav_e2}), constitutes a single
nonlinear, algebraic equation for the anisotropy ratio $k$ that must be solved numerically. (Alternatively, since the right hand side of (\ref{eqk}) depends on $k$ but not on $m$, we can assign values to $k$ and obtain from (\ref{eqk}) the corresponding values of $m$.)

Finally, the `second-order' estimate (\ref{SOW2}) can be written as (\ref{power_law_eff})$_1$ with the effective flow stress given by
\begin{equation}
    \frac{\sef}{\s0} = (1-f) \left[ \left( \frac{\heeq^{(1)}}{\oeeq} \right)^{1+m}
        -(1+m)\left( \frac{\hepa^{(1)}}{\oeeq} - \frac{\overline{\varepsilon}_{L\,e}^{(1)}}{\oeeq} \right) \right],
  \label{sef}
\end{equation}
where $\hepa^{(1)}$, $\hepe^{(1)}$ and $\overline{\eps}_{L\,e}^{(1)}$ are given in terms of $k$ by relations (\ref{eav_e2}). The corresponding estimates for the field statistics follow from relations (\ref{sav})-(\ref{ssec}). Since the phase averages should be of the form
(\ref{averages}), $\obfe^{(1)}$ is completely determined by (\ref{eav_e}), and $\obfe^{(2)}$ can be readily obtained from the macroscopic balance $(1-f) \obfe^{(1)} + f \obfe^{(2)}=\obfe$. In turn, the estimates for the standard deviations of the `parallel' and `perpendicular' components of the strain in the matrix phase, as defined by (\ref{Traces}), read:
\begin{equation}
    SD^{(1)}(\eps_{\pr,\pp}) = |1+\tau| \ \hat{r}_{\pr,\pp}^{1/2}.
    \label{sdeps}
\end{equation}
In addition, the tensor $\bfrho^{(1)}$ required to compute the stress statistics simplifies to
\begin{equation}
        \bfrho^{(1)} = 2\left( \l0 - \lambda_t \right) \left(\hepa^{(1)} - \overline{\eps}_{Le}^{(1)}\right) (\obfe_d/\oeeq),
        \label{rhoiso}
\end{equation}
where $2 \lambda_t = \bfE_{\pr} \cdot \bfL_t^{(1)} = (2/3)\phi'(\oeeq)$. Since phase 2 is vacuous, the phase averages of the stress are trivial. On the other hand, the estimates for the standard deviations of the stress components, which follow from (\ref{ssec}), are rather complicated and are omitted for conciseness.

In the ideally plastic limit ($m \rightarrow 0$), Eq.\ (\ref{eqk}) may admit more than one solution for $k$ varying with $f$. The solution of interest is the positive one (for $f>0$) continuously connected to $k=1$ for $m=1$, which solves
\begin{equation}
        (1-k) \left(\hepa^{(1)}/\oeeq\right) + k = \heeq^{(1)}/\oeeq.
        \label{eqk0}
\end{equation}
The corresponding effective flow stress follows from (\ref{sef}) with $m=0$.

\vspace{2mm}
\noindent\textit{Dilute limit.} We now evaluate the above expressions as $f \rightarrow 0$. We begin by noting that from numerical solutions of (\ref{eqk}) it can be inferred that $k \rightarrow m$ for both PS and SS loadings. 

When $m>0$, the anisotropy ratio $k$ remains strictly positive. In this case, \cite{Willot08a} have shown that the effective modulus $\leff$ appearing in expressions (\ref{hav}) has an expansion of the form (see expression (25) in that Ref.)
\begin{equation}
	\leff/\l0 = 1 + \mathcal{C}_1(k) f + \mathcal{C}_2(k) f^2 + ...
	\label{leff_dil}
\end{equation}
where $\mathcal{C}_1(k)=-(1+\sqrt{k})$, and $\mathcal{C}_2(k)$ is assumed to have bounded first derivative and presumably depends on the specific loading. To first order in $f$, the strain variables (\ref{eav_e2}) are then
\begin{equation}
    \frac{\hepa^{(1)}}{\oeeq} = 1-\frac{1}{\sqrt{2}k^{3/4}} f^{1/2} + \frac{1-k}{2k^{3/2}}f, \quad
    \left(\frac{\hepe^{(1)}}{\oeeq}\right)^2 = \frac{1}{2k^{1/2}} f, \quad
    \frac{\overline{\eps}_{L\,e}^{(1)}}{\oeeq} = 1- \frac{1}{k^{1/2}} f,
    \label{eav_dil}
\end{equation}
independent of $\mathcal{C}_2(k)$. Making use of these expressions, it is easy to verify that the zeroth-order term of the left-hand side of equation (\ref{eqk}) is $(1-k)$, which confirms that $k = m + O(f)$ as $f\rightarrow 0$. Finally, it is easy to verify that the expansion of expression (\ref{sef}) for $\sef$, to first order in $f$, is given by (\ref{so_dilute_m}) with (\ref{alpha}).

When $m=0$, the anisotropy ratio $k \rightarrow 0$ as $f \rightarrow 0$. Numerical solutions of equation (\ref{eqk0}) show that $k\approx b f^{2/3}$. In this case, \cite{Willot08a} have shown that the expansion (\ref{leff_dil}) for the effective modulus $\leff$ remains valid. To order $f^{2/3}$, the strain variables (\ref{eav_e2}) are then
\begin{equation}
    \frac{\hepa^{(1)}}{\oeeq} = \frac{\sqrt{2}b^{3/4}}{1+\sqrt{2}b^{3/4}} + b^{1/4} \frac{\sqrt{2} \mathcal{C}_2(0) - b^{3/4}}{(1+\sqrt{2}b^{3/4})^2} f^{2/3},   \quad
    \left(\frac{\hepe^{(1)}}{\oeeq}\right)^2 = \frac{b}{(1+\sqrt{2}b^{3/4})^2} f^{2/3}, \quad
    \frac{\overline{\eps}_{L\,e}^{(1)}}{\oeeq} = 1- \frac{\sqrt{2} b^{1/4}}{1+\sqrt{2}b^{3/4}} f^{2/3}.
    \label{eav0_dil}
\end{equation}
The coefficient $b$ is obtained by expanding equation (\ref{eqk0}) to order $f^{2/3}$; the solution is $b=1/4$, independent of $\mathcal{C}_2(0)$ and in agreement with numerical results. Having determined the strain variables (\ref{eav0_dil}), it is easy to show that the expansion of expression (\ref{sef}) for $\sef$, to order $f^{2/3}$, is given by (\ref{so_dilute}) with $\alpha_{so}=1/2$. Note that the result is independent of the second-order coefficient $\mathcal{C}_2(k)$.

\subsection{Stress formulation}
\label{sec:sou}

A dual version of the `second-order' method follows from exactly analogous arguments. When specialized to porous materials with an isotropic matrix phase, the `second-order' variational estimate for the effective stress potential is given by \citep{PC02a,Idiart06a}
\begin{equation}
    \Ueff(\obfs)\!=\!\!\mathop{\rm stat}_{\l0,\m0}\! \left\{\Ueff_L(\obfs;\cbfs^{(1)}\!\!,\bfM^{(1)})\!-\!(1\!-\!f) v^{(1)}(\cbfs^{(1)}\!\!, \bfM^{(1)})\right\}.
    \label{SOU}
\end{equation}
In this expression, $\Ueff_L$ is the effective potential of a porous LCC with a matrix characterized by a second-order, Taylor-type expansion of the nonlinear potential (\ref{mat_const_u})$_2$ about a reference stress tensor $\cbfs^{(1)}$, given by
\begin{equation}
    u^{(1)}_L(\bfs; \cbfs^{(1)}, \bfM^{(1)})\!\! = \!\!u^{(1)}(\cbfs^{(1)})\!+\!\frac{\p u^{(1)}}{\p\bfs}(\cbfs^{(1)})\!:\!(\bfs\!-\!\cbfs^{(1)})
    +\frac{1}{2}(\bfs\!-\!\cbfs^{(1)})\!:\!\bfM^{(1)}\!:\!(\bfs\!-\!\cbfs^{(1)}),
    \label{uL}
\end{equation}
where $\bfM^{(1)}$ is a symmetric, fourth-order tensor (of compliances) of the form
\begin{equation}
        \bfMo^{(1)} = \frac{1}{2\l0} \bfE_{\pr} + \frac{1}{2\m0}\bfE_{\pp}, \quad k = \l0/\m0.
        \label{M0}
\end{equation}
Here, $\l0$ and $\m0$ are two shear moduli, and
\begin{equation}
    \bfE_{\pr} = \frac{3}{2} \frac{\cbfs_d^{(1)}}{\cseq^{(1)}} \otimes \frac{\cbfs_d^{(1)}}{\cseq^{(1)}}, \quad \bfE_{\pp} = \bfK - \bfE_{\pr},
    \label{Etensors_u}
\end{equation}
are two orthogonal, fourth-order, projection tensors with principal axes `aligned' with $\cbfs^{(1)}$.

In turn, the `error' function $v^{(1)}$ in (\ref{SOU}) is defined as
\begin{equation}
    v^{(1)}\!(\cbfs^{(1)}\!\!,\bfM^{(1)})\!=\!\mathop{\rm stat}_{\hbfs^{(1)}}\!
    \left\{u_L^{(1)}\!(\hbfs^{(1)};\cbfs^{(1)}\!,\bfM^{(1)})\!-\!u^{(1)}(\hbfs^{(1)})\right\},
    \label{VFunct_U}
\end{equation}
and is a measure of the degree of nonlinearity of $u^{(1)}$.

Given that the matrix potential (\ref{uL}) corresponds to a `thermoelastic' material with a `thermal strain' and a `specific heat' given by
\begin{eqnarray}
    \bfeta^{(1)} &=& \frac{\p u^{(1)}}{\p\bfs}(\cbfs^{(1)}) - \bfMo^{(1)}\!:\!\cbfs^{(1)},\label{eta0} \\
    h^{(1)}\! &=& \!u^{(1)}(\cbfs^{(1)})\!-\!\bfeta^{(1)}\!:\!\cbfs^{(1)}\!\!-\!\frac{1}{2}\,\cbfs^{(1)} \!:\!\bfMo^{(1)}\!:\! \cbfs^{(1)}\!, \label{g0}
\end{eqnarray}
the effective potential $\Ueff_L$ can be written as \citep{Willis81}
\begin{equation}
    \Ueff_L(\obfs) = \frac{1}{2} \, \obfs \!:\!\Meff\!:\! \obfs + \etaeff\!:\!\obfs + \geff,
    \label{UeffL}
\end{equation}
where $\etaeff = \bfeta^{(1)}$, $\geff=(1-f)\, h^{(1)}$, and $\Meff$ denotes the effective compliance tensor of a porous material with a purely elastic matrix with compliance tensor (\ref{M0}). It is useful to recall at this point that the first and second moments of the stress field $\bfs_L$ in the matrix of the porous LCC can be obtained from relations completely analogous to (\ref{eeL}) \citep{Idiart07}.

In addition, the reference stress $\cbfs^{(1)}$ must also be specified in order to characterize the potential $u^{(1)}_L$. As already mentioned, an
`optimality' condition for this variable, if it exists, is not yet available, and an ad hoc prescription is therefore required. Following \citet{Idiart06a} we identify $\cbfs^{(1)}$ with the macroscopic deviatoric stress, i.e.,
\begin{equation}
        \cbfs^{(1)} = \obfs_d.
        \label{sref}
\end{equation}
Then, the stationarity conditions in (\ref{SOU}) and (\ref{VFunct_U}) yield a system of nonlinear algebraic equations for the variables $\hbfs^{(1)}$, $\l0$ and $\m0$. With (\ref{mat_const_u})$_2$, (\ref{M0}) and (\ref{sref}), the conditions resulting from the stationary operation in (\ref{VFunct_U}) read:
\begin{equation}
    \left(\hspa^{(1)}-\oseq \right)/(3\l0)\!=\!\psi'(\hseq^{(1)})(\hspa^{(1)}/\hseq^{(1)})\!-\!\psi'(\oseq), \quad
    (1/3\m0) = \psi'(\hseq^{(1)})/\hseq^{(1)}.
\label{gen_comp_u}
\end{equation}
where $\hat{\sigma}_{\parallel,\perp}^{(1)} = \sqrt{(3/2) \hbfs^{(1)}\!:\!\bfE_{\pr,\pp}\!:\!\hbfs^{(1)}}$ represent the components of $\hbfs^{(1)}$ that
are `parallel' and `perpendicular' to the macroscopic stress $\obfs$. The equivalent part of $\hbfs^{(1)}$ is then $(\hseq^{(1)})^2 = (\hspa^{(1)})^2+(\hspe^{(1)})^2$. In turn, the conditions resulting from the stationarity operation in (\ref{SOU}) are
\begin{equation}
    \hspa^{(1)} = \oseq + \sqrt{ \frac{6}{1-f}\, \frac{\p \Ueff_L}{\p \l0^{-1}} }, \quad
    \hspe^{(1)} = \sqrt{ \frac{6}{1-f}\, \frac{\p \Ueff_L}{\p \m0^{-1}} },
    \label{spe}
\end{equation}
where a choice of roots has been made.\cite{Idiart06a} Finally, making use of the relations (\ref{gen_comp_u})-(\ref{spe}), the estimate (\ref{SOU}) can be written in the simplified form
\begin{equation}
    \Ueff(\obfs) = (1-f) \left[\psi(\hseq^{(1)})-\psi'(\oseq)\, (\hspa^{(1)}-\overline{\sigma}_{L\,e}^{(1)}) \right],
  \label{SOU2}
\end{equation}
where $\os_{Le}^{(1)}$ is the equivalent part of the average stress in the matrix of the LCC.

\vspace{2mm}
\noindent\textit{Effective behavior.} Differentiating expression (\ref{SOU}) with respect to $\obfs$ yields \citep{Idiart07}
\begin{equation}
        \obfe = \frac{\p \Ueff}{\p \obfs}(\obfs) = \obfe_L + (1-f) \ \bfg^{(1)},
        \label{somac_u}
\end{equation}
where $\obfe_L = \Meff \obfs + \etaeff$ is the macroscopic strain in the LCC, and the (incompressible) tensor $\bfg^{(1)}$ is given by
\begin{eqnarray}
    \bfg^{(1)} &=& \left[\bfM^{(1)}-\bfM_t^{(1)}\right]\!:\!(\hbfs^{(1)}-\obfs_L^{(1)})+\frac{3}{4}\frac{\l0^{-1}-\m0^{-1}}{\oseq^2}\\
                     & & \!\times\!\!\left[\left\langle(\bfs_{L_d}\!\!\!-\!\obfs_d)\!\otimes\!(\bfs_{L_d}\!\!\!-\!\obfs_d)\right\rangle^{(1)} \!\!\!-\!(\hbfs_d^{(1)}\!\!\!-\!\!\obfs_d)\!\otimes\!(\hbfs_d^{(1)}\!\!\!-\!\obfs_d) \right]\!:\!\obfs_d.\nonumber
\label{gammaso}
\end{eqnarray}
Here, $\bfM_t^{(1)} = \p^2 u^{(1)}(\obfs)/\p \bfs \p \bfs$ is the tangent compliance tensor of the matrix phase evaluated at $\obfs$.

\vspace{2mm}
\noindent\textit{Field statistics.} Corresponding estimates for the first and second moments of the fields in each phase corresponding to (\ref{SOU2}) are given by \citep{Idiart07}
\begin{eqnarray}
     \obfe\r=\obfe_L\r + \bfg\r, && \obfs\r=\obfs\r_L, \label{eav_u} \\
     \langle \bfe \otimes \bfe \rangle\r = \langle \bfe_L \otimes \bfe_L \rangle\r + 2\,\cbfe\r \otimes_s \bfg\r, &&
\langle \bfs \otimes \bfs \rangle\r = \langle \bfs_L \otimes \bfs_L \rangle\r, \label{ssec_u}
\end{eqnarray}
where $\cbfe^{(1)} = \p u^{(1)}(\obfs)/\p \bfs$, in the porous phase $\cbfe^{(2)}=\bfg^{(2)}=\bf0$, and the subscript $s$ in the tensor product denotes symmetrization. Thus, while the nonlinear estimates for the stress statistics in the matrix coincide with those quantities in the LCC, the estimates for the strain statistics exhibit certain correction terms which depend on the matrix nonlinearity through the tensor (\ref{gammaso}).

\vspace{2mm}
\noindent\textit{Expressions for power-law materials under `aligned' loadings.} When the macroscopic loading is `aligned' the tensor $\Meff$ is given in terms of $\l0$ and $\m0$ by the inverse of (\ref{Leff}). Then, evaluating the derivatives of $\Ueff_L$ in (\ref{spe}), we arrive at the following relations:
\begin{equation}
    \hspa^{(1)}/\oseq = 1 + \sqrt{\hat{r}_{\pr}+\left(1-\os_{L\,e}^{(1)}/\oseq\right)^2}, \quad \hspe^{(1)}/\oseq = \hat{r}_{\pp}^{1/2}, \quad
  \os_{L\,e}^{(1)}/\oseq = 1/(1-f),
  \label{sav_e}
\end{equation}
where the functions $\hat{r}_{\pr}$ and $\hat{r}_{\pp}$ are given by
\begin{equation}
    \hat{r}_{\pr} = \frac{1}{(1-f)^2} \left\{\!(1-f)\!\left[ \frac{\leff^{-1}}{\l0^{-1}} - k \left(\frac{\leff^{-1}}{\l0^{-1}}\right)'\right] \!-1\!\right\}, \quad \hat{r}_{\pp} = \frac{1}{1-f} \left( \frac{\leff^{-1}}{\l0^{-1}} \right)'.
    \label{hpe_u}
\end{equation}
Again, the prime in these expressions denotes differentiation with respect to $k$, and the ratio $\leff/\l0$ is a function of $k$ and $f$, and depends on the specific loading. Explicit expressions have been given in \cite{Willot08a}. Then, by taking the ratio between the two conditions (\ref{gen_comp_u}) with $\psi$ given by (\ref{power_law})$_2$, we obtain the relation
\begin{equation}
       \frac{\ln\left[ \left( 1-1/k\right)\left(\hspa^{(1)}/\oseq\right) + (1/k)\right]}{\ln\left(\hseq^{(1)}/\oseq\right)} = 1 - n,
        \label{eq_u}
\end{equation}
where $\hseq^{(1)} = \sqrt{(\hspa^{(1)})^2+(\hspe^{(1)})^2}$. This relation, together with (\ref{sav_e}), constitutes a single
nonlinear, algebraic equation for the anisotropy ratio $k$ that must be solved numerically.

Finally, the `second-order' estimate (\ref{SOU2}) can be written as (\ref{power_law_eff})$_2$ with the effective flow stress given by
\begin{equation}
        \frac{\sef}{\s0}=(1-f)^{-\frac{1}{n}}\!\left[\left(\frac{\hseq^{(1)}}{\oseq} \right)^{1+n}\hspace{-1.5em}-(1+n)\!\left(\frac{\hspa^{(1)}}{\oseq} - \frac{\overline{\sigma}_{L\,e}^{(1)}}{\oseq} \right) \right]^{-\frac{1}{n}},
        \label{sef_u}
\end{equation}
where $\hspa^{(1)}$, $\hspe^{(1)}$ and $\os_{L\,e}^{(1)}$ are given in terms of $k$ by relations (\ref{sav_e}). The corresponding estimates for the field statistics follow from relations (\ref{eav_u})-(\ref{ssec_u}). The estimates for the standard deviations of the stress components in the matrix phase, as defined by (\ref{Traces}), are given by
\begin{equation}
    SD^{(1)}(\sigma_{\pr,\pp}) = \hat{r}_{\pr,\pp}^{1/2},
    \label{sdsig}
\end{equation}
and the tensor $\bfg^{(1)}$ required to compute the strain statistics simplifies to
\begin{equation}
        \bfg^{(1)} = \frac{1}{2} \left( \l0^{-1} - \lambda_t^{-1} \right) \left(\hspa^{(1)} - \overline{\sigma}_{Le}^{(1)}\right) \obfs_d/\oseq,
        \label{gammaiso}
\end{equation}
where $(2 \lambda_t)^{-1} = \bfE_{\pr} \cdot \bfM_t^{(1)} = (3/2)\psi'(\oseq)$. However, the estimates for the strain statistics, which follow from (\ref{eav_u}) and (\ref{ssec_u}), are rather complicated and are omitted for conciseness.

In the ideally plastic limit ($n \rightarrow \infty$), equation (\ref{eq_u}) simplifies further to
\begin{equation}
        \hspa^{(1)}/\oseq = 1/(1-k),
        \label{eq_u_0}
\end{equation}
and the `second-order' estimate (\ref{sef_u}) reduces to
\begin{equation}
        \frac{\sef}{\s0} = \left(\hseq^{(1)}/\oseq\right)^{-1}.
        \label{sef_u_0}
\end{equation}

\vspace{5mm}
\noindent\textit{Dilute limit.} We now evaluate the above expressions as $f \rightarrow 0$. As in the strain version, numerical solutions of (\ref{eq_u}) indicate that $k \rightarrow m$ for both PS and SS loadings.

When $m>0$, the anisotropy ratio $k$ remains strictly positive and the effective modulus $\leff$ is given by (\ref{leff_dil}) to second order in $f$ \citep{Willot08a}. To first order in $f$, the stress variables (\ref{sav_e}) are then
\begin{equation}
    \frac{\hspa^{(1)}}{\oseq} = 1 + \frac{k^{1/4}}{\sqrt{2}} f^{1/2} + O(f^{3/2}), \quad
    \left(\frac{\hspe^{(1)}}{\oseq}\right)^2 = \frac{1}{2k^{1/2}} f, \quad
    \label{sav_dil}
\end{equation}
independent of $\mathcal{C}_2(k)$. Making use of these expressions, it is easy to verify that the zeroth-order term of the left-hand side of equation (\ref{eq_u}) is $(1-1/k)$, which confirms that $k = m + O(f)$ as $f\rightarrow 0$. Finally, it is easy to verify that the expansion of expression (\ref{sef_u}) for $\sef$, to first order in $f$, is given by (\ref{so_dilute_m}) with (\ref{alpha}).

When $m=0$, the anisotropy ratio $k \rightarrow 0$ as $f \rightarrow 0$. As in the strain version, numerical solutions of equation (\ref{eq_u_0}) show that $k\approx b f^{2/3}$, and the expansion (\ref{leff_dil}) for the effective modulus $\leff$ remains valid. To order $f^{2/3}$, the stress variables (\ref{sav_e}) are then
\begin{equation}
    \frac{\hspa^{(1)}}{\oseq} = 1 + \frac{b^{1/4}}{\sqrt{2}} f^{2/3},   \quad
    \left(\frac{\hspe^{(1)}}{\oseq}\right)^2 = \frac{1}{2 b^{1/2}} f^{2/3}. \quad
    \label{sav0_dil}
\end{equation}
The coefficient $b$ is obtained by expanding equation (\ref{eq_u_0}) to order $f^{2/3}$; the solution is $b=1/2^{2/3}$, in agreement with numerical results. Having determined the stress variables (\ref{sav0_dil}), it is easy to show that the expansion of expression (\ref{sef_u}) for $\sef$, to order $f^{2/3}$, is given by (\ref{so_dilute}) with $\alpha_{so}=3/2^{5/3}$.

\subsection{Estimates for rigidly-reinforced composites}

Making use of the duality relations of Section \ref{sec:dual}, second-order estimates for rigidly-reinforced composites can be generated directly from the above estimates for porous materials. According to those relations, see Table \ref{table:duality}, we have the following correspondences between the two sets of estimates: $1/m \rightarrow m$, $(\sef/\s0)^{-1/m}\rightarrow (\sef/\s0)$, SS (PS) loading $\rightarrow$ PS (SS) loading, $1/k\rightarrow k$, strain (stress) version $\rightarrow$ stress (strain) version, strain (stress) statistics $\rightarrow$ stress (strain) statistics.

\section{Limit analysis bounds for periodic materials}
\label{app:la}

\begin{figure*}[ht]
\centerline{
\begin{tabular}{ccccc}
    \includegraphics*[width=0.95in]{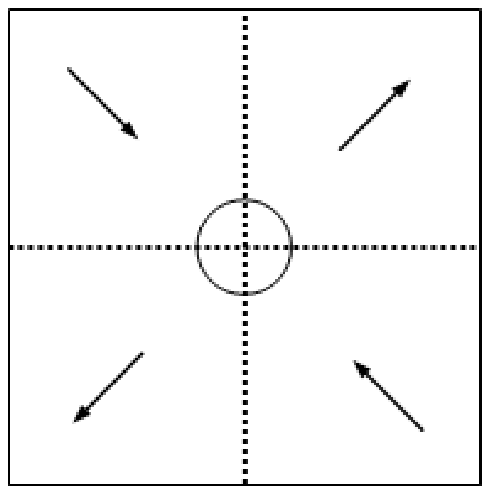} &
    \includegraphics*[width=0.95in]{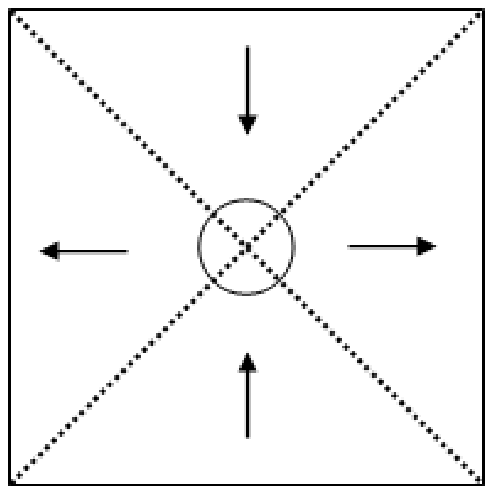} &
    \includegraphics*[width=0.95in]{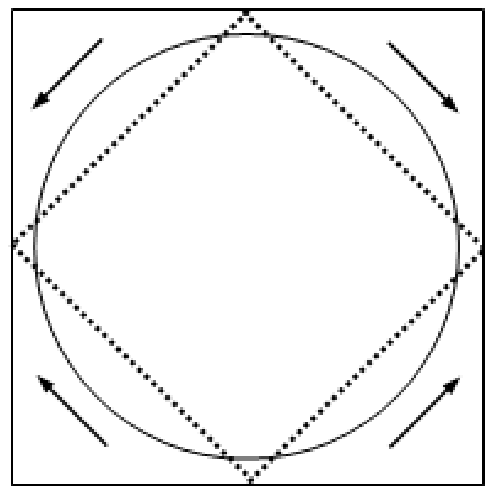} &
    \includegraphics*[width=0.95in]{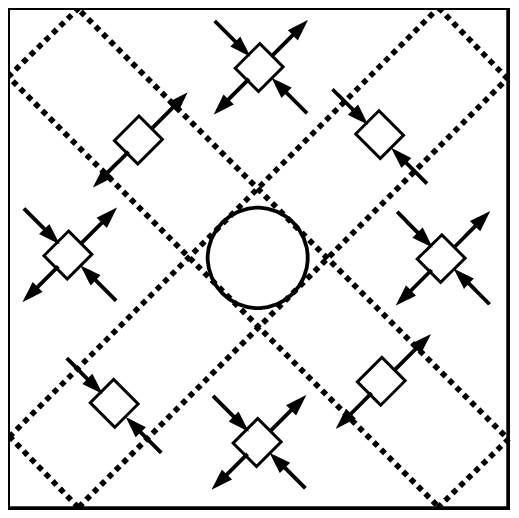} &
    \includegraphics*[width=0.95in]{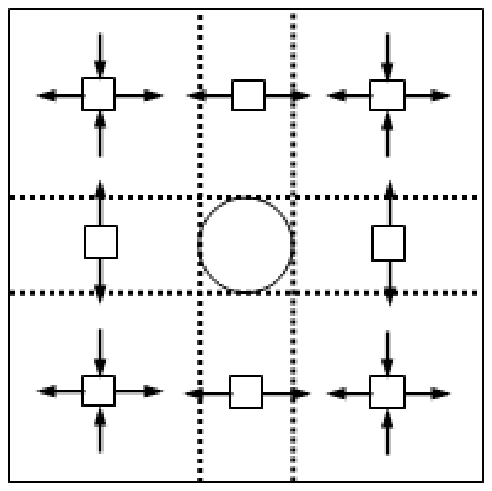} \\
    (a) & (b) & (c) & (d) & (e)
\end{tabular}
}
\caption{Trial fields used to obtain limit analysis bounds for porous materials. (a)-(c) Arrows denote displacement vectors, all of the same
magnitude;  (d)-(e) Arrows denote traction vectors associated with the stress field, all of the same magnitude. Dotted lines indicate surfaces of
discontinuity.}
        \label{fig:la_displ}
\end{figure*}

In the limiting case of a rigid-ideally plastic matrix, use can be made of the kinematic and static theorems of classical limit analysis to compute upper and lower bounds, respectively, for the effective flow stress of the composite \citep{Bouchitte91}. The computation of such bounds amounts to evaluating the effective potentials (\ref{EffConst}) at (simple) trial fields $\bfe(\bfx)$ (or displacement $\bfu(\bfx)$) and $\bfs(\bfx)$, which in perfect plasticity include piecewise continuous fields with surfaces of discontinuity not only at material interfaces but also within the individual phases. Thus, this allows the use of piecewise uniform trial fields, which can lead to simple analytical bounds, as shown below. For consiceness, the procedure for computing these bounds is only briefly addressed here, and the interested reader is referred to \citet{Salencon83,Taliercio92,Maghous91}.

\subsection{Porous materials}

\noindent\textit{Simple shear.} Making use of the trial displacement and stress fields shown in parts (a) and (d) of Fig. \ref{fig:la_displ}, the following bounds for $\sef$ are obtained:
\begin{equation}
    1 - 2 \sqrt{2/\pi}\, f^{1/2} \leq \ \sef/\s0 \ \leq 1 - (2/\sqrt{\pi}) f^{1/2}.
    \label{bounds_ss}
\end{equation}

\noindent\textit{Pure shear.} Making use of the trial displacement and stress fields shown in parts (b), (c) and (e) of Fig. \ref{fig:la_displ}, the following bounds for $\sef$ are obtained:
\begin{equation}
    1 - \frac{2}{\sqrt{\pi}} f^{1/2} \leq \frac{\sef}{\s0} \leq
    \begin{cases}
        \displaystyle{ 1 - \sqrt{\frac{2}{\pi}}\,f^{1/2}}, & 0  \leq f \leq \frac{\pi}{6} \cr
        \displaystyle{ 1 - \sqrt{\frac{8 f}{\pi} - 1} },  & \frac{\pi}{6} < f \leq
        \frac{\pi}{4}
    \end{cases}
    \label{bounds_ps}
\end{equation}
Unlike in the case of simple shear loading, two different displacement trial fields have been used in this case, motivated by the FFT results shown in Table \ref{table:maps}. The trial field shown in Fig. \ref{fig:la_displ}b yields the smaller bound for low to moderate porosities, while that of Fig.
\ref{fig:la_displ}c yields the smaller bound for porosities closer to the close-packing limit. Both bounds coincide for $f = \pi/6 \approx 0.524$.

\subsection{Rigidly-reinforced materials}
\label{app-rigidly-reinforced}

\begin{figure}[t]
\centerline{
\begin{tabular}{c@{\hspace{5mm}}c}
    \includegraphics[width=1in]{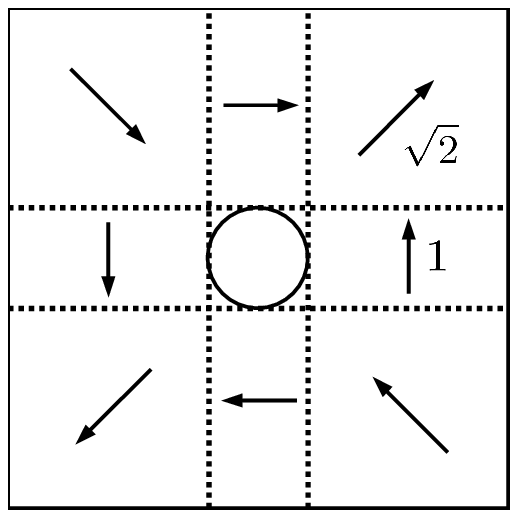} &
    \includegraphics[width=1in]{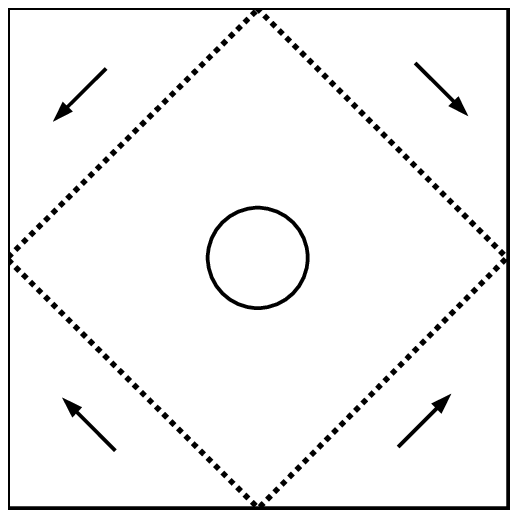} \\
     (a) SS: $0 \leq f \leq \frac{\pi}{4}$ & (b) PS: $0 \leq f \leq \frac{\pi}{8}$
\end{tabular}
}
\caption{Trial displacement fields used to obtain limit analysis bounds for rigidly-reinforced materials. Dotted lines indicate surfaces of discontinuity.}
\label{fig:la_displ_rig}
\end{figure}

Figure \ref{fig:la_displ_rig} shows two displacement trial fields for SS and PS loadings. Both cases involve straight shear bands passing entirely through the matrix, from which it follows immediately that the resulting upper bound for $\sef$ under both loading conditions is
\begin{equation}
    \sef/\s0 \leq 1.
    \label{bounds_rig}
\end{equation}
This upper bound agrees exactly with the corresponding Reuss lower bound, and consequently it is the \textit{exact} result. It should be emphasized, however, that in the PS case, the simple trial field of Fig. \ref{fig:la_displ_rig}b is valid for $f \leq \pi/8$. For larger values of $f$, trial fields involving straight shear bands are no longer possible, and therefore the resulting upper bound on $\sef$ will exhibit a finite reinforcement effect.

\bibliographystyle{elsart-harv.bst}

\begin{thebibliography}{}

\bibitem[Bilger et al.(2005)]{Bilger05} Bilger, N., Auslender, F., Bornert, M., Michel, J.-C., Moulinec, H., Suquet, P., Zaoui, A., 2005.
\newblock Effect of a nonuniform distribution of voids on the plastic response of voided materials: a computational and statistical analysis.
\newblock Int. J. Solids Struct. 42, 517--538.

\bibitem[Bouchitt\'{e} and Suquet(1991)]{Bouchitte91} Bouchitt\'{e}, G., Suquet, P., 1991.
\newblock Homogenization, plasticity and yield design.
\newblock Proc. Workshop on Composite Media and Homogenization Theory, 107--133.

\bibitem[Chenchiah and Bhattacharya(2005)]{CHEN05}
Chenchiah, I.V. and Bhattacharya, K., 2005.
\newblock Examples of non-linear homogenization involving degenerate energies. I. Plane strain.
\newblock Proc.\ R.\ Soc.\ A 461, 3681--3703.

\bibitem[Danas et al.(2008b)]{DIP08b} Danas, K., Idiart, M.I., Ponte Casta\~neda, P., 2008.
\newblock  Homogenization-based constitutive model for isotropic viscoplastic porous media.
\newblock {Int. J. Solids Struct.} 45, 3392--3409.

\bibitem[Drucker(1966)]{Drucker66} Drucker, D. C., 1966.
\newblock The continuum theory of plasticity on the macroscale and the microscale.
\newblock J. Mater. 1, 873--910.

\bibitem[Duva and Hutchinson(1984)]{Duva84} Duva, J. M., Hutchinson, J. W., 1984.
\newblock Constitutive potentials for dilutely voided nonlinear materials.
\newblock Mech. Mater. 3, 41--54.

\bibitem[Duxbury et al.(2006)]{Duxbury06} Duxbury, P. M., McGarrity, E. S., Holm, E. A., 2006.
\newblock Critical manifolds in non-linear response of complex materials.
\newblock Mech. Mater. 38, 757–-771.

\bibitem[Eshelby(1957)]{Eshelby57} Eshelby, J. D., 1957.
\newblock The determination of the elastic field of an ellipsoidal inclusion, and related problems.
\newblock Proc. R. Soc. Lond. A 241, 376--396.

\bibitem[Fleck and Hutchinson(1986)]{Fleck86} Fleck, N. A., Hutchinson, J. W., 1986.
\newblock Void growth in shear.
\newblock Proc. R. Soc. Lond. A 407, 435--458.

\bibitem[Francfort and Suquet(2001)]{Francfort01} Francfort, G. A., Suquet, P., 2001.
\newblock Duality relations for nonlinear incompressible two dimensional elasticity.
\newblock Proc. R. Soc. Edimb. A 131, 351--369.

\bibitem[Hill(1965)]{Hill65} Hill, R., 1965.
\newblock Continuum micro-mechanics of elastoplastic polycrystals.
\newblock J. Mech. Phys. Solids 13, 89--10.

\bibitem[Hutchinson(1976)]{Hutch76} Hutchinson, J. W., 1976.
\newblock Bounds and self-consistent estimates for creep of polycrystalline
  materials.
\newblock Proc. R. Soc. Lond. A 348, 101--127.

\bibitem[Idiart and Ponte Casta\~neda(2007a)]{Idiart07} Idiart, M. I., Ponte Casta\~neda, P., 2007a.
\newblock Field statistics in nonlinear composites. I. Theory.
\newblock Proc. R. Soc. Lond. A 463, 183--202.

\bibitem[Idiart and Ponte Casta\~neda(2007b)]{Idiart07b} Idiart, M. I., Ponte Casta\~neda, P., 2007b.
\newblock Field statistics in nonlinear composites. II. Applications.
\newblock Proc. R. Soc. Lond. A 463, 203--222.

\bibitem[Idiart et al.(2006a)]{Idiart06a} Idiart, M. I., Danas, K., Ponte Casta\~neda, P., 2006a.
\newblock Second-order theory for nonlinear composites and application to isotropic constituents.
\newblock C. R. Mecanique 334, 575--581.

\bibitem[Idiart et al.(2006b)]{Idiart06b} Idiart, M. I., Moulinec, H., Ponte Casta\~neda, P., Suquet, P., 2006b.
\newblock Macroscopic behavior and field fluctuations in viscoplastic composites: second-order estimates vs full-field simulations.
\newblock J. Mech. Phys. Solids 54, 1029--1063.

\bibitem[Kachanov(2004)]{Kachanov04} Kachanov, L. M., 2004.
\newblock \textit{Fundamentals of the Theory of Plasticity}
\newblock Dover, New York.

\bibitem[Lebensohn et al.(2004a)]{Lebensohn04a} Lebensohn, R. A., Liu, Y., Ponte Casta\~neda, P., 2004.
\newblock Macroscopic properties and field fluctuations in model power-law polycrystals: full-field solutions versus self-consistent estimates.
\newblock Proc. R. Soc. Lond. 460, 1381--1405.

\bibitem[Lebensohn et al.(2004b)]{Lebensohn04b} Lebensohn, R. A., Liu, Y., Ponte Casta\~neda, P., 2004.
\newblock On the accuracy of the self-consistent approximation for polycrystals: Comparisons with full-field numerical simulations.
\newblock Acta Mat. 52, 5347--5361.

\bibitem[Lebensohn et al.(2007)]{Lebensohn07} Lebensohn, R. A., Tom\'e, C.N., Ponte Casta\~neda, P., 2007.
\newblock Self-consistent modelling of the mechanical behaviour of viscoplastic polycrystals incorporating intragranular field fluctuations.
\newblock Phil. Mag. 87, 4287--4322.


\bibitem[Lee and Mear(1992)]{Lee92} Lee, B.J., Mear, M.E., 1992.
\newblock Effective properties of power-law solids containing elliptical inhomogeneities. Part I: Rigid inclusions.
\newblock Mech. Mater. 13, 313--335.

\bibitem[Maghous(1991)]{Maghous91} Maghous, S., 1991.
\newblock D\'{e}termination du Crit\`{e}re de R\'{e}sistance Macroscopique d'un Mat\'{e}riau H\'{e}t\'{e}rog\`{e}ne \`{a} Structure P\'{e}riodique.
\newblock Ph.D. Thesis, \'{E}cole Nationale des Ponts et Chauss\'{e}es.

\bibitem[Michel et al.(2001)]{Michel01} Michel, J.-C., Moulinec, H., Suquet, P., 2001.
\newblock A computational scheme for linear and non-linear composites with arbitrary phase contrast.
\newblock Int. J. Numer. Meth. Engng. 52, 139--158.

\bibitem[Moulinec and Suquet(1994)]{Moulinec94} Moulinec, H., Suquet, P., 1994.
\newblock A fast numerical method for computing the linear and nonlinear properties of composites.
\newblock C. R. Acad. Sci. Paris II 318, 1417--1423.

\bibitem[Moulinec and Suquet(1998)]{Moulinec98} Moulinec, H., Suquet, P., 1998.
\newblock A numerical method for computing the overall response of nonlinear composites with complex microstructure.
\newblock Comp. Meth. Appl. Mech. Engng. 157, 69--94.

\bibitem[Moulinec and Suquet(2003)]{Moulinec03} Moulinec, H., Suquet, P., 2003.
\newblock Intraphase strain heterogeneity in nonlinear composites: a computational approach.
\newblock Eur. J. Mech. A 22, 751--770.

\bibitem[Moulinec and Suquet(2004)]{Moulinec04}  Moulinec, H., Suquet, P., 2004.
\newblock Homogenization for nonlinear composites in the light of numerical simulations.
\newblock In: Ponte Casta\~neda, P., et al. (eds.), \textit{Nonlinear Homogenization and its Applications to Composites, Polycrystals and Smart Materials} (Kluwer Academic Publishers, The Netherlands), 193--223.

\bibitem[Pastor and Ponte Casta\~neda(2002)]{Pastor02} Pastor, J., Ponte Casta\~neda, P., 2002.
\newblock Yield Criteria for porous media in plane strain: second-order estimates versus numerical results.
\newblock C. R. Mecanique 330, 741--747.

\bibitem[Pellegrini(2001a)]{Pell01} Pellegrini, Y.P., 2001a.
Self-consistent effective-medium approximation for strongly non-linear media. Phys.\ Rev.\ B 64, 134211.

\bibitem[Pellegrini(2001b)]{Pellegrini01} Pellegrini, Y.P., 2001b.
Unpublished work.

\bibitem[Ponte Casta\~neda(1991)]{PC91} Ponte Casta\~neda, P., 1991.
The effective mechanical properties of nonlinear isotropic composites. J.\ Mech.\ Phys.\ Solids. 39, 45--71.

\bibitem[Ponte Casta\~neda(1992)]{PC92} Ponte Casta\~neda, P., 1992.
New variational principles in plasticity and their application to
  composite materials. J.\ Mech.\ Phys.\ Solids. 40, 1757--1788.

\bibitem[Ponte Casta\~neda(1996)]{PC96} Ponte Casta\~neda, P., 1996.
Exact second-order estimates for the effective mechanical properties
  of nonlinear composite materials. J.\ Mech.\ Phys.\ Solids. 44, 827--862.

\bibitem[Ponte Casta\~neda(2002a)]{PC02a} Ponte Casta\~neda, P., 2002a.
\newblock Second-order homogenization estimates for nonlinear composites incorporating field fluctuations. I. Theory.
\newblock J. Mech. Phys. Solids 50, 737--757.

\bibitem[Ponte Casta\~neda(2002b)]{PC02b} Ponte Casta\~neda, P., 2002b.
\newblock Second-order homogenization estimates for nonlinear composites incorporating field fluctuations. II. Applications.
\newblock J. Mech. Phys. Solids 50, 759--782.

\bibitem[Ponte Casta\~neda and Suquet(1998)]{PCS98} Ponte Casta\~neda, P., Suquet, P., 1998.
\newblock Nonlinear composites.
\newblock Adv. Appl. Mech. 34, 171--302.

\bibitem[Rekik et al.(2007)]{Rekik07} Rekik, A., Auslender, F., Bornert,M., Zaoui, A., 2007.
\newblock Objective evaluation of linearization procedures in nonlinear homogenization: A methodology and some implications on the accuracy of micromechanical schemes.
\newblock Int. J. Solids Struct. 44, 3468–-3496.

\bibitem[Salen\c{c}on(1983)]{Salencon83} Salen\c{c}on, J., 1983.
\newblock \textit{Calcul \`{a} la rupture et analyse limite}.
\newblock Presses \'{E}cole Nationale des Ponts et Chauss\'{e}es, Paris.

\bibitem[Suquet(1981)]{Suquet81} Suquet, P., 1981.
\newblock Sur les \'{e}quations de la plasticit\'{e}: existence et r\'{e}gularit\'{e} des solutions.
\newblock J. M\'{e}canique 20, 3--39.

\bibitem[Suquet(1995)]{Suquet95} Suquet, P., 1995.
\newblock Overall properties of nonlinear composites~: a modified secant moduli
  theory and its link with Ponte Casta\~neda's nonlinear variational procedure.
\newblock C.R. Acad. Sc. Paris II  320, 563--571.

\bibitem[Taliercio(1992)]{Taliercio92} Taliercio, A., 1992.
\newblock Lower and upper bounds for the macroscopic strength domain of a fiber-reinforced composite material.
\newblock Int. J. Plasticity 8, 741--762.

\bibitem[Willot(2007)]{Willot07a} Willot, F., 2007.
\newblock Contribution \`{a} l'\'{e}tude th\'{e}orique de la localisation plastique dans les poreux.
\newblock Ph.D. thesis, \'{E}cole Polytechnique and Commissariat \`{a} l'\'{E}nergie Atomique, France.

\bibitem[Willot and Pellegrini(2008)]{Willot07} Willot, F., Pellegrini, Y. P., 2008.
\newblock Numerical FFT computations in 2D perfectly plastic pixelwise disordered porous media.
\newblock In Procs.\ 11$^\text{th}$ Int.\ Symp.\ On Continuum
Models and Discrete Systems (Jul.\ 30 - Aug.\ 3, Paris, 2007)
(\'Ecole des Mines, Paris, 2008), pp.\ 443--449. Also,
\texttt{arXiv:0802.2488v1}.

\bibitem[Willot et al.(2008a)]{Willot08a} Willot, F., Pellegrini, Y. P., Idiart, M. I., Ponte Casta\~neda, P., 2008.
\newblock Effective-medium theory for infinite-contrast two-dimensionally periodic linear composites with strongly anisotropic matrix behavior: dilute limit and cross-over behavior.
\newblock Phys. Rev. B 78, 104111.

\bibitem[Willot et al.(2008b)]{Willot08b} Willot, F., Pellegrini, Y. P., Ponte Casta\~neda, P., 2008.
\newblock Localization of elastic deformation in strongly anisotropic, porous, linear materials with periodic microstructures: exact solutions and dilute expansions.
\newblock J. Mech. Phys. Solids 56, 1245--1268.

\bibitem[Willis(1981)]{Willis81} Willis, J. R., 1981.
\newblock Variational and related methods for the overall properties of composites.
\newblock Adv. App. Mech. 21, 1--78.

\bibitem[Willis(2000)]{Willis00} Willis, J. R., 2000.
\newblock The overall response of nonlinear composite media.
\newblock Eur. J. Mech. A/Solids 19, S165--S184.

\end{thebibliography}

\end{document}